\documentclass[reprint,amsmath,amsfonts,amssymb,aps,prx,preprintnumbers,superscriptaddress]{revtex4-2}

\usepackage[utf8]{inputenc}
\usepackage[T1]{fontenc}
\usepackage{lmodern}
\usepackage{graphicx}
\usepackage{dcolumn}
\usepackage{bm}
\usepackage{textcomp}
\usepackage{ulem}
\usepackage{ifpdf}
\usepackage[squaren,Gray]{SIunits}
\usepackage{color}
\definecolor{red}{rgb}{1,0,0}
\definecolor{blue}{rgb}{0,0,1}
\definecolor{darkred}{rgb}{0.6,0,0}
\definecolor{darkblue}{rgb}{0,0,0.6}
\definecolor{darkgreen}{rgb}{0,0.5,0}
\definecolor{grey}{rgb}{0.7,0.5,0.8}

\ifpdf
\usepackage{epstopdf}
\usepackage[pdftex,unicode,pdfstartview={FitH},pdfborder={0 0 0}]{hyperref}
\usepackage{hypcap}
\else
\usepackage[hypertex]{hyperref}
\fi
\hypersetup{
    bookmarksnumbered = true,
    colorlinks = true, linkcolor = darkblue,
    citecolor = darkblue, filecolor = darkblue,
    menucolor = darkblue, urlcolor = darkblue
}

\newcolumntype{R}{>{$\displaystyle}r<{$}}
\newcolumntype{C}{>{$\displaystyle}c<{$}}

\newcommand{\Rh}{$R_h^h\phantom{.}$}
\newcommand{\RX}{$R_h^X\phantom{.}$}
\newcommand{\RM}{$R_h^M\phantom{.}$}
\newcommand{\Hh}{$H_h^h\phantom{.}$}
\newcommand{\HX}{$H_h^X\phantom{.}$}
\newcommand{\HM}{$H_h^M\phantom{.}$}

\begin{document}
    
\title{Excitons in mesoscopically reconstructed moir\'e heterostructures}

\author{Shen Zhao}
\affiliation{Fakult\"at f\"ur Physik, Munich Quantum Center, and Center for NanoScience (CeNS), Ludwig-Maximilians-Universit\"at M\"unchen, Geschwister-Scholl-Platz 1, 80539 M\"unchen, Germany}
\author{Zhijie Li}
\affiliation{Fakult\"at f\"ur Physik, Munich Quantum Center, and Center for NanoScience (CeNS), Ludwig-Maximilians-Universit\"at M\"unchen, Geschwister-Scholl-Platz 1, 80539 M\"unchen, Germany}
\author{Xin Huang}
\affiliation{Fakult\"at f\"ur Physik, Munich Quantum Center, and Center for NanoScience (CeNS), Ludwig-Maximilians-Universit\"at M\"unchen, Geschwister-Scholl-Platz 1, 80539 M\"unchen, Germany}
\affiliation{Present affiliations: Beijing National Laboratory for Condensed Matter Physics, Institute of Physics, Chinese Academy of Sciences, Beijing 100190, People's Republic of China \\ School of Physical Sciences, CAS Key Laboratory of Vacuum Physics, University of Chinese Academy of Sciences, Beijing 100190, People's Republic of China}
\author{Anna Rupp}
\affiliation{Fakult\"at f\"ur Physik, Munich Quantum Center, and Center for NanoScience (CeNS), Ludwig-Maximilians-Universit\"at M\"unchen, Geschwister-Scholl-Platz 1, 80539 M\"unchen, Germany}
\author{Jonas G{\"o}ser}
\affiliation{Fakult\"at f\"ur Physik, Munich Quantum Center, and Center for NanoScience (CeNS), Ludwig-Maximilians-Universit\"at M\"unchen, Geschwister-Scholl-Platz 1, 80539 M\"unchen, Germany}
\author{Ilia~A.~Vovk}
\affiliation{PhysNano Department, ITMO University, Saint Petersburg 197101, Russia}
\author{Stanislav~Yu.~Kruchinin}
\affiliation{Center for Computational Materials Sciences, Faculty of Physics, University of Vienna, Sensengasse 8/12, 1090 Vienna, Austria}
\affiliation{Nuance Communications Austria GmbH, Technologiestra\ss{} 8, 1120 Wien}
\author{Kenji Watanabe}
\affiliation{Research Center for Functional Materials, National Institute for Materials Science, 1-1 Namiki, Tsukuba 305-0044, Japan}
\author{Takashi Taniguchi}
\affiliation{International Center for Materials Nanoarchitectonics, National Institute for Materials Science, 1-1 Namiki, Tsukuba 305-0044, Japan}
\author{Ismail Bilgin}
\affiliation{Fakult\"at f\"ur Physik, Munich Quantum Center, and Center for NanoScience (CeNS), Ludwig-Maximilians-Universit\"at M\"unchen, Geschwister-Scholl-Platz 1, 80539 M\"unchen, Germany}
\author{Anvar~S.~Baimuratov}
\affiliation{Fakult\"at f\"ur Physik, Munich Quantum Center, and Center for NanoScience (CeNS), Ludwig-Maximilians-Universit\"at M\"unchen, Geschwister-Scholl-Platz 1, 80539 M\"unchen, Germany}
\author{Alexander H{\"o}gele}
\affiliation{Fakult\"at f\"ur Physik, Munich Quantum Center, and Center for NanoScience (CeNS), Ludwig-Maximilians-Universit\"at M\"unchen, Geschwister-Scholl-Platz 1, 80539 M\"unchen, Germany}
\affiliation{Munich Center for Quantum Science and Technology (MCQST), Schellingtra\ss{}e 4, 80799 M\"unchen, Germany}
       
\date{\today}

\begin{abstract}
Moir\'e effects in twisted or lattice-incommensurate vertical assemblies of two-dimensional crystals give rise to a new class of quantum materials with rich transport and optical phenomena, including correlated electron physics in flat bands of bilayer graphene \cite{Bistritzer2011} and moir\'e excitons in semiconductor heterostructures \cite{WuTopo2017,Yu2017,WuExciton2018}. These phenomena arise from modulations of interlayer interactions on the nanoscale of spatially varying atomic registries of moir\'e supercells \cite{Ponomarenko2013,Dean2013,Hunt2013,ZhangInterlayer2017}. Due to finite elasticity, however, lattices of marginally-twisted homobilayers and heterostructures can transform from moir\'e to periodically reconstructed patterns with triangular or hexagonal tiling \cite{Wijk2014,Woods2014,Wijk2015,Carr2018,Sunku2018,Yoo2019,Halbertal2021,Enaldiev2020,Weston2020,McGilly2020,Sung2020,Andersen2021,Rosenberger2020,Enaldiev2021,Shabani2021,Weston2022}. Here, we expand the notion of nanoscale lattice reconstruction to the mesoscopic scale of extended samples and demonstrate rich consequences in optical studies of excitons in MoSe$_2$-WSe$_2$ heterostructures with parallel and antiparallel alignment. Our results provide a unified perspective on diverse and partly controversial signatures of moir\'e excitons in semiconductor heterostructures \cite{Seyler2019,Tran2019,Jin2019,Alexeev2019,Bai2020,Tartakovskii2020} by identifying domains with exciton properties of distinct effective dimensionality and establish mesoscopic reconstruction as a compelling feature of real samples and devices with inherent finite-size effects and disorder. Generalized to stacks of other two-dimensional materials, this notion of mesoscale domain formation with emergent topological defects \cite{Alden2013} and percolation networks will instructively expand our understanding of fundamental electronic, optical, and magnetic properties of van der Waals heterostructures.
\end{abstract}
   
\maketitle

\clearpage

Vertical assemblies of twisted or lattice-mismatched heterobilayers of two-dimensional transition metal dichalcogenides (TMDs) with modulated van der Waals interlayer coupling in moir\'e superlattices give rise to correlated Hubbard model physics \cite{WuHubbard2018,Tang2020} with signatures of collective phases in transport \cite{Wang2020a,Huang2021,Ghiotto2021,Li-Mak2021} and optical experiments \cite{Tang2020,Regan2020,Shimazaki2020,Xu2020}. Periodic moir\'e interference patterns on the length scale of a few nanometers \cite{ZhangInterlayer2017} have profound effects on the electronic band structure via the formation of flat mini-bands which enhance many-body correlations and induce emergent magnetism \cite{Tang2020}, correlated insulating states \cite{Regan2020,Shimazaki2020,Wang2020a,Huang2021,Ghiotto2021,Xu2020} or Wigner crystals \cite{Regan2020} with periodic order in spatial charge distribution \cite{Li-Wang2021}. Moir\'e effects also result in rich optical signatures of intralayer \cite{ZhangPlochocka2018} and interlayer \cite{Seyler2019,Tran2019,Jin2019,Alexeev2019} excitons formed by Coulomb attraction among layer-locked and layer-separated electrons and holes, with angle-controlled exciton valley coherence and dynamics~\cite{Gong2013,Scuri2020,Andersen2021}, optical nonlinearities~\cite{LinNonlinear2021} or correlated excitonic insulating states \cite{Ma2021}.

Despite extensive optical studies of moir\'e interference effects in TMD heterobilayers as in MoSe$_2$-WSe$_2$ \cite{Wilson2021}, a consolidated picture of the rich and partly conflicting experimental features remains elusive \cite{Tartakovskii2020}. The experiments report inconsistent results for peak energies of interlayer exciton photoluminescence (PL) \cite{Joe2021,Ciarrocchi2019}, $g$-factors \cite{Ciarrocchi2019,Seyler2019}, and degree of polarization \cite{Seyler2019,Tran2019}, and even in the same sample the PL spectra can differ substantially from spot to spot \cite{Brotons-Gisbert2020}. The diversity of models invoked to explain the plethora of experimental signatures is not inherent to the theory of moir\'e excitons~\cite{WuTopo2017,Yu2017,WuExciton2018} but rather related to variations in actual real-world samples. For TMD bilayers with small twist angles away from parallel and antiparallel configurations, in particular, canonical moir\'e superlattices are known to transform into periodic domains of distinct atomic registries in triangular and hexagonal tiling \cite{Weston2020,Sung2020,Andersen2021,McGilly2020,Rosenberger2020,Shabani2021,Weston2022},  as dictated by the competition between intralayer strain and interlayer adhesion energies \cite{Carr2018,Enaldiev2020,Enaldiev2021}. However, correlative studies of exciton features and reconstructed domains are mainly limited to twisted MoSe$_2$ homobilayers \cite{Sung2020}, providing only limited insight into exciton landscapes of mesoscopically reconstructed moir\'e heterostructures.

\begin{figure*}[tb!]  
\includegraphics[scale=0.979]{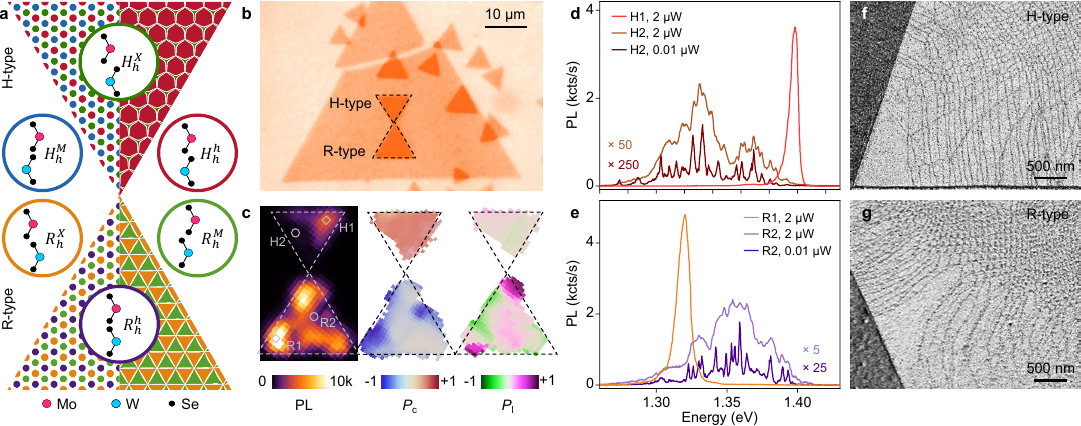}
\caption{\textbf{Characteristics of MoSe$_2$-WSe$_2$ HBLs in H- and R-type stacking.} \textbf{a}, Schematics of H and R HBLs (upper and lower triangles) with ideal moir\'e (left) and periodically reconstructed (right) patterns. The colored regions represent high-symmetry atomic registries illustrated in the respective circles. \textbf{b}, Optical micrograph of sample 1 with H and R HBLs (indicated by dashed lines) obtained as stacks of CVD-grown MoSe$_2$ (small triangles) and WSe$_2$ (large triangle). \textbf{c}, From left to right: maps of interlayer exciton PL, degrees of circular and linear polarization, $P_\mathrm{c}$ and $P_\mathrm{l}$, of the H and R HBLs in \textbf{b}. For each HBL, one bright and one dark spots are marked by diamond (H1,R1) and circle (H2,R2), respectively, in the PL map. \textbf{d,e}, PL spectra at the bright and dark spots marked in \textbf{c}. At a excitation power of $2~\mu$W, the spectra of H1 and R1 are representative for regions with a single bright peak, whereas the spectra of H2 and R2 (scaled by $50$ and $5$) are characteristics of dark regions with broad and structured PL which evolves into narrow peaks at low excitation with $0.01~\mu$W (scaled by $250$ and $25$). All spectroscopy data were recorded on sample 1. \textbf{f,g}, Scanning electron micrographs of H and R HBLs recorded with secondary electron imaging.}
\label{fig1}
\end{figure*}

Here, we demonstrate theoretically and experimentally how atomic reconstruction in extended heterobilayer samples with inhomogeneous strain and finite-size boundary conditions impacts the optical signatures of excitons. We begin by considering two high-symmetry configurations of heterobilayers (HBLs) assembled as vertical stacks of distinct TMD monolayers (MLs), each without inversion center, with parallel alignment at strictly $0$° twist in rhombohedral R-type stacking and antiparallel alignment at $180$° twist in hexagonal H-type stacking. Due to the lattice mismatch of the ML crystals, both R and H heterostacks form a two-dimensional moir\'e interference pattern, with a superlattice constant of $\sim 100$~nm in aligned MoSe$_2$-WSe$_2$ HBL, which reduces asymptotically to the ML lattice constant with increasing twist angle away from high-symmetry configurations. Upon lateral translation through a moir\'e supercell, each stacking modulates through points of highly symmetric atomic registries ($H_h^h$, $H_h^M$, $H_h^X$ and $R_h^h$, $R_h^M$, $R_h^X$) shown schematically in the colored circles of Fig.~\ref{fig1}a for a vertical heterostack of top MoSe$_2$ and bottom WSe$_2$ MLs. 

In the limit of rigid lattices, the resulting ideal moir\'e pattern is shown as the left side of H (top) and R (bottom) triangles. The positions of high-symmetry registries with identical areas are represented by their respective colors, spanning a periodic lattice of laterally alternating registries with gradual interconversion according to the geometrical interference condition. In the presence of local lattice deformation, however, the ideal moir\'e pattern transforms into periodically reconstructed patterns shown in the right halves of the H and R triangles in Fig.~\ref{fig1}a. Driven by competition between intralayer strain and interlayer adhesion energy, energetically favored registries expand at the expense of unfavorable registries into periodic domains with areas inversely proportional to the twist angle $\theta$. Theoretical and experimental results~\cite{Carr2018,Weston2020} indicate for H-type HBLs that only the \Hh stacking prevails in hexagonal domains after reconstruction, whereas \RX and \RM consolidate in tessellated triangular domains as two equally optimal registries in R-type heterostacks.

\begin{table*}[t!]
\caption{\textbf{Theoretical parameters of interlayer excitons in distinct atomic registries of MoSe$_2$-WSe$_2$ HBLs.} Transition energy (in eV, estimated from Ref.~\cite{Enaldiev2021}), oscillator strength (proportional to the square of dipole moment $|\boldsymbol{\mu}|$ in Debye), polarization selection rule and $g$-factor for zero-momentum $KK$ or $K'K$ interlayer excitons in R- and H-stacking in spin-singlet (electron and hole with antiparallel spin) and spin-triplet (electron and hole with parallel spin) configurations. Note that singlet (triplet) interlayer excitons are lowest-energy states in R (H).}
    \begin{ruledtabular}
        \begin{tabular}{lRRRRRRRR}
            & \multicolumn{4}{c}{Singlet} & \multicolumn{4}{c}{Triplet}\\
            \cline{2-5} \cline{6-9}
            Stacking & \text{Energy}  & |\boldsymbol{\mu}| & \text{Polarization} & g\text{-factor}& \text{Energy} & |\boldsymbol{\mu}| & \text{Polarization} & g\text{-factor}  \\
            \hline
            \RX (optimal) & 1.33  & 1.47 & \sigma_-     &  +5.8 & 1.35  & 0.70 & \sigma_+ & -10.5    \\
            \Rh           & 1.38  & 2.06 & \sigma_+     &  -6.4  & 1.40 & 0.19  & z\phantom{+}  &  11.0    \\
            \RM (optimal) & 1.48  & 1.09 & z\phantom{+} &   6.3 & 1.50 & 0.02  & \sigma_- & +10.9    \\
            \hline
            \HM           & 1.39 & 0.40 & z\phantom{+}          &  13.1  & 1.37 & 0.10  & \sigma_- & +17.6     \\
            \HX           & 1.41 & 0.69 & \sigma_+ & -13.2 & 1.39  & 0.40  & z\phantom{+} &  17.7   \\
            \Hh (optimal) & 1.42 & 2.14 &  \sigma_- & +12.9 & 1.40 & 0.42& \sigma_+ & -17.6
        \end{tabular}
    \end{ruledtabular}
    \label{tab_gfactor}
\end{table*}

The two scenarios of ideal and reconstructed moir\'e landscapes of Fig.~\ref{fig1}a can be in principle discerned with optical spectroscopy despite the length scale mismatch between the optical spot size of a few hundred nm and the domain dimensions well below $100$~nm. Utilizing distinct PL characteristics of interlayer excitons in atomic registries of MoSe$_2$-WSe$_2$ HBLs summarized in Table~\ref{tab_gfactor}, the observer could infer the contribution from each domain present in the optical spot. By virtue of different spin-valley configurations, stacking symmetries and related degrees of interlayer coupling, R and H interlayer excitons exhibit distinct transition energies~\cite{Yu2017,WuExciton2018,Gillen2018,Forg2021,Enaldiev2021}, oscillator strengths~\cite{Gillen2018,Forg2021} and dipolar selection rules~\cite{Yu2017,WuExciton2018,Forg2019} accessible with optical spectroscopy. Moreover, magneto-luminescence experiments allow the assignment of interlayer exciton PL to domains  of specific registries using first-principles calculations of exciton Land\'e $g$-factors~\cite{Wozniak2020,Forg2021,Xuan2020} and experimental values with sample-to-sample variations in the order of $10 - 20\%$.

Despite this conceptual clarity, a unified understanding of moir\'e excitons in MoSe$_2$-WSe$_2$ HBLs represents a challenge, which we address by studying laterally extended samples with complementary optical spectroscopy techniques. The HBL samples were fabricated from MoSe$_2$ and WSe$_2$ MLs exfoliated from native crystals or synthesized by chemical vapor deposition (CVD) (see the Methods section for fabrication details). For the sample in Fig.~\ref{fig1}b based on CVD-grown MLs, we placed single-crystal MoSe$_2$ triangles on top of a large WSe$_2$ triangular ML by standard dry-transfer. The resulting HBL triangles with small-twist H and R stacking configurations are delimited by dashed lines in the optical micrograph of Fig.~\ref{fig1}b. To access narrow exciton linewidths in cryogenic spectroscopy, all HBLs were encapsulated in hexagonal boron nitride (hBN)~\cite{Cadiz2017}.

\begin{figure*}[ht!]
\includegraphics[scale=1.04]{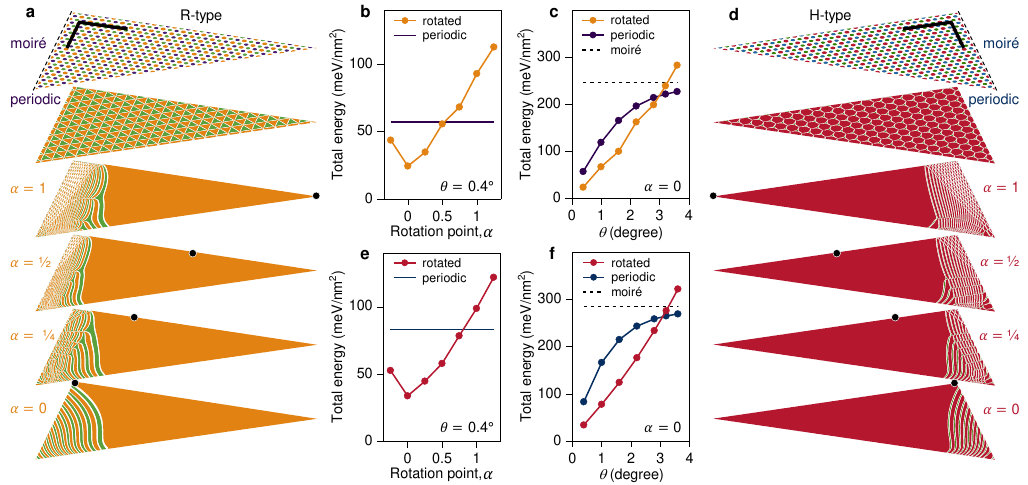}
\caption{\textbf{Mesoscopic reconstruction in finite-size simulations.} \textbf{a}, and \textbf{d}, Maps of reconstructed domains in triangular tips of R and H HBLs with a twist angle $\theta=0.4^\circ$ (only triangle halves are shown in the projections; the scale bars are $200$~nm). The topmost maps show moir\'e patterns without reconstruction, delimited by dashed lines from the moir\'e core of the HBL. The maps below show periodic reconstruction and mesoscopically reconstructed domain patterns obtained for different zero-twist deformations around points marked by black dots at dimensionless positions $\alpha$ (note that orange $R_h^X$ and green $R_h^M$ domains can interconvert because of same adhesion energy~\cite{Enaldiev2020}). \textbf{b} and \textbf{e}, Total areal energy for R and H at $\theta = 0.4^\circ$ and different untwisting points $\alpha$ (the energy of the respective periodic patterns is shown by solid lines). \textbf{c} and \textbf{f}, Total areal energy of periodic and optimally reconstructed (for $\alpha = 0$) patterns for different twist angles $\theta$ in R and H (the energy of the respective moir\'e patterns is shown by dashed lines).}
\label{fig2}
\end{figure*}

Figure~\ref{fig1}c shows the characteristics of cryogenic PL recorded in the spectral bandwidth of interlayer excitons. The laterally extended maps recorded at $3.2$~K show from left to right the integrated PL intensity and the degrees of its circular and linear polarizations, $P_\mathrm{c}$ and $P_\mathrm{\,l}$ \cite{Neumann2017}. The PL map exhibits sizable intensity variations on micrometer scales across the H and R stacking regions, with an overall much brighter emission from R stacking, consistent with the reversed energetic ordering of singlet and triplet interlayer exciton states sizable variations in the PL intensity across a given stacking, on the other hand, could be attributed to sample imperfections yielding dark spots as regions of suppressed layer interactions. This scenario, however,  conflicts with absorption characteristics of intralayer excitons in the regions of low PL with multi-peak resonances reminiscent of layer hybridization \cite{Jin2019,Alexeev2019}. Thus, sample imperfections alone are insufficient to explain the pronounced inhomogeneities in the PL map of Fig.~\ref{fig1}c.  

The variations in PL intensity upon lateral displacement of the observation spot by a few microns are accompanied by changes in the spectral characteristics shown representatively in Fig.~\ref{fig1}d,e. In the upper right and lower left bright corners of H- and R-type triangles marked by diamonds (H1 and R1), the respective PL spectra feature only one peak at $1.40$ and $1.33$~eV, with highest degrees of circular polarization and opposite signs in the $P_\mathrm{c}$ map of Fig.~\ref{fig1}c. These spectral signatures jointly suggest that both H and R triangles exhibit areas that can be at least as large as the optical spot yet entirely dominated by the respective triplet and singlet interlayer excitons in \Hh and \RX registries. Given the finite twist angle in our sample, the absence of PL contributions from all other registries is striking. 

The signatures of bright spots are contrasted on sample positions with low PL, labelled by circles in the PL map (H2 and R2) and shown as brown and purple traces in Fig.~\ref{fig1}d,e. At the expense of simple single-peak spectra, the PL is structured and spectrally dispersed over $100$~meV on the low and high energy sides of the solitary peaks of triplet \Hh and singlet \RX interlayer excitons. On dark spots and under identical excitation conditions, the integrated PL is typically much lower (note the scaling factor of $50$ and $5$ for H- and R-type spectra), and $P_\mathrm{c}$ is reduced in its absolute value though preserved in sign. These characteristics, indicative of moir\'e effects \cite{Tran2019,Brotons-Gisbert2020,Forg2021}, evolve into a series of spectrally narrow peaks at reduced excitation powers, signifying interlayer exciton localization in moir\'e quantum dots \cite{Seyler2019,Brotons-Gisbert2020}. This scenario of energy-reducing moir\'e pockets in H-type HBL is in stark conflict with quantum-dot features on the high energy side of $R_h^X$.

Additional confusion arises from the observation of finite degree of linear polarization with variations across the $P_\mathrm{\,l}$ map of Fig.~\ref{fig1}c. According to Table~\ref{tab_gfactor}, the dipolar selection rules of spin-valley specific states of $KK$ interlayer excitons dictate valley contrasting circularly polarized optical transitions perpendicular to the HBL plane as well as $z$-polarized in-plane transitions \cite{Yu2017,WuExciton2018,Forg2019}. Whereas the former constitute predominantly positive and negative $P_\mathrm{c}$ values of H- and R-type stackings in the map of Fig.~\ref{fig1}c, the latter should exhibit neither circular nor linear degrees of polarization when probed in back-scattering configuration \cite{Forg2021}. In contrast to this expectation, the $P_\mathrm{\,l}$ map of Fig.~\ref{fig1}c features regions with very high degrees of linear polarization (note the upper and lower left corner of the R-type triangle), reminiscent of uniaxially strained moir\'e landscapes~\cite{Bai2020} or HBL samples with transfer-induced layer corrugation \cite{Alexeev2020}. This observation implies the presence of moir\'e quantum wires in our sample.

All the above features manifest consistently across all samples of our studies (see Supplementary Notes~1 and 3 for other samples). The key to the understanding of such drastic spatial variations in the PL characteristics of interlayer excitons is provided by mesoscopic reconstruction. To visualize the phenomena of micron-scale reconstruction in our samples, we performed secondary electron imaging in scanning electron microscopy (SEM)~\cite{Andersen2021}. The respective SEM images of H- and R-stacked HBL near the triangle tip and edge with stacking-sensitive contrast (Fig.~\ref{fig1}f,g and see also Supplementary Note~1) provide evidence for the formation of large domains of one atomic registry separated by thin lines of domain walls. The domain networks observed in different samples exhibit common patterns on the mesoscopic scale: large, micron-sized two-dimensional (2D) domains at HBL tips and edges are surrounded by elongated one-dimensional (1D) stripes that merge in the sample core into a network of finely structured domains with dimensions well below $100$~nm, forming quasi zero-dimensional (0D) arrays with hexagonal and triangular tiling for H and R stackings.

The mesoscopic reconstruction is driven by the interplay of intralayer strain and interlayer adhesion energy. Recent theoretical work \cite{Carr2018,Enaldiev2020,Enaldiev2021} pointed out that moir\'e lattices of marginally twisted or incommensurate bilayers relax on the nanoscale into periodic domains by rearranging lattice atoms in each layer according to a vectorial 2D displacement field that overcompensates the associated strain cost by the gain in interlayer adhesion. To provide intuition for the domain formation on scales ranging from a few nanometers to a few microns, we adopt the theoretical model of lattice reconstruction \cite{Enaldiev2020} and account for finite-size effects as well as singular point rotations that can grow to large-area 2D domains of optimal stacking. In brief (for details see Supplementary Note~2), we model the tip of a marginally twisted HBL by an equilateral triangle with its micron-sized base connecting to the rest of the HBL triangle with ideal moir\'e periodicity. On the line bisecting the triangular tip into two equal halves, we place a point of zero-twist deformation to create an initial lattice displacement field which is modified in consecutive iterations to obtain a stacking configuration from the final displacement field that minimizes the sum of intralayer strain and interlayer adhesion energies in the tip area. This procedure yields series of reconstructed landscapes, characterized by their respective initial displacement fields.

The results of our numerical simulations are shown in Fig.~\ref{fig2}a,d for R- and H-type HBLs with a twist of $\theta = 0.4^\circ$. The two topmost maps illustrate the ideal moir\'e and periodically reconstructed patterns. The four maps below show reconstruction patterns obtained after optimization of initial displacement fields that untwist the HBL around a rotation center indicated by black points and labelled by a dimensionless coordinate $\alpha = 1,0.5,0.25,0$. In all cases, optimization yields mesoscopic reconstruction into large 2D domains of energetically favored stackings at the triangle tip (\RX or \RM and \Hh in R- and H-type HBLs) with characteristic domain sizes on the micron scale and orders of magnitude larger areas than in periodically reconstructed lattices. The extended domains are flanked by 1D stripes which merge into 0D domain arrays. These different domain types and length scales predicted by our model comply with the global reconstruction picture observed experimentally in SEM images as in Fig.~\ref{fig1}f,g, suggesting that mesoscale domain formation is most favorable at sample edges.

\begin{figure*}[ht!]
\includegraphics[scale=1]{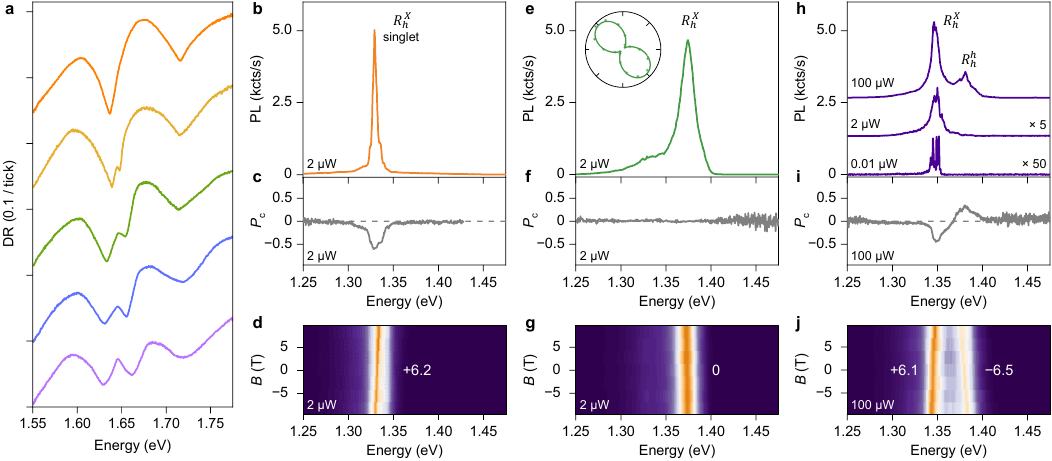}
\caption{\textbf{Spectral characteristics of excitons in reconstructed R-type MoSe$_2$-WSe$_2$ heterobilayers.} \textbf{a}, Evolution of differential reflection (DR) spectra of intralayer excitons upon gradual displacement from a bright to a dark region in R-type HBL (shown from top to bottom). The peak multiplicity is a hallmark of nanoscale reconstructed domains. The positions of spectra are indicated by black dots in Supplementary Fig.~9a. \textbf{b-d}, Interlayer exciton PL (\textbf{b}), degree of circular polarization $P_\mathrm{c}$ (\textbf{c}), and dispersion in perpendicular magnetic field $B$ (\textbf{d}) characteristics of bright sample regions with low degree of linear polarization, corresponding to large 2D domains. The magneto-luminescence data were recorded under linearly polarized excitation with $\sigma_+$ detection to determine the $g$-factor values from linear slopes. \textbf{e-g}, Same for regions with a large degree of linear polarization (as shown in the inset), corresponding to regions with 1D stripes. \textbf{h-j}, Spectral characteristics in a dark sample region with 0D domains (shown with offsets and different scaling for excitation powers of $100$, $2$ and $0.01~\mu$W. All data were recorded on sample 2; $g$-factor values with least-square error bars were obtained from linear fits to the data shown in Supplementary Fig.~16.}
\label{fig3}
\end{figure*}

As expected, optimality reconstructed patterns minimize the total energy of the system according to our simulation results shown in Fig.~\ref{fig2}b,c for R-type and Fig.~\ref{fig2}e,f for H-type HBLs. For $\theta = 0.4^\circ$, the total energy (normalized by the triangle area) in both stackings is reduced by a factor of $10$ and $2$ below the energies of ideal moir\'e and periodic limits (solid lines in Fig.~\ref{fig2}b,e) upon reconstruction after optimal rotation at $\alpha=0$. The global energy minimum, obtained for the rotation point at the borderline between the HBL moir\'e core and the triangle base, features the most direct transition from 2D domain through 1D stripes to the core. As the rotation point is moved towards the tip (via $0<\alpha<1$), 0D regions emerge at the border and the energy gain decreases, passing the threshold of periodic reconstruction at $\alpha \simeq 0.5$ ($0.75$) in R (H) twisted HBL but remaining well below the moir\'e lattice energy throughout. Remarkably, for the optimally relaxed tip around $\alpha=0$, mesoscopic reconstruction remains energetically favorable for twist angles up to $\theta = 3^\circ$ for both stackings (Fig.~\ref{fig2}c,f), in accord with the recent Raman spectroscopy results~\cite{Holler2020}. 

With this intuition for mesoscopic reconstruction from theory and SEM experiment, we revisit the main spectroscopic features of distinct sample regions in R- and H-stacking to provide a unifying interpretation of diverse optical signatures observed in MoSe$_2$-WSe$_2$ HBLs. The set of data in Fig.~\ref{fig3} shows the characteristics of R-type HBL observed in the spectral region of intralayer excitons with DR and interlayer excitons with PL and magneto-luminescence. The DR spectra in Fig.~\ref{fig3}a probe the evolution of intralayer exciton absorption as the observation spot is displaced from bright to dark PL areas. Upon displacement, the DR resonances of intralayer excitons at $1.64$ and $1.71$~eV on bright spots (top spectrum) gradually develop into two well-resolved resonances and a broadening around MoSe$_2$ and WSe$_2$ intralayer exciton transitions which become most pronounced in the darkest areas (bottom spectrum). This evolution reflects the presence of only one registry in bright spots and modulation of the intralayer exciton energy by alternating stackings in dark-most spots. This splitting of intralayer exciton transitions is not to be confused with band hybridization effects in homobilayers \cite{Sung2020,Andersen2021} or heterobilayers with near-resonant conduction \cite{Alexeev2019} or valence \cite{Jin2019} bands. As resonant hybridization in MoSe$_2$-WSe$_2$ heterostacks is strongly inhibited by large band offsets, multi-peak DR spectra result from different intralayer exciton energies within the different registries present in the optical spot, convincingly assigning sample areas with dark PL to areas of 0D arrays and not to regions of local layer-separating imperfections such as bubbles or folds.

Bearing in mind the absence of peak multiplicity in the DR spectra of bright spots, the corresponding characteristics of interlayer exciton PL (Fig.~\ref{fig3}b-d) are straightforwardly explained. Due to complete local reconstruction into domains exceeding the optical spot size, only lowest-energy spin-singlet excitons of \RX registry with sizable oscillator strength contribute to PL (top panel) with a single peak at $1.33$~eV and full-width at half-maximum (FWHM) linewidth of $6$~meV, negative $P_\mathrm{c}$ (central panel) and a positive $g$-factor of $\sim 6$ (bottom panel) in agreement with previous report for aligned HBLs~\cite{Joe2021}. Such characteristics are frequently observed at sample edges and tips (as in Fig.~\ref{fig1}c) where large-scale reconstruction is energetically most favorable (as in reconstructed maps of Fig.~\ref{fig2}a).

The PL emission from spatially neighboring regions exhibits blue-shifted emission around $1.36$~eV paired with a high degree of linear polarization (as signified by the inset in the top panel of Fig.~\ref{fig3}e) and vanishing $P_\mathrm{c}$ as well as $g$-factor values (Fig.~\ref{fig3}f,g), which are indicatives of quantum wires~\cite{Bai2020}. As suggested by our theory, these quantum wires are formed by alternating optically bright \RX and dark \RM domains. This 1D confinement not only breaks the threefold rotational symmetry of the exciton wave functions, thereby admixing $K$ and $K'$ valleys and obliterating both $P_\mathrm{c}$ and $g$-factors in perpendicular magnetic field, it is also responsible for the blue-shift in the PL energy as a consequence of quantum confinement of interlayer exciton states in stripes of lower-energy \RX domains flanked by potential walls of higher-energy \RM states. Spot-to-spot variations in the PL characteristics of sample regions with high $P_\mathrm{\,l}$ and varying orientations of the respective polarization axes are consistent with the diversity in stripe geometries observed both in SEM and our mesoscale reconstruction model. A prominent example is the left corner of the R-type triangle in Fig.~\ref{fig1}c where the bright spot of a large \RX domain with $P_\mathrm{c}\simeq -1$ is encompassed by regions with $P_\mathrm{\,l} \simeq \pm1$ stemming from quantum wire ensembles with nearly orthogonal orientation.  

Quantum confinement is also prominent in R-type regions of 0D arrays (Fig.~\ref{fig3}h) with much reduced PL intensity and spectrally narrow lines of quantum dots at low excitation power (note the scaling factors of $5$ and $50$ for the spectra at $2$ and $0.01~\mu$W excitation powers) with characteristic negative $P_\mathrm{c}$ and positive $g$-factors of $6$ \cite{Seyler2019}. Depending on the actual sample spot of dark PL, such quantum dot lines with a FWHM well below $1$~meV can be observed within spectrally narrow or broad windows of $10$ and $100$~meV (as in Fig.~\ref{fig3}h and Fig.~\ref{fig1}e, respectively) energetically above the single peak of extended \RX domains at $1.33$~eV. These blue-shifts are in stark contrast to the common notion of exciton localization in energy-reducing moir\'e potentials \cite{Seyler2019,Tran2019}. In fact, regions of 0D reconstructed arrays increase the energy of interlayer excitons by quantum confinement in triangular boxes of \RX stacking with potential barriers formed by adjacent \RM~triangles. Unlike in ideal moir\'e patterns, the potential boundaries in periodically reconstructed R-type arrays are steep, and the variations in the quantum dot PL energies reflect different strengths of confinement in quantum boxes of varying size in the triangular tiling.

\begin{figure*}[ht!]
\includegraphics[scale=1]{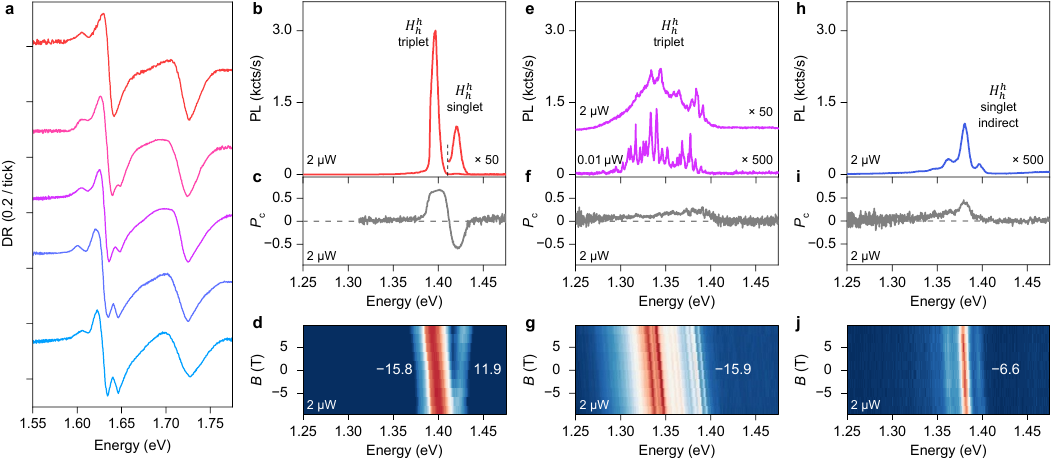}
\caption{\textbf{Spectral characteristics of excitons in reconstructed H-type MoSe$_2$-WSe$_2$ heterobilayers.} \textbf{a}, Evolution of differential reflection (DR) spectra of intralayer excitons upon displacement from a bright to a dark region in H-type HBL (shown from top to bottom). As in the R-type case, the peak multiplicity signifies regions of reconstructed nano-domains. The positions of spectra are indicated by red dots in Supplementary Fig.~10a. \textbf{b-j}, Interlayer exciton PL (\textbf{b,e,h}), degree of circular polarization $P_\mathrm{c}$ (\textbf{c,f,i}) and dispersion in perpendicular magnetic field $B$ (\textbf{d,g,j}) for three representative positions. The magneto-luminescence data were recorded under linearly polarized excitation with $\sigma_+$ detection to determine the $g$-factor values from linear slopes. Spots with bright PL as in \textbf{b} feature triplet and singlet peaks with opposite $P_\mathrm{c}$ signs and characteristic $g$-factors of about $-16$ and $12$. Sample positions with low PL intensity exhibit structured spectra as in \textbf{e} and \textbf{h} with reduced $P_\mathrm{c}$. Their characteristics differ both in spectral profiles and $g$-factors. Data in \textbf{e-g} are from sample 2, all other data are from sample 3; $g$-factor values with least-square error bars were obtained from linear fits to the data shown in Supplementary Fig.~17.}
\label{fig4}
\end{figure*}

Since the optical spot samples a large number of non-identically reconstructed arrays, the PL from quantum-confined interlayer excitons is spectrally dispersed and merges at higher excitation powers (spectra at $2~\mu$W in Fig.~\ref{fig3}h and Fig.~\ref{fig1}e) into structured PL peaks of sub-ensembles grouped by similar characteristic length scales. Periodically reconstructed 0D arrays with lateral homogeneity on the scale of the optical spot give rise to a narrow ensemble of similar emission energies (within $10$~meV as in Fig.~\ref{fig3}h) which allow to observe a hot luminescence peak with positive $P_c$ and negative $g$-factor of $-6.5$ (Fig.~\ref{fig3}i,j at $100~\mu$W excitation power). These features, absent in the spectra of 1D and 2D domains, correspond to spin-singlet states of interlayer excitons in \Rh staking $\sim 50$~meV above \RX states with similar oscillator strength. Such a prominent contribution of \Rh stacking to the PL is surprising as theoretical considerations predict vanishingly small point-like areas for this non-optimal stacking~\cite{Carr2018,Enaldiev2020,Enaldiev2021}. Recent experiments observed related PL features of \Rh energetically below \RX exciton emission in MoSe$_2$-WSe$_2$ HBLs with a twist angle of $4^\circ$~\cite{Forg2021} where piezo- and ferroelectric effects are predicted to swap the energetic ordering of \Rh and \RX exciton states \cite{Enaldiev2021}. However, in all present considerations of reconstruction \cite{Carr2018,Enaldiev2020,Enaldiev2021}, \Rh domains are predicted to exhibit vanishingly small areas, such that sizable PL with \Rh characteristics from small-twist MoSe$_2$-WSe$_2$ HBLs implies either areas of \Rh domains that are larger than anticipated from theory, or strongly imbalanced exciton population feeding of \Rh and \RX states via polarization-contrasting relaxation pathways.

Despite many similarities, the picture of mesoscopic reconstruction in H-type MoSe$_2$-WSe$_2$ HBLs and the consequences for local optical properties is different and thus requires a dedicated consideration. The DR spectra of intralayer exciton transitions in bright and dark PL regions of H-type HBL shown in Fig.~\ref{fig4}a (note that the feature at $1.61$~eV is due to residual doping in this sample) lead to the same conclusion as for R-type spectra of Fig.~\ref{fig3}a: multi-peak DR resonances of intralayer excitons in dark PL regions are absent in regions of bright PL. For the latter, the atomic registry of large-scale reconstructed 2D domains is \Hh, unrivaled by other registries in terms of energy-optimizing stacking \cite{Carr2018,Enaldiev2020,Rosenberger2020}. Thus, the respective PL spectrum of interlayer excitons (Fig.~\ref{fig4}b) is rather simple, featuring only the emission peaks of spin-triplet and spin-singlet configurations with respective positive and negative $P_\mathrm{c}$ values (Fig.~\ref{fig4}c) and $g$-factors of $-16$ and $12$ (Fig.~\ref{fig4}d) \cite{Zhang2019,Wang2020,Joe2021}. In agreement with our reconstruction model, such characteristics were observed in the top-right corner of the H-type triangle in Fig.~\ref{fig1}c and at HBL edges of other samples prone to reconstruction. Note that the double-peak PL structure of \Hh reconstructed domains has no counterpart in the \RX case (Fig.~\ref{fig3}b) due to reversed energetic ordering of singlet and triplet states: despite a factor of $25$ smaller oscillator strength estimated from Table~\ref{tab_gfactor}, the PL from lower-energy spin-triplet exciton transition in \Hh is two orders of magnitude brighter than the hot-luminescence from spin-singlet state due to thermal population imbalance dictated by the spin-orbit splitting of $25$~meV.

As for R-type HBLs, the DR spectra exhibit an increasingly pronounced splitting of the MoSe$_2$ intralayer exciton resonance in dark regions of H-type samples (bottom spectra in Fig.~\ref{fig4}a). In the respective PL spectra (Fig.~\ref{fig4}e) we again observe for low and moderate excitation powers ($0.01$ and $2~\mu$W) that sharp quantum dot peaks develop into a multi-peak ensemble spectrum characteristic of 0D domain arrays with an inhomogeneous size distribution of hexagonal domains in optimal \Hh stacking illustrated in Fig.~\ref{fig1}a. The positive degree of circular polarization (Fig.~\ref{fig4}f) and the $g$-factor values (Fig.~\ref{fig4}g) confirm the \Hh stacking as the origin of the quantum-dot PL. For this stacking, theoretical estimates (Table~\ref{tab_gfactor}) place its optically bright interlayer exciton states at the top of the energy hierarchy, well above optically dark \HM and \HX states. 

This energetic ordering of interlayer excitons in H stacking is responsible for the absence of PL characteristics ascribed to stripes in R-type HBLs: Excitons in reconstructed \Hh domains are not bound by potential barriers that would mix $K$ and $K'$ states as in 1D quantum wires of R-type samples (note the absence of areas with high degree of in-plane linear polarization throughout the $P_\mathrm{l}$ map of H-type triangles in Fig.~\ref{fig1}c). This is contrasted by reports of quantum wire signatures in H-type stacks subjected to sizable uniaxial tensile strain \cite{Bai2020} which could give rise to reversal of energetic ordering of interlayer excitons in different atomic registries of H-type stacking, facilitating quantum wire formation similar to R-type heterostructures. In the absence of such strain-induced effects, our theory predicts that reconstructed \Hh domains preserve exciton population on spatially extended optically bright plateaus, whereas in regions of 0D arrays the PL intensity decreases (note the scaling factors of $50$ and $500$ in Fig.~\ref{fig4}e) due to population drain into lower-energy valleys of the surrounding \HM and \HX domain network with much reduced optical activity of vanishingly small domain areas. Consistently, nanoscale domain formation in regions of 0D arrays is accompanied by size-dependent PL red-shifts of up to $100$~meV below the \Hh triplet peak at $1.4$~eV. 

Direct evidence for all these features and conclusions are provided by one-to-one correlations between spectral characteristics and the underlying sample morphology visualized by secondary electron imaging (see Supplementary Note 5).

For completeness, we show in Fig.~\ref{fig4}h-j the PL characteristics of a dark area in an H-type sample of about $3^\circ$ twist. The heterostack is not entirely prone to reconstruction, with tips and edges exhibiting signatures of bright \Hh domains (see Supplementary Note~3). However, the dark areas of the sample exhibit very different features than reconstructed 0D arrays, implying that the HBL largely maintains its moir\'e structure. As shown in Fig.~\ref{fig4}h, the PL at $2~\mu$W excitation power is reduced by another factor of $10$ (note the scaling factor of $500$), with positive $P_\mathrm{c}$ (Fig.~\ref{fig4}i) and negative $g$-factor of $-6.6$ (Fig.~\ref{fig4}j) that has no counterpart in the realm of zero-momentum $K'K$ interlayer excitons of H-type registries yet is characteristic of finite-momentum $KK$ exciton states (see Supplementary Table~2). The peaks with equidistant energy spacing of $16$~meV up to $6$th order (see Supplementary Note~3) are indicative of interlayer exciton-polaron formation \cite{Semina2020}. In this regime, the heterostructure with layer and valley-separated excitons dressed by strong exciton-phonon coupling is momentum-dark, with population decay of exciton-polaron states mediated by a series of phonon replica in luminescence.  

Our insight into mesoscopic reconstruction in MoSe$_2$-WSe$_2$ HBLs identifies coexisting domains of different dimensionality with distinct exciton regimes. Extended HBLs with small twist exhibit brightly luminescent micron-sized 2D domains of energetically favorable \RX and \Hh registries which occur mainly at sample edges or line defects and feature signatures of 2D singlet and triplet interlayer excitons. Stripes of 1D domains connect 2D domain areas to arrays of nanometer-sized 0D domains characterized by split intralayer exciton resonances (see Supplementary Note~6) and spectrally narrow lines of interlayer excitons. In contrast, HBLs with relatively large twist angles locked in moir\'e patterns, have received only limited attention in optical spectroscopy due to the dark nature of exciton-polarons. The broad evidence for mesoscopic reconstruction in bilayer graphene \cite{Alden2013,Sunku2018,Halbertal2021,McGilly2020}
hBN \cite{Vizner2021,Woods2021}, CrI$_3$ ferromagnets \cite{Tiancheng2021,Xie2022} and TMD semiconductors \cite{Weston2020,Sung2020,Rosenberger2020,Andersen2021,Shabani2021,Weston2022} suggests that the phenomenon is universal across samples and devices of layered van der Waals heterostructures. Ultimately, both understanding and control of mesoscale reconstruction will facilitate further developments of van der Waals quantum materials with tailored electronic, optical and magnetic properties.

\section*{Methods}
\textbf{Sample fabrication:} MLs of MoSe$_2$ and WSe$_2$ were either mechanically exfoliated from bulk crystals (HQ Graphene) or obtained from chemical vapor deposition (CVD) synthesis. Thin flakes of hBN were exfoliated from bulk crystals (NIMS). Fully hBN encapsulated MoSe$_2$-WSe$_2$ heterobilayers were prepared with PC/PDMS stamp by dry-transfer~\cite{Pizzocchero2016}. First a layer of hBN was picked up, followed by the MoSe$_2$ and WSe$_2$ MLs and a capping layer of hBN. The pick-up temperatures for hBN flakes, WSe$_2$ and MoSe$_2$ MLs was around 50~°C, 130~°C, and 100~°C, respectively. The MLs were aligned to 0° (R-type) or 60° (H-type) by selecting adjacent straight edges with angles of either 60° or 120°. The precision of alignment was limited to below 1°. The heterostacks were finally released onto 300~nm SiO$_2$/Si substrates at a temperature of 180~°C. To avoid thermally activated rotation of TMD layers, no thermal annealing was performed unless specified otherwise.
    
For direct comparison of R- and H-stackings, two MoSe$_2$-WSe$_2$ samples with both alignment configurations were fabricated. One sample was stacked from CVD-grown triangular single-layer crystals with two MoSe$_2$ MLs of opposite orientation placed onto a large WSe$_2$ ML to form two HBL regions of R- and H-type. It has been shown that zigzag edges are predominant in CVD-grown triangular TMD MLs~\cite{Zhu2017}. Therefore, in the sample of Fig.~\ref{fig1}b, the regions with parallel and anti-parallel edges of MoSe$_2$ ML triangles to the edge of the large WSe$_2$ triangle are R- and H-type, respectively. Other samples were obtained from mechanical exfoliation-stacking by the tear-and-stack method~\cite{Kim2016} using a large WSe$_2$ ML to pick up a part of a MoSe$_2$ ML in a first step and the remaining part in a second step after a rotation by 60°.

\textbf{Optical spectroscopy:} Cryogenic PL and DR measurements were conducted using a home-built confocal microscope in back-scattering geometry. The samples were loaded into a closed-cycle cryostat (attocube systems, attoDRY1000) with a base temperature of 3.2~K. The cryostat was equipped with a superconducting magnet providing magnetic fields of up to 9~T in Faraday configuration. Piezo-stepping and scanning units (attocube systems, ANPxyz and ANSxy100) were used for sample positioning with respect to a low-temperature apochromatic objective. For PL measurements, a Ti:sapphire laser (Coherent, Mira) in continuous-wave mode was employed to excite the samples and was tuned to the resonance of intralayer exciton transition in WSe$_2$ ML at 725~nm. For DR measurements, a stabilized Tungsten-Halogen lamp (Thorlabs, SLS201L) was used as a broadband light source. The PL or reflection signal were spectrally dispersed by a monochromator (Roper Scientific, Acton SP2500 or Acton SpectraPro 300i with a 300 grooves/mm grating) and detected by a liquid nitrogen cooled or Peltier cooled charge-coupled device (Roper Scientific, Spec-10:100BR or Andor, iDus 416). A set of linear polarizers (Thorlabs, LPVIS), half- and quarter-waveplates (B. Halle, $310-1100$~nm achromatic) mounted on piezo-rotators (attocube systems, ANR240) were used to control the polarization in excitation and detection. The DR spectra were obtained by normalizing the reflected spectra from the HBL region ($R$) to that from the sample region without MoSe$_2$ and WSe$_2$ layers ($R_0$) as $\textrm{DR} = (R-R_0)/R_0$. Time-resolved PL was excited with a wavelength-tunable supercontinuum laser (NKT Photonics, SuperK Extreme and SuperK Varia) at 725~nm with a pulse duration of 6~ps and repetition rates down to 0.625~MHz, detected with a silicon avalanche photodiode (PerkinElmer, SPCM-AQRH-15) and correlated with a time-correlating single-photon counting module (PicoQuant, PicoHarp 300).
    
\textbf{SEM imaging:} Scanning electron microscopy (SEM) imaging of reconstruction in MoSe$_2$-WSe$_2$ HBLs was performed with a Raith eLine system. Further technical details are provided in Supplementary Note~1.

\textbf{Theoretical modeling:} Mesoscopic reconstruction was modelled in numerical simulations by discretizing the displacement field of the HBL lattice with a square mesh, followed by optimization of the total energy with the trust-region algorithm implemented in MATLAB. Density functional theory calculations of high symmetry HBL stackings with relaxed lattices were performed with the Vienna ab-initio simulation package with the PBEsol exchange-correlation functional. Technical details are provided in Supplementary Notes~2 and 7.\\



\textbf{Acknowledgements:}\\
This research was funded by the European Research Council (ERC) under the Grant Agreement No.~772195 as well as the Deutsche Forschungsgemeinschaft (DFG, German Research Foundation) within the Priority Programme SPP~2244 2DMP and the Germany's Excellence Strategy EXC-2111-390814868. Theoretical work was financially supported by the Foundation for the Advancement of Theoretical Physics and Mathematics "BASIS". S.\,Z. and I.\,B. acknowledge support from the Alexander von Humboldt Foundation. X.\,H. and A.\,S.\,B. received funding from the European Union's Framework Programme for Research and Innovation Horizon 2020 (2014--2020) under the Marie Sk{\l}odowska-Curie Grant Agreement No.~754388 (LMUResearchFellows) and from LMUexcellent, funded by the Federal Ministry of Education and Research (BMBF) and the Free State of Bavaria under the Excellence Strategy of the German Federal Government and the L{\"a}nder. Z.\,Li. was supported by China Scholarship Council (CSC), No. 201808140196. K.\,W. and T.\,T. acknowledge support from the Elemental Strategy Initiative conducted by the MEXT, Japan, Grant Number JPMXP0112101001, JSPS KAKENHI Grant Numbers JP20H00354 and the CREST(JPMJCR15F3), JST. The authors gratefully acknowledge the Leibniz Supercomputing Centre for funding this project by providing computing time and support on its Linux-Cluster.
\\
\\
\textbf{Contributions:}\\
S.\,Z., Z.\,Li., X.\,H. and A.\,R. equally contributed to this work. A.\,H. conceived and supervised the project. X.\,H. and J.\,G. fabricated heterobilayer samples from native crystals by exfoliation-stacking. Z.\,Li. and I.\,B. synthesized monolayers and fabricated heterobilayer samples from CVD-grown crystals. K.\,W. and T.\,T. provided high-quality hBN crystals. S.\,Z., Z.\,Li. and J.\,G. performed optical spectroscopy. A.\,R., S.\,Z. and Z.\,Li. performed scanning electron microscopy imaging. A.\,S.\,B. developed the theoretical model and together with I.\,A.\,V. performed numerical calculations. S.\,Yu.\,K. carried out DFT calculations. S.\,Z., A.\,S.\,B., and A.\,H. analyzed the data and wrote the manuscript. All authors commented on the manuscript.
\\
\\
\textbf{Corresponding authors:}\\
S.\,Z. (shen.zhao@physik.uni-muenchen.de), A.\,S.\,B. (anvar.baimuratov@lmu.de) and A.\,H. (alexander.hoegele@lmu.de). 
    
%


\begin{thebibliography}{69}%
\makeatletter
\providecommand \@ifxundefined [1]{%
 \@ifx{#1\undefined}
}%
\providecommand \@ifnum [1]{%
 \ifnum #1\expandafter \@firstoftwo
 \else \expandafter \@secondoftwo
 \fi
}%
\providecommand \@ifx [1]{%
 \ifx #1\expandafter \@firstoftwo
 \else \expandafter \@secondoftwo
 \fi
}%
\providecommand \natexlab [1]{#1}%
\providecommand \enquote  [1]{``#1''}%
\providecommand \bibnamefont  [1]{#1}%
\providecommand \bibfnamefont [1]{#1}%
\providecommand \citenamefont [1]{#1}%
\providecommand \href@noop [0]{\@secondoftwo}%
\providecommand \href [0]{\begingroup \@sanitize@url \@href}%
\providecommand \@href[1]{\@@startlink{#1}\@@href}%
\providecommand \@@href[1]{\endgroup#1\@@endlink}%
\providecommand \@sanitize@url [0]{\catcode `\\12\catcode `\$12\catcode
  `\&12\catcode `\#12\catcode `\^12\catcode `\_12\catcode `\%12\relax}%
\providecommand \@@startlink[1]{}%
\providecommand \@@endlink[0]{}%
\providecommand \url  [0]{\begingroup\@sanitize@url \@url }%
\providecommand \@url [1]{\endgroup\@href {#1}{\urlprefix }}%
\providecommand \urlprefix  [0]{URL }%
\providecommand \Eprint [0]{\href }%
\providecommand \doibase [0]{https://doi.org/}%
\providecommand \selectlanguage [0]{\@gobble}%
\providecommand \bibinfo  [0]{\@secondoftwo}%
\providecommand \bibfield  [0]{\@secondoftwo}%
\providecommand \translation [1]{[#1]}%
\providecommand \BibitemOpen [0]{}%
\providecommand \bibitemStop [0]{}%
\providecommand \bibitemNoStop [0]{.\EOS\space}%
\providecommand \EOS [0]{\spacefactor3000\relax}%
\providecommand \BibitemShut  [1]{\csname bibitem#1\endcsname}%
\let\auto@bib@innerbib\@empty
\bibitem [{\citenamefont {Bistritzer}\ and\ \citenamefont
  {MacDonald}(2011)}]{Bistritzer2011}%
  \BibitemOpen
  \bibfield  {author} {\bibinfo {author} {\bibfnamefont {R.}~\bibnamefont
  {Bistritzer}}\ and\ \bibinfo {author} {\bibfnamefont {A.~H.}\ \bibnamefont
  {MacDonald}},\ }\bibfield  {title} {\bibinfo {title} {Moir\'e bands in
  twisted double-layer graphene},\ }\href
  {https://doi.org/10.1073/pnas.1108174108} {\bibfield  {journal} {\bibinfo
  {journal} {Proc Natl Acad Sci USA}\ }\textbf {\bibinfo {volume} {108}},\
  \bibinfo {pages} {12233} (\bibinfo {year} {2011})}\BibitemShut {NoStop}%
\bibitem [{\citenamefont {Wu}\ \emph {et~al.}(2017)\citenamefont {Wu},
  \citenamefont {Lovorn},\ and\ \citenamefont {MacDonald}}]{WuTopo2017}%
  \BibitemOpen
  \bibfield  {author} {\bibinfo {author} {\bibfnamefont {F.}~\bibnamefont
  {Wu}}, \bibinfo {author} {\bibfnamefont {T.}~\bibnamefont {Lovorn}},\ and\
  \bibinfo {author} {\bibfnamefont {A.~H.}\ \bibnamefont {MacDonald}},\
  }\bibfield  {title} {\bibinfo {title} {{Topological Exciton Bands in Moir\'e
  Heterojunctions}},\ }\href {https://doi.org/10.1103/PhysRevLett.118.147401}
  {\bibfield  {journal} {\bibinfo  {journal} {Phys. Rev. Lett.}\ }\textbf
  {\bibinfo {volume} {118}},\ \bibinfo {pages} {147401} (\bibinfo {year}
  {2017})}\BibitemShut {NoStop}%
\bibitem [{\citenamefont {Yu}\ \emph {et~al.}(2017)\citenamefont {Yu},
  \citenamefont {Liu}, \citenamefont {Tang}, \citenamefont {Xu},\ and\
  \citenamefont {Yao}}]{Yu2017}%
  \BibitemOpen
  \bibfield  {author} {\bibinfo {author} {\bibfnamefont {H.}~\bibnamefont
  {Yu}}, \bibinfo {author} {\bibfnamefont {G.-B.}\ \bibnamefont {Liu}},
  \bibinfo {author} {\bibfnamefont {J.}~\bibnamefont {Tang}}, \bibinfo {author}
  {\bibfnamefont {X.}~\bibnamefont {Xu}},\ and\ \bibinfo {author}
  {\bibfnamefont {W.}~\bibnamefont {Yao}},\ }\bibfield  {title} {\bibinfo
  {title} {Moiré excitons: From programmable quantum emitter arrays to
  spin-orbit-coupled artificial lattices},\ }\href
  {http://advances.sciencemag.org/content/3/11/e1701696.abstract} {\bibfield
  {journal} {\bibinfo  {journal} {Sci. Adv.}\ }\textbf {\bibinfo {volume}
  {3}},\ \bibinfo {pages} {e1701696} (\bibinfo {year} {2017})}\BibitemShut
  {NoStop}%
\bibitem [{\citenamefont {Wu}\ \emph {et~al.}(2018{\natexlab{a}})\citenamefont
  {Wu}, \citenamefont {Lovorn},\ and\ \citenamefont
  {MacDonald}}]{WuExciton2018}%
  \BibitemOpen
  \bibfield  {author} {\bibinfo {author} {\bibfnamefont {F.}~\bibnamefont
  {Wu}}, \bibinfo {author} {\bibfnamefont {T.}~\bibnamefont {Lovorn}},\ and\
  \bibinfo {author} {\bibfnamefont {A.~H.}\ \bibnamefont {MacDonald}},\
  }\bibfield  {title} {\bibinfo {title} {Theory of optical absorption by
  interlayer excitons in transition metal dichalcogenide heterobilayers},\
  }\href {https://doi.org/10.1103/PhysRevB.97.035306} {\bibfield  {journal}
  {\bibinfo  {journal} {Phys. Rev. B}\ }\textbf {\bibinfo {volume} {97}},\
  \bibinfo {pages} {035306} (\bibinfo {year} {2018}{\natexlab{a}})}\BibitemShut
  {NoStop}%
\bibitem [{\citenamefont {Ponomarenko}\ \emph {et~al.}(2013)\citenamefont
  {Ponomarenko}, \citenamefont {Gorbachev}, \citenamefont {Yu}, \citenamefont
  {Elias}, \citenamefont {Jalil}, \citenamefont {Patel}, \citenamefont
  {Mishchenko}, \citenamefont {Mayorov}, \citenamefont {Woods}, \citenamefont
  {Wallbank}, \citenamefont {Mucha-Kruczynski}, \citenamefont {Piot},
  \citenamefont {Potemski}, \citenamefont {Grigorieva}, \citenamefont
  {Novoselov}, \citenamefont {Guinea}, \citenamefont {Fal’ko},\ and\
  \citenamefont {Geim}}]{Ponomarenko2013}%
  \BibitemOpen
  \bibfield  {author} {\bibinfo {author} {\bibfnamefont {L.~A.}\ \bibnamefont
  {Ponomarenko}}, \bibinfo {author} {\bibfnamefont {R.~V.}\ \bibnamefont
  {Gorbachev}}, \bibinfo {author} {\bibfnamefont {G.~L.}\ \bibnamefont {Yu}},
  \bibinfo {author} {\bibfnamefont {D.~C.}\ \bibnamefont {Elias}}, \bibinfo
  {author} {\bibfnamefont {R.}~\bibnamefont {Jalil}}, \bibinfo {author}
  {\bibfnamefont {A.~A.}\ \bibnamefont {Patel}}, \bibinfo {author}
  {\bibfnamefont {A.}~\bibnamefont {Mishchenko}}, \bibinfo {author}
  {\bibfnamefont {A.~S.}\ \bibnamefont {Mayorov}}, \bibinfo {author}
  {\bibfnamefont {C.~R.}\ \bibnamefont {Woods}}, \bibinfo {author}
  {\bibfnamefont {J.~R.}\ \bibnamefont {Wallbank}}, \bibinfo {author}
  {\bibfnamefont {M.}~\bibnamefont {Mucha-Kruczynski}}, \bibinfo {author}
  {\bibfnamefont {B.~A.}\ \bibnamefont {Piot}}, \bibinfo {author}
  {\bibfnamefont {M.}~\bibnamefont {Potemski}}, \bibinfo {author}
  {\bibfnamefont {I.~V.}\ \bibnamefont {Grigorieva}}, \bibinfo {author}
  {\bibfnamefont {K.~S.}\ \bibnamefont {Novoselov}}, \bibinfo {author}
  {\bibfnamefont {F.}~\bibnamefont {Guinea}}, \bibinfo {author} {\bibfnamefont
  {V.~I.}\ \bibnamefont {Fal’ko}},\ and\ \bibinfo {author} {\bibfnamefont
  {A.~K.}\ \bibnamefont {Geim}},\ }\bibfield  {title} {\bibinfo {title}
  {Cloning of dirac fermions in graphene superlattices},\ }\href
  {https://doi.org/10.1038/nature12187} {\bibfield  {journal} {\bibinfo
  {journal} {Nature}\ }\textbf {\bibinfo {volume} {497}},\ \bibinfo {pages}
  {594} (\bibinfo {year} {2013})}\BibitemShut {NoStop}%
\bibitem [{\citenamefont {Dean}\ \emph {et~al.}(2013)\citenamefont {Dean},
  \citenamefont {Wang}, \citenamefont {Maher}, \citenamefont {Forsythe},
  \citenamefont {Ghahari}, \citenamefont {Gao}, \citenamefont {Katoch},
  \citenamefont {Ishigami}, \citenamefont {Moon}, \citenamefont {Koshino},
  \citenamefont {Taniguchi}, \citenamefont {Watanabe}, \citenamefont {Shepard},
  \citenamefont {Hone},\ and\ \citenamefont {Kim}}]{Dean2013}%
  \BibitemOpen
  \bibfield  {author} {\bibinfo {author} {\bibfnamefont {C.~R.}\ \bibnamefont
  {Dean}}, \bibinfo {author} {\bibfnamefont {L.}~\bibnamefont {Wang}}, \bibinfo
  {author} {\bibfnamefont {P.}~\bibnamefont {Maher}}, \bibinfo {author}
  {\bibfnamefont {C.}~\bibnamefont {Forsythe}}, \bibinfo {author}
  {\bibfnamefont {F.}~\bibnamefont {Ghahari}}, \bibinfo {author} {\bibfnamefont
  {Y.}~\bibnamefont {Gao}}, \bibinfo {author} {\bibfnamefont {J.}~\bibnamefont
  {Katoch}}, \bibinfo {author} {\bibfnamefont {M.}~\bibnamefont {Ishigami}},
  \bibinfo {author} {\bibfnamefont {P.}~\bibnamefont {Moon}}, \bibinfo {author}
  {\bibfnamefont {M.}~\bibnamefont {Koshino}}, \bibinfo {author} {\bibfnamefont
  {T.}~\bibnamefont {Taniguchi}}, \bibinfo {author} {\bibfnamefont
  {K.}~\bibnamefont {Watanabe}}, \bibinfo {author} {\bibfnamefont {K.~L.}\
  \bibnamefont {Shepard}}, \bibinfo {author} {\bibfnamefont {J.}~\bibnamefont
  {Hone}},\ and\ \bibinfo {author} {\bibfnamefont {P.}~\bibnamefont {Kim}},\
  }\bibfield  {title} {\bibinfo {title} {Hofstadter’s butterfly and the
  fractal quantum hall effect in moiré superlattices},\ }\href
  {https://doi.org/10.1038/nature12186} {\bibfield  {journal} {\bibinfo
  {journal} {Nature}\ }\textbf {\bibinfo {volume} {497}},\ \bibinfo {pages}
  {598} (\bibinfo {year} {2013})}\BibitemShut {NoStop}%
\bibitem [{\citenamefont {Hunt}\ \emph {et~al.}(2013)\citenamefont {Hunt},
  \citenamefont {Sanchez-Yamagishi~J.}, \citenamefont {Young~A.}, \citenamefont
  {Yankowitz}, \citenamefont {LeRoy~B.}, \citenamefont {Watanabe},
  \citenamefont {Taniguchi}, \citenamefont {Moon}, \citenamefont {Koshino},
  \citenamefont {Jarillo-Herrero},\ and\ \citenamefont
  {Ashoori~R.}}]{Hunt2013}%
  \BibitemOpen
  \bibfield  {author} {\bibinfo {author} {\bibfnamefont {B.}~\bibnamefont
  {Hunt}}, \bibinfo {author} {\bibfnamefont {D.}~\bibnamefont
  {Sanchez-Yamagishi~J.}}, \bibinfo {author} {\bibfnamefont {F.}~\bibnamefont
  {Young~A.}}, \bibinfo {author} {\bibfnamefont {M.}~\bibnamefont {Yankowitz}},
  \bibinfo {author} {\bibfnamefont {J.}~\bibnamefont {LeRoy~B.}}, \bibinfo
  {author} {\bibfnamefont {K.}~\bibnamefont {Watanabe}}, \bibinfo {author}
  {\bibfnamefont {T.}~\bibnamefont {Taniguchi}}, \bibinfo {author}
  {\bibfnamefont {P.}~\bibnamefont {Moon}}, \bibinfo {author} {\bibfnamefont
  {M.}~\bibnamefont {Koshino}}, \bibinfo {author} {\bibfnamefont
  {P.}~\bibnamefont {Jarillo-Herrero}},\ and\ \bibinfo {author} {\bibfnamefont
  {C.}~\bibnamefont {Ashoori~R.}},\ }\bibfield  {title} {\bibinfo {title}
  {{Massive Dirac Fermions and Hofstadter Butterfly in a van der Waals
  Heterostructure}},\ }\href {https://doi.org/10.1126/science.1237240}
  {\bibfield  {journal} {\bibinfo  {journal} {Science}\ }\textbf {\bibinfo
  {volume} {340}},\ \bibinfo {pages} {1427} (\bibinfo {year}
  {2013})}\BibitemShut {NoStop}%
\bibitem [{\citenamefont {Zhang}\ \emph {et~al.}(2017)\citenamefont {Zhang},
  \citenamefont {Chuu}, \citenamefont {Ren}, \citenamefont {Li}, \citenamefont
  {Li}, \citenamefont {Jin}, \citenamefont {Chou},\ and\ \citenamefont
  {Shih}}]{ZhangInterlayer2017}%
  \BibitemOpen
  \bibfield  {author} {\bibinfo {author} {\bibfnamefont {C.}~\bibnamefont
  {Zhang}}, \bibinfo {author} {\bibfnamefont {C.-P.}\ \bibnamefont {Chuu}},
  \bibinfo {author} {\bibfnamefont {X.}~\bibnamefont {Ren}}, \bibinfo {author}
  {\bibfnamefont {M.-Y.}\ \bibnamefont {Li}}, \bibinfo {author} {\bibfnamefont
  {L.-J.}\ \bibnamefont {Li}}, \bibinfo {author} {\bibfnamefont
  {C.}~\bibnamefont {Jin}}, \bibinfo {author} {\bibfnamefont {M.-Y.}\
  \bibnamefont {Chou}},\ and\ \bibinfo {author} {\bibfnamefont {C.-K.}\
  \bibnamefont {Shih}},\ }\bibfield  {title} {\bibinfo {title} {{Interlayer
  couplings, Moir{\'e} patterns, and 2D electronic superlattices in
  MoS$_2$/WSe$_2$ hetero-bilayers}},\ }\href
  {https://www.science.org/doi/10.1126/sciadv.1601459} {\bibfield  {journal}
  {\bibinfo  {journal} {Sci. Adv.}\ }\textbf {\bibinfo {volume} {3}},\ \bibinfo
  {pages} {e1601459} (\bibinfo {year} {2017})}\BibitemShut {NoStop}%
\bibitem [{\citenamefont {van Wijk}\ \emph {et~al.}(2014)\citenamefont {van
  Wijk}, \citenamefont {Schuring}, \citenamefont {Katsnelson},\ and\
  \citenamefont {Fasolino}}]{Wijk2014}%
  \BibitemOpen
  \bibfield  {author} {\bibinfo {author} {\bibfnamefont {M.~M.}\ \bibnamefont
  {van Wijk}}, \bibinfo {author} {\bibfnamefont {A.}~\bibnamefont {Schuring}},
  \bibinfo {author} {\bibfnamefont {M.~I.}\ \bibnamefont {Katsnelson}},\ and\
  \bibinfo {author} {\bibfnamefont {A.}~\bibnamefont {Fasolino}},\ }\bibfield
  {title} {\bibinfo {title} {{Moir\'e Patterns as a Probe of Interplanar
  Interactions for Graphene on h-BN}},\ }\href
  {https://doi.org/10.1103/PhysRevLett.113.135504} {\bibfield  {journal}
  {\bibinfo  {journal} {Phys. Rev. Lett.}\ }\textbf {\bibinfo {volume} {113}},\
  \bibinfo {pages} {135504} (\bibinfo {year} {2014})}\BibitemShut {NoStop}%
\bibitem [{\citenamefont {Woods}\ \emph {et~al.}(2014)\citenamefont {Woods},
  \citenamefont {Britnell}, \citenamefont {Eckmann}, \citenamefont {Ma},
  \citenamefont {Lu}, \citenamefont {Guo}, \citenamefont {Lin}, \citenamefont
  {Yu}, \citenamefont {Cao}, \citenamefont {Gorbachev}, \citenamefont
  {Kretinin}, \citenamefont {Park}, \citenamefont {Ponomarenko}, \citenamefont
  {Katsnelson}, \citenamefont {Gornostyrev}, \citenamefont {Watanabe},
  \citenamefont {Taniguchi}, \citenamefont {Casiraghi}, \citenamefont {Gao},
  \citenamefont {Geim},\ and\ \citenamefont {Novoselov}}]{Woods2014}%
  \BibitemOpen
  \bibfield  {author} {\bibinfo {author} {\bibfnamefont {C.~R.}\ \bibnamefont
  {Woods}}, \bibinfo {author} {\bibfnamefont {L.}~\bibnamefont {Britnell}},
  \bibinfo {author} {\bibfnamefont {A.}~\bibnamefont {Eckmann}}, \bibinfo
  {author} {\bibfnamefont {R.~S.}\ \bibnamefont {Ma}}, \bibinfo {author}
  {\bibfnamefont {J.~C.}\ \bibnamefont {Lu}}, \bibinfo {author} {\bibfnamefont
  {H.~M.}\ \bibnamefont {Guo}}, \bibinfo {author} {\bibfnamefont
  {X.}~\bibnamefont {Lin}}, \bibinfo {author} {\bibfnamefont {G.~L.}\
  \bibnamefont {Yu}}, \bibinfo {author} {\bibfnamefont {Y.}~\bibnamefont
  {Cao}}, \bibinfo {author} {\bibfnamefont {R.}~\bibnamefont {Gorbachev}},
  \bibinfo {author} {\bibfnamefont {A.~V.}\ \bibnamefont {Kretinin}}, \bibinfo
  {author} {\bibfnamefont {J.}~\bibnamefont {Park}}, \bibinfo {author}
  {\bibfnamefont {L.~A.}\ \bibnamefont {Ponomarenko}}, \bibinfo {author}
  {\bibfnamefont {M.~I.}\ \bibnamefont {Katsnelson}}, \bibinfo {author}
  {\bibfnamefont {Y.}~\bibnamefont {Gornostyrev}}, \bibinfo {author}
  {\bibfnamefont {K.}~\bibnamefont {Watanabe}}, \bibinfo {author}
  {\bibfnamefont {T.}~\bibnamefont {Taniguchi}}, \bibinfo {author}
  {\bibfnamefont {C.}~\bibnamefont {Casiraghi}}, \bibinfo {author}
  {\bibfnamefont {H.-J.}\ \bibnamefont {Gao}}, \bibinfo {author} {\bibfnamefont
  {A.~K.}\ \bibnamefont {Geim}},\ and\ \bibinfo {author} {\bibfnamefont
  {K.}~\bibnamefont {Novoselov}},\ }\bibfield  {title} {\bibinfo {title}
  {Commensurate-incommensurate transition in graphene on hexagonal boron
  nitride},\ }\href {https://doi.org/10.1038/nphys2954} {\bibfield  {journal}
  {\bibinfo  {journal} {Nat. Phys.}\ }\textbf {\bibinfo {volume} {10}},\
  \bibinfo {pages} {451} (\bibinfo {year} {2014})}\BibitemShut {NoStop}%
\bibitem [{\citenamefont {van Wijk}\ \emph {et~al.}(2015)\citenamefont {van
  Wijk}, \citenamefont {Schuring}, \citenamefont {Katsnelson},\ and\
  \citenamefont {Fasolino}}]{Wijk2015}%
  \BibitemOpen
  \bibfield  {author} {\bibinfo {author} {\bibfnamefont {M.~M.}\ \bibnamefont
  {van Wijk}}, \bibinfo {author} {\bibfnamefont {A.}~\bibnamefont {Schuring}},
  \bibinfo {author} {\bibfnamefont {M.~I.}\ \bibnamefont {Katsnelson}},\ and\
  \bibinfo {author} {\bibfnamefont {A.}~\bibnamefont {Fasolino}},\ }\bibfield
  {title} {\bibinfo {title} {Relaxation of moiré patterns for slightly
  misaligned identical lattices: graphene on graphite},\ }\href
  {http://dx.doi.org/10.1088/2053-1583/2/3/034010} {\bibfield  {journal}
  {\bibinfo  {journal} {2D Mater.}\ }\textbf {\bibinfo {volume} {2}},\ \bibinfo
  {pages} {034010} (\bibinfo {year} {2015})}\BibitemShut {NoStop}%
\bibitem [{\citenamefont {Carr}\ \emph {et~al.}(2018)\citenamefont {Carr},
  \citenamefont {Massatt}, \citenamefont {Torrisi}, \citenamefont {Cazeaux},
  \citenamefont {Luskin},\ and\ \citenamefont {Kaxiras}}]{Carr2018}%
  \BibitemOpen
  \bibfield  {author} {\bibinfo {author} {\bibfnamefont {S.}~\bibnamefont
  {Carr}}, \bibinfo {author} {\bibfnamefont {D.}~\bibnamefont {Massatt}},
  \bibinfo {author} {\bibfnamefont {S.~B.}\ \bibnamefont {Torrisi}}, \bibinfo
  {author} {\bibfnamefont {P.}~\bibnamefont {Cazeaux}}, \bibinfo {author}
  {\bibfnamefont {M.}~\bibnamefont {Luskin}},\ and\ \bibinfo {author}
  {\bibfnamefont {E.}~\bibnamefont {Kaxiras}},\ }\bibfield  {title} {\bibinfo
  {title} {Relaxation and domain formation in incommensurate two-dimensional
  heterostructures},\ }\href {https://doi.org/10.1103/PhysRevB.98.224102}
  {\bibfield  {journal} {\bibinfo  {journal} {Phys. Rev. B}\ }\textbf {\bibinfo
  {volume} {98}},\ \bibinfo {pages} {224102} (\bibinfo {year}
  {2018})}\BibitemShut {NoStop}%
\bibitem [{\citenamefont {Sunku~S.}\ \emph {et~al.}(2018)\citenamefont
  {Sunku~S.}, \citenamefont {Ni~G.}, \citenamefont {Jiang~B.}, \citenamefont
  {Yoo}, \citenamefont {Sternbach}, \citenamefont {McLeod~A.}, \citenamefont
  {Stauber}, \citenamefont {Xiong}, \citenamefont {Taniguchi}, \citenamefont
  {Watanabe}, \citenamefont {Kim}, \citenamefont {Fogler~M.},\ and\
  \citenamefont {Basov~D.}}]{Sunku2018}%
  \BibitemOpen
  \bibfield  {author} {\bibinfo {author} {\bibfnamefont {S.}~\bibnamefont
  {Sunku~S.}}, \bibinfo {author} {\bibfnamefont {X.}~\bibnamefont {Ni~G.}},
  \bibinfo {author} {\bibfnamefont {Y.}~\bibnamefont {Jiang~B.}}, \bibinfo
  {author} {\bibfnamefont {H.}~\bibnamefont {Yoo}}, \bibinfo {author}
  {\bibfnamefont {A.}~\bibnamefont {Sternbach}}, \bibinfo {author}
  {\bibfnamefont {S.}~\bibnamefont {McLeod~A.}}, \bibinfo {author}
  {\bibfnamefont {T.}~\bibnamefont {Stauber}}, \bibinfo {author} {\bibfnamefont
  {L.}~\bibnamefont {Xiong}}, \bibinfo {author} {\bibfnamefont
  {T.}~\bibnamefont {Taniguchi}}, \bibinfo {author} {\bibfnamefont
  {K.}~\bibnamefont {Watanabe}}, \bibinfo {author} {\bibfnamefont
  {P.}~\bibnamefont {Kim}}, \bibinfo {author} {\bibfnamefont {M.}~\bibnamefont
  {Fogler~M.}},\ and\ \bibinfo {author} {\bibfnamefont {N.}~\bibnamefont
  {Basov~D.}},\ }\bibfield  {title} {\bibinfo {title} {Photonic crystals for
  nano-light in moiré graphene superlattices},\ }\href
  {https://doi.org/10.1126/science.aau5144} {\bibfield  {journal} {\bibinfo
  {journal} {Science}\ }\textbf {\bibinfo {volume} {362}},\ \bibinfo {pages}
  {1153} (\bibinfo {year} {2018})}\BibitemShut {NoStop}%
\bibitem [{\citenamefont {Yoo}\ \emph {et~al.}(2019)\citenamefont {Yoo},
  \citenamefont {Engelke}, \citenamefont {Carr}, \citenamefont {Fang},
  \citenamefont {Zhang}, \citenamefont {Cazeaux}, \citenamefont {Sung},
  \citenamefont {Hovden}, \citenamefont {Tsen}, \citenamefont {Taniguchi},
  \citenamefont {Watanabe}, \citenamefont {Yi}, \citenamefont {Kim},
  \citenamefont {Luskin}, \citenamefont {Tadmor}, \citenamefont {Kaxiras},\
  and\ \citenamefont {Kim}}]{Yoo2019}%
  \BibitemOpen
  \bibfield  {author} {\bibinfo {author} {\bibfnamefont {H.}~\bibnamefont
  {Yoo}}, \bibinfo {author} {\bibfnamefont {R.}~\bibnamefont {Engelke}},
  \bibinfo {author} {\bibfnamefont {S.}~\bibnamefont {Carr}}, \bibinfo {author}
  {\bibfnamefont {S.}~\bibnamefont {Fang}}, \bibinfo {author} {\bibfnamefont
  {K.}~\bibnamefont {Zhang}}, \bibinfo {author} {\bibfnamefont
  {P.}~\bibnamefont {Cazeaux}}, \bibinfo {author} {\bibfnamefont {S.~H.}\
  \bibnamefont {Sung}}, \bibinfo {author} {\bibfnamefont {R.}~\bibnamefont
  {Hovden}}, \bibinfo {author} {\bibfnamefont {A.~W.}\ \bibnamefont {Tsen}},
  \bibinfo {author} {\bibfnamefont {T.}~\bibnamefont {Taniguchi}}, \bibinfo
  {author} {\bibfnamefont {K.}~\bibnamefont {Watanabe}}, \bibinfo {author}
  {\bibfnamefont {G.-C.}\ \bibnamefont {Yi}}, \bibinfo {author} {\bibfnamefont
  {M.}~\bibnamefont {Kim}}, \bibinfo {author} {\bibfnamefont {M.}~\bibnamefont
  {Luskin}}, \bibinfo {author} {\bibfnamefont {E.~B.}\ \bibnamefont {Tadmor}},
  \bibinfo {author} {\bibfnamefont {E.}~\bibnamefont {Kaxiras}},\ and\ \bibinfo
  {author} {\bibfnamefont {P.}~\bibnamefont {Kim}},\ }\bibfield  {title}
  {\bibinfo {title} {Atomic and electronic reconstruction at the van der
  {W}aals interface in twisted bilayer graphene},\ }\href
  {https://doi.org/10.1038/s41563-019-0346-z} {\bibfield  {journal} {\bibinfo
  {journal} {Nat. Mater.}\ }\textbf {\bibinfo {volume} {18}},\ \bibinfo {pages}
  {448} (\bibinfo {year} {2019})}\BibitemShut {NoStop}%
\bibitem [{\citenamefont {Halbertal}\ \emph {et~al.}(2021)\citenamefont
  {Halbertal}, \citenamefont {Finney}, \citenamefont {Sunku}, \citenamefont
  {Kerelsky}, \citenamefont {Rubio-Verd{\ifmmode\acute{u}\else\'{u}\fi}},
  \citenamefont {Shabani}, \citenamefont {Xian}, \citenamefont {Carr},
  \citenamefont {Chen}, \citenamefont {Zhang}, \citenamefont {Wang},
  \citenamefont {Gonzalez-Acevedo}, \citenamefont {McLeod}, \citenamefont
  {Rhodes}, \citenamefont {Watanabe}, \citenamefont {Taniguchi}, \citenamefont
  {Kaxiras}, \citenamefont {Dean}, \citenamefont {Hone}, \citenamefont
  {Pasupathy}, \citenamefont {Kennes}, \citenamefont {Rubio},\ and\
  \citenamefont {Basov}}]{Halbertal2021}%
  \BibitemOpen
  \bibfield  {author} {\bibinfo {author} {\bibfnamefont {D.}~\bibnamefont
  {Halbertal}}, \bibinfo {author} {\bibfnamefont {N.~R.}\ \bibnamefont
  {Finney}}, \bibinfo {author} {\bibfnamefont {S.~S.}\ \bibnamefont {Sunku}},
  \bibinfo {author} {\bibfnamefont {A.}~\bibnamefont {Kerelsky}}, \bibinfo
  {author} {\bibfnamefont {C.}~\bibnamefont
  {Rubio-Verd{\ifmmode\acute{u}\else\'{u}\fi}}}, \bibinfo {author}
  {\bibfnamefont {S.}~\bibnamefont {Shabani}}, \bibinfo {author} {\bibfnamefont
  {L.}~\bibnamefont {Xian}}, \bibinfo {author} {\bibfnamefont {S.}~\bibnamefont
  {Carr}}, \bibinfo {author} {\bibfnamefont {S.}~\bibnamefont {Chen}}, \bibinfo
  {author} {\bibfnamefont {C.}~\bibnamefont {Zhang}}, \bibinfo {author}
  {\bibfnamefont {L.}~\bibnamefont {Wang}}, \bibinfo {author} {\bibfnamefont
  {D.}~\bibnamefont {Gonzalez-Acevedo}}, \bibinfo {author} {\bibfnamefont
  {A.~S.}\ \bibnamefont {McLeod}}, \bibinfo {author} {\bibfnamefont
  {D.}~\bibnamefont {Rhodes}}, \bibinfo {author} {\bibfnamefont
  {K.}~\bibnamefont {Watanabe}}, \bibinfo {author} {\bibfnamefont
  {T.}~\bibnamefont {Taniguchi}}, \bibinfo {author} {\bibfnamefont
  {E.}~\bibnamefont {Kaxiras}}, \bibinfo {author} {\bibfnamefont {C.~R.}\
  \bibnamefont {Dean}}, \bibinfo {author} {\bibfnamefont {J.~C.}\ \bibnamefont
  {Hone}}, \bibinfo {author} {\bibfnamefont {A.~N.}\ \bibnamefont {Pasupathy}},
  \bibinfo {author} {\bibfnamefont {D.~M.}\ \bibnamefont {Kennes}}, \bibinfo
  {author} {\bibfnamefont {A.}~\bibnamefont {Rubio}},\ and\ \bibinfo {author}
  {\bibfnamefont {D.~N.}\ \bibnamefont {Basov}},\ }\bibfield  {title} {\bibinfo
  {title} {{Moir{\ifmmode\acute{e}\else\'{e}\fi} metrology of energy landscapes
  in van der Waals heterostructures}},\ }\href
  {https://doi.org/10.1038/s41467-020-20428-1} {\bibfield  {journal} {\bibinfo
  {journal} {Nat. Commun.}\ }\textbf {\bibinfo {volume} {12}},\ \bibinfo
  {pages} {1} (\bibinfo {year} {2021})}\BibitemShut {NoStop}%
\bibitem [{\citenamefont {Enaldiev}\ \emph {et~al.}(2020)\citenamefont
  {Enaldiev}, \citenamefont {Z\'olyomi}, \citenamefont {Yelgel}, \citenamefont
  {Magorrian},\ and\ \citenamefont {Fal'ko}}]{Enaldiev2020}%
  \BibitemOpen
  \bibfield  {author} {\bibinfo {author} {\bibfnamefont {V.~V.}\ \bibnamefont
  {Enaldiev}}, \bibinfo {author} {\bibfnamefont {V.}~\bibnamefont {Z\'olyomi}},
  \bibinfo {author} {\bibfnamefont {C.}~\bibnamefont {Yelgel}}, \bibinfo
  {author} {\bibfnamefont {S.~J.}\ \bibnamefont {Magorrian}},\ and\ \bibinfo
  {author} {\bibfnamefont {V.~I.}\ \bibnamefont {Fal'ko}},\ }\bibfield  {title}
  {\bibinfo {title} {Stacking domains and dislocation networks in marginally
  twisted bilayers of transition metal dichalcogenides},\ }\href
  {https://doi.org/10.1103/PhysRevLett.124.206101} {\bibfield  {journal}
  {\bibinfo  {journal} {Phys. Rev. Lett.}\ }\textbf {\bibinfo {volume} {124}},\
  \bibinfo {pages} {206101} (\bibinfo {year} {2020})}\BibitemShut {NoStop}%
\bibitem [{\citenamefont {Weston}\ \emph {et~al.}(2020)\citenamefont {Weston},
  \citenamefont {Zou}, \citenamefont {Enaldiev}, \citenamefont {Summerfield},
  \citenamefont {Clark}, \citenamefont
  {Z{\ifmmode\acute{o}\else\'{o}\fi}lyomi}, \citenamefont {Graham},
  \citenamefont {Yelgel}, \citenamefont {Magorrian}, \citenamefont {Zhou},
  \citenamefont {Zultak}, \citenamefont {Hopkinson}, \citenamefont {Barinov},
  \citenamefont {Bointon}, \citenamefont {Kretinin}, \citenamefont {Wilson},
  \citenamefont {Beton}, \citenamefont {Fal{'}ko}, \citenamefont {Haigh},\ and\
  \citenamefont {Gorbachev}}]{Weston2020}%
  \BibitemOpen
  \bibfield  {author} {\bibinfo {author} {\bibfnamefont {A.}~\bibnamefont
  {Weston}}, \bibinfo {author} {\bibfnamefont {Y.}~\bibnamefont {Zou}},
  \bibinfo {author} {\bibfnamefont {V.}~\bibnamefont {Enaldiev}}, \bibinfo
  {author} {\bibfnamefont {A.}~\bibnamefont {Summerfield}}, \bibinfo {author}
  {\bibfnamefont {N.}~\bibnamefont {Clark}}, \bibinfo {author} {\bibfnamefont
  {V.}~\bibnamefont {Z{\ifmmode\acute{o}\else\'{o}\fi}lyomi}}, \bibinfo
  {author} {\bibfnamefont {A.}~\bibnamefont {Graham}}, \bibinfo {author}
  {\bibfnamefont {C.}~\bibnamefont {Yelgel}}, \bibinfo {author} {\bibfnamefont
  {S.}~\bibnamefont {Magorrian}}, \bibinfo {author} {\bibfnamefont
  {M.}~\bibnamefont {Zhou}}, \bibinfo {author} {\bibfnamefont {J.}~\bibnamefont
  {Zultak}}, \bibinfo {author} {\bibfnamefont {D.}~\bibnamefont {Hopkinson}},
  \bibinfo {author} {\bibfnamefont {A.}~\bibnamefont {Barinov}}, \bibinfo
  {author} {\bibfnamefont {T.~H.}\ \bibnamefont {Bointon}}, \bibinfo {author}
  {\bibfnamefont {A.}~\bibnamefont {Kretinin}}, \bibinfo {author}
  {\bibfnamefont {N.~R.}\ \bibnamefont {Wilson}}, \bibinfo {author}
  {\bibfnamefont {P.~H.}\ \bibnamefont {Beton}}, \bibinfo {author}
  {\bibfnamefont {V.~I.}\ \bibnamefont {Fal{'}ko}}, \bibinfo {author}
  {\bibfnamefont {S.~J.}\ \bibnamefont {Haigh}},\ and\ \bibinfo {author}
  {\bibfnamefont {R.}~\bibnamefont {Gorbachev}},\ }\bibfield  {title} {\bibinfo
  {title} {{Atomic reconstruction in twisted bilayers of transition metal
  dichalcogenides}},\ }\href {https://doi.org/10.1038/s41565-020-0682-9}
  {\bibfield  {journal} {\bibinfo  {journal} {Nat. Nanotechnol.}\ }\textbf
  {\bibinfo {volume} {15}},\ \bibinfo {pages} {592} (\bibinfo {year}
  {2020})}\BibitemShut {NoStop}%
\bibitem [{\citenamefont {McGilly}\ \emph {et~al.}(2020)\citenamefont
  {McGilly}, \citenamefont {Kerelsky}, \citenamefont {Finney}, \citenamefont
  {Shapovalov}, \citenamefont {Shih}, \citenamefont {Ghiotto}, \citenamefont
  {Zeng}, \citenamefont {Moore}, \citenamefont {Wu}, \citenamefont {Bai},
  \citenamefont {Watanabe}, \citenamefont {Taniguchi}, \citenamefont {Stengel},
  \citenamefont {Zhou}, \citenamefont {Hone}, \citenamefont {Zhu},
  \citenamefont {Basov}, \citenamefont {Dean}, \citenamefont {Dreyer},\ and\
  \citenamefont {Pasupathy}}]{McGilly2020}%
  \BibitemOpen
  \bibfield  {author} {\bibinfo {author} {\bibfnamefont {L.~J.}\ \bibnamefont
  {McGilly}}, \bibinfo {author} {\bibfnamefont {A.}~\bibnamefont {Kerelsky}},
  \bibinfo {author} {\bibfnamefont {N.~R.}\ \bibnamefont {Finney}}, \bibinfo
  {author} {\bibfnamefont {K.}~\bibnamefont {Shapovalov}}, \bibinfo {author}
  {\bibfnamefont {E.-M.}\ \bibnamefont {Shih}}, \bibinfo {author}
  {\bibfnamefont {A.}~\bibnamefont {Ghiotto}}, \bibinfo {author} {\bibfnamefont
  {Y.}~\bibnamefont {Zeng}}, \bibinfo {author} {\bibfnamefont {S.~L.}\
  \bibnamefont {Moore}}, \bibinfo {author} {\bibfnamefont {W.}~\bibnamefont
  {Wu}}, \bibinfo {author} {\bibfnamefont {Y.}~\bibnamefont {Bai}}, \bibinfo
  {author} {\bibfnamefont {K.}~\bibnamefont {Watanabe}}, \bibinfo {author}
  {\bibfnamefont {T.}~\bibnamefont {Taniguchi}}, \bibinfo {author}
  {\bibfnamefont {M.}~\bibnamefont {Stengel}}, \bibinfo {author} {\bibfnamefont
  {L.}~\bibnamefont {Zhou}}, \bibinfo {author} {\bibfnamefont {J.}~\bibnamefont
  {Hone}}, \bibinfo {author} {\bibfnamefont {X.}~\bibnamefont {Zhu}}, \bibinfo
  {author} {\bibfnamefont {D.~N.}\ \bibnamefont {Basov}}, \bibinfo {author}
  {\bibfnamefont {C.}~\bibnamefont {Dean}}, \bibinfo {author} {\bibfnamefont
  {C.~E.}\ \bibnamefont {Dreyer}},\ and\ \bibinfo {author} {\bibfnamefont
  {A.~N.}\ \bibnamefont {Pasupathy}},\ }\bibfield  {title} {\bibinfo {title}
  {Visualization of moir\'e superlattices},\ }\href
  {https://doi.org/10.1038/s41565-020-0708-3} {\bibfield  {journal} {\bibinfo
  {journal} {Nat. Nanotechnol.}\ }\textbf {\bibinfo {volume} {15}},\ \bibinfo
  {pages} {580} (\bibinfo {year} {2020})}\BibitemShut {NoStop}%
\bibitem [{\citenamefont {Sung}\ \emph {et~al.}(2020)\citenamefont {Sung},
  \citenamefont {Zhou}, \citenamefont {Scuri}, \citenamefont {Zólyomi},
  \citenamefont {Andersen}, \citenamefont {Yoo}, \citenamefont {Wild},
  \citenamefont {Joe}, \citenamefont {Gelly}, \citenamefont {Heo},
  \citenamefont {Magorrian}, \citenamefont {Bérubé}, \citenamefont
  {Valdivia}, \citenamefont {Taniguchi}, \citenamefont {Watanabe},
  \citenamefont {Lukin}, \citenamefont {Kim}, \citenamefont {Fal’ko},\ and\
  \citenamefont {Park}}]{Sung2020}%
  \BibitemOpen
  \bibfield  {author} {\bibinfo {author} {\bibfnamefont {J.}~\bibnamefont
  {Sung}}, \bibinfo {author} {\bibfnamefont {Y.}~\bibnamefont {Zhou}}, \bibinfo
  {author} {\bibfnamefont {G.}~\bibnamefont {Scuri}}, \bibinfo {author}
  {\bibfnamefont {V.}~\bibnamefont {Zólyomi}}, \bibinfo {author}
  {\bibfnamefont {T.~I.}\ \bibnamefont {Andersen}}, \bibinfo {author}
  {\bibfnamefont {H.}~\bibnamefont {Yoo}}, \bibinfo {author} {\bibfnamefont
  {D.~S.}\ \bibnamefont {Wild}}, \bibinfo {author} {\bibfnamefont {A.~Y.}\
  \bibnamefont {Joe}}, \bibinfo {author} {\bibfnamefont {R.~J.}\ \bibnamefont
  {Gelly}}, \bibinfo {author} {\bibfnamefont {H.}~\bibnamefont {Heo}}, \bibinfo
  {author} {\bibfnamefont {S.~J.}\ \bibnamefont {Magorrian}}, \bibinfo {author}
  {\bibfnamefont {D.}~\bibnamefont {Bérubé}}, \bibinfo {author}
  {\bibfnamefont {A.~M.~M.}\ \bibnamefont {Valdivia}}, \bibinfo {author}
  {\bibfnamefont {T.}~\bibnamefont {Taniguchi}}, \bibinfo {author}
  {\bibfnamefont {K.}~\bibnamefont {Watanabe}}, \bibinfo {author}
  {\bibfnamefont {M.~D.}\ \bibnamefont {Lukin}}, \bibinfo {author}
  {\bibfnamefont {P.}~\bibnamefont {Kim}}, \bibinfo {author} {\bibfnamefont
  {V.~I.}\ \bibnamefont {Fal’ko}},\ and\ \bibinfo {author} {\bibfnamefont
  {H.}~\bibnamefont {Park}},\ }\bibfield  {title} {\bibinfo {title} {Broken
  mirror symmetry in excitonic response of reconstructed domains in twisted
  {MoSe}$_2$/{MoSe}$_2$ bilayers},\ }\href
  {https://doi.org/10.1038/s41565-020-0728-z} {\bibfield  {journal} {\bibinfo
  {journal} {Nat. Nanotechnol.}\ }\textbf {\bibinfo {volume} {15}},\ \bibinfo
  {pages} {750} (\bibinfo {year} {2020})}\BibitemShut {NoStop}%
\bibitem [{\citenamefont {Andersen}\ \emph {et~al.}(2021)\citenamefont
  {Andersen}, \citenamefont {Scuri}, \citenamefont {Sushko}, \citenamefont
  {De~Greve}, \citenamefont {Sung}, \citenamefont {Zhou}, \citenamefont {Wild},
  \citenamefont {Gelly}, \citenamefont {Heo}, \citenamefont {Bérubé},
  \citenamefont {Joe}, \citenamefont {Jauregui}, \citenamefont {Watanabe},
  \citenamefont {Taniguchi}, \citenamefont {Kim}, \citenamefont {Park},\ and\
  \citenamefont {Lukin}}]{Andersen2021}%
  \BibitemOpen
  \bibfield  {author} {\bibinfo {author} {\bibfnamefont {T.~I.}\ \bibnamefont
  {Andersen}}, \bibinfo {author} {\bibfnamefont {G.}~\bibnamefont {Scuri}},
  \bibinfo {author} {\bibfnamefont {A.}~\bibnamefont {Sushko}}, \bibinfo
  {author} {\bibfnamefont {K.}~\bibnamefont {De~Greve}}, \bibinfo {author}
  {\bibfnamefont {J.}~\bibnamefont {Sung}}, \bibinfo {author} {\bibfnamefont
  {Y.}~\bibnamefont {Zhou}}, \bibinfo {author} {\bibfnamefont {D.~S.}\
  \bibnamefont {Wild}}, \bibinfo {author} {\bibfnamefont {R.~J.}\ \bibnamefont
  {Gelly}}, \bibinfo {author} {\bibfnamefont {H.}~\bibnamefont {Heo}}, \bibinfo
  {author} {\bibfnamefont {D.}~\bibnamefont {Bérubé}}, \bibinfo {author}
  {\bibfnamefont {A.~Y.}\ \bibnamefont {Joe}}, \bibinfo {author} {\bibfnamefont
  {L.~A.}\ \bibnamefont {Jauregui}}, \bibinfo {author} {\bibfnamefont
  {K.}~\bibnamefont {Watanabe}}, \bibinfo {author} {\bibfnamefont
  {T.}~\bibnamefont {Taniguchi}}, \bibinfo {author} {\bibfnamefont
  {P.}~\bibnamefont {Kim}}, \bibinfo {author} {\bibfnamefont {H.}~\bibnamefont
  {Park}},\ and\ \bibinfo {author} {\bibfnamefont {M.~D.}\ \bibnamefont
  {Lukin}},\ }\bibfield  {title} {\bibinfo {title} {Excitons in a reconstructed
  moiré potential in twisted {WSe}$_2$/{WSe}$_2$ homobilayers},\ }\href
  {https://doi.org/10.1038/s41563-020-00873-5} {\bibfield  {journal} {\bibinfo
  {journal} {Nat. Mater.}\ }\textbf {\bibinfo {volume} {20}},\ \bibinfo {pages}
  {480} (\bibinfo {year} {2021})}\BibitemShut {NoStop}%
\bibitem [{\citenamefont {Rosenberger}\ \emph {et~al.}(2020)\citenamefont
  {Rosenberger}, \citenamefont {Chuang}, \citenamefont {Phillips},
  \citenamefont {Oleshko}, \citenamefont {McCreary}, \citenamefont {Sivaram},
  \citenamefont {Hellberg},\ and\ \citenamefont {Jonker}}]{Rosenberger2020}%
  \BibitemOpen
  \bibfield  {author} {\bibinfo {author} {\bibfnamefont {M.~R.}\ \bibnamefont
  {Rosenberger}}, \bibinfo {author} {\bibfnamefont {H.-J.}\ \bibnamefont
  {Chuang}}, \bibinfo {author} {\bibfnamefont {M.}~\bibnamefont {Phillips}},
  \bibinfo {author} {\bibfnamefont {V.~P.}\ \bibnamefont {Oleshko}}, \bibinfo
  {author} {\bibfnamefont {K.~M.}\ \bibnamefont {McCreary}}, \bibinfo {author}
  {\bibfnamefont {S.~V.}\ \bibnamefont {Sivaram}}, \bibinfo {author}
  {\bibfnamefont {C.~S.}\ \bibnamefont {Hellberg}},\ and\ \bibinfo {author}
  {\bibfnamefont {B.~T.}\ \bibnamefont {Jonker}},\ }\bibfield  {title}
  {\bibinfo {title} {Twist angle-dependent atomic reconstruction and moiré
  patterns in transition metal dichalcogenide heterostructures},\ }\href
  {https://doi.org/10.1021/acsnano.0c00088} {\bibfield  {journal} {\bibinfo
  {journal} {ACS Nano}\ }\textbf {\bibinfo {volume} {14}},\ \bibinfo {pages}
  {4550} (\bibinfo {year} {2020})}\BibitemShut {NoStop}%
\bibitem [{\citenamefont {Enaldiev}\ \emph {et~al.}(2021)\citenamefont
  {Enaldiev}, \citenamefont {Ferreira}, \citenamefont {Magorrian},\ and\
  \citenamefont {Fal'ko}}]{Enaldiev2021}%
  \BibitemOpen
  \bibfield  {author} {\bibinfo {author} {\bibfnamefont {V.~V.}\ \bibnamefont
  {Enaldiev}}, \bibinfo {author} {\bibfnamefont {F.}~\bibnamefont {Ferreira}},
  \bibinfo {author} {\bibfnamefont {S.~J.}\ \bibnamefont {Magorrian}},\ and\
  \bibinfo {author} {\bibfnamefont {V.~I.}\ \bibnamefont {Fal'ko}},\ }\bibfield
   {title} {\bibinfo {title} {Piezoelectric networks and ferroelectric domains
  in twistronic superlattices in {WS}$_2$/{MoS}$_2$ and {WSe}$_2$/{MoSe}$_2$
  bilayers},\ }\href {https://doi.org/10.1088/2053-1583/abdd92} {\bibfield
  {journal} {\bibinfo  {journal} {2D Mater.}\ }\textbf {\bibinfo {volume}
  {8}},\ \bibinfo {pages} {025030} (\bibinfo {year} {2021})}\BibitemShut
  {NoStop}%
\bibitem [{\citenamefont {Shabani}\ \emph {et~al.}(2021)\citenamefont
  {Shabani}, \citenamefont {Halbertal}, \citenamefont {Wu}, \citenamefont
  {Chen}, \citenamefont {Liu}, \citenamefont {Hone}, \citenamefont {Yao},
  \citenamefont {Basov}, \citenamefont {Zhu},\ and\ \citenamefont
  {Pasupathy}}]{Shabani2021}%
  \BibitemOpen
  \bibfield  {author} {\bibinfo {author} {\bibfnamefont {S.}~\bibnamefont
  {Shabani}}, \bibinfo {author} {\bibfnamefont {D.}~\bibnamefont {Halbertal}},
  \bibinfo {author} {\bibfnamefont {W.}~\bibnamefont {Wu}}, \bibinfo {author}
  {\bibfnamefont {M.}~\bibnamefont {Chen}}, \bibinfo {author} {\bibfnamefont
  {S.}~\bibnamefont {Liu}}, \bibinfo {author} {\bibfnamefont {J.}~\bibnamefont
  {Hone}}, \bibinfo {author} {\bibfnamefont {W.}~\bibnamefont {Yao}}, \bibinfo
  {author} {\bibfnamefont {D.~N.}\ \bibnamefont {Basov}}, \bibinfo {author}
  {\bibfnamefont {X.}~\bibnamefont {Zhu}},\ and\ \bibinfo {author}
  {\bibfnamefont {A.~N.}\ \bibnamefont {Pasupathy}},\ }\bibfield  {title}
  {\bibinfo {title} {{Deep moir{\ifmmode\acute{e}\else\'{e}\fi} potentials in
  twisted transition metal dichalcogenide bilayers}},\ }\href
  {https://doi.org/10.1038/s41567-021-01174-7} {\bibfield  {journal} {\bibinfo
  {journal} {Nat. Phys.}\ }\textbf {\bibinfo {volume} {17}},\ \bibinfo {pages}
  {720} (\bibinfo {year} {2021})}\BibitemShut {NoStop}%
\bibitem [{\citenamefont {Weston}\ \emph {et~al.}(2022)\citenamefont {Weston},
  \citenamefont {Castanon}, \citenamefont {Enaldiev}, \citenamefont {Ferreira},
  \citenamefont {Bhattacharjee}, \citenamefont {Xu}, \citenamefont
  {Corte-León}, \citenamefont {Wu}, \citenamefont {Clark}, \citenamefont
  {Summerfield}, \citenamefont {Hashimoto}, \citenamefont {Gao}, \citenamefont
  {Wang}, \citenamefont {Hamer}, \citenamefont {Read}, \citenamefont
  {Fumagalli}, \citenamefont {Kretinin}, \citenamefont {Haigh}, \citenamefont
  {Kazakova}, \citenamefont {Geim}, \citenamefont {Fal’ko},\ and\
  \citenamefont {Gorbachev}}]{Weston2022}%
  \BibitemOpen
  \bibfield  {author} {\bibinfo {author} {\bibfnamefont {A.}~\bibnamefont
  {Weston}}, \bibinfo {author} {\bibfnamefont {E.~G.}\ \bibnamefont
  {Castanon}}, \bibinfo {author} {\bibfnamefont {V.}~\bibnamefont {Enaldiev}},
  \bibinfo {author} {\bibfnamefont {F.}~\bibnamefont {Ferreira}}, \bibinfo
  {author} {\bibfnamefont {S.}~\bibnamefont {Bhattacharjee}}, \bibinfo {author}
  {\bibfnamefont {S.}~\bibnamefont {Xu}}, \bibinfo {author} {\bibfnamefont
  {H.}~\bibnamefont {Corte-León}}, \bibinfo {author} {\bibfnamefont
  {Z.}~\bibnamefont {Wu}}, \bibinfo {author} {\bibfnamefont {N.}~\bibnamefont
  {Clark}}, \bibinfo {author} {\bibfnamefont {A.}~\bibnamefont {Summerfield}},
  \bibinfo {author} {\bibfnamefont {T.}~\bibnamefont {Hashimoto}}, \bibinfo
  {author} {\bibfnamefont {Y.}~\bibnamefont {Gao}}, \bibinfo {author}
  {\bibfnamefont {W.}~\bibnamefont {Wang}}, \bibinfo {author} {\bibfnamefont
  {M.}~\bibnamefont {Hamer}}, \bibinfo {author} {\bibfnamefont
  {H.}~\bibnamefont {Read}}, \bibinfo {author} {\bibfnamefont {L.}~\bibnamefont
  {Fumagalli}}, \bibinfo {author} {\bibfnamefont {A.~V.}\ \bibnamefont
  {Kretinin}}, \bibinfo {author} {\bibfnamefont {S.~J.}\ \bibnamefont {Haigh}},
  \bibinfo {author} {\bibfnamefont {O.}~\bibnamefont {Kazakova}}, \bibinfo
  {author} {\bibfnamefont {A.~K.}\ \bibnamefont {Geim}}, \bibinfo {author}
  {\bibfnamefont {V.~I.}\ \bibnamefont {Fal’ko}},\ and\ \bibinfo {author}
  {\bibfnamefont {R.}~\bibnamefont {Gorbachev}},\ }\bibfield  {title} {\bibinfo
  {title} {Interfacial ferroelectricity in marginally twisted 2d
  semiconductors},\ }\href {https://doi.org/10.1038/s41565-022-01072-w}
  {\bibfield  {journal} {\bibinfo  {journal} {Nat. Nanotechnol.}\ } (\bibinfo
  {year} {2022})}\BibitemShut {NoStop}%
\bibitem [{\citenamefont {Seyler}\ \emph {et~al.}(2019)\citenamefont {Seyler},
  \citenamefont {Rivera}, \citenamefont {Yu}, \citenamefont {Wilson},
  \citenamefont {Ray}, \citenamefont {Mandrus}, \citenamefont {Yan},
  \citenamefont {Yao},\ and\ \citenamefont {Xu}}]{Seyler2019}%
  \BibitemOpen
  \bibfield  {author} {\bibinfo {author} {\bibfnamefont {K.~L.}\ \bibnamefont
  {Seyler}}, \bibinfo {author} {\bibfnamefont {P.}~\bibnamefont {Rivera}},
  \bibinfo {author} {\bibfnamefont {H.}~\bibnamefont {Yu}}, \bibinfo {author}
  {\bibfnamefont {N.~P.}\ \bibnamefont {Wilson}}, \bibinfo {author}
  {\bibfnamefont {E.~L.}\ \bibnamefont {Ray}}, \bibinfo {author} {\bibfnamefont
  {D.~G.}\ \bibnamefont {Mandrus}}, \bibinfo {author} {\bibfnamefont
  {J.}~\bibnamefont {Yan}}, \bibinfo {author} {\bibfnamefont {W.}~\bibnamefont
  {Yao}},\ and\ \bibinfo {author} {\bibfnamefont {X.}~\bibnamefont {Xu}},\
  }\bibfield  {title} {\bibinfo {title} {Signatures of moiré-trapped valley
  excitons in {MoSe}$_2$/{WSe}$_2$ heterobilayers},\ }\href
  {https://doi.org/10.1038/s41586-019-0957-1} {\bibfield  {journal} {\bibinfo
  {journal} {Nature}\ }\textbf {\bibinfo {volume} {567}},\ \bibinfo {pages}
  {66} (\bibinfo {year} {2019})}\BibitemShut {NoStop}%
\bibitem [{\citenamefont {Tran}\ \emph {et~al.}(2019)\citenamefont {Tran},
  \citenamefont {Moody}, \citenamefont {Wu}, \citenamefont {Lu}, \citenamefont
  {Choi}, \citenamefont {Kim}, \citenamefont {Rai}, \citenamefont {Sanchez},
  \citenamefont {Quan}, \citenamefont {Singh}, \citenamefont {Embley},
  \citenamefont {Zepeda}, \citenamefont {Campbell}, \citenamefont {Autry},
  \citenamefont {Taniguchi}, \citenamefont {Watanabe}, \citenamefont {Lu},
  \citenamefont {Banerjee}, \citenamefont {Silverman}, \citenamefont {Kim},
  \citenamefont {Tutuc}, \citenamefont {Yang}, \citenamefont {MacDonald},\ and\
  \citenamefont {Li}}]{Tran2019}%
  \BibitemOpen
  \bibfield  {author} {\bibinfo {author} {\bibfnamefont {K.}~\bibnamefont
  {Tran}}, \bibinfo {author} {\bibfnamefont {G.}~\bibnamefont {Moody}},
  \bibinfo {author} {\bibfnamefont {F.}~\bibnamefont {Wu}}, \bibinfo {author}
  {\bibfnamefont {X.}~\bibnamefont {Lu}}, \bibinfo {author} {\bibfnamefont
  {J.}~\bibnamefont {Choi}}, \bibinfo {author} {\bibfnamefont {K.}~\bibnamefont
  {Kim}}, \bibinfo {author} {\bibfnamefont {A.}~\bibnamefont {Rai}}, \bibinfo
  {author} {\bibfnamefont {D.~A.}\ \bibnamefont {Sanchez}}, \bibinfo {author}
  {\bibfnamefont {J.}~\bibnamefont {Quan}}, \bibinfo {author} {\bibfnamefont
  {A.}~\bibnamefont {Singh}}, \bibinfo {author} {\bibfnamefont
  {J.}~\bibnamefont {Embley}}, \bibinfo {author} {\bibfnamefont
  {A.}~\bibnamefont {Zepeda}}, \bibinfo {author} {\bibfnamefont
  {M.}~\bibnamefont {Campbell}}, \bibinfo {author} {\bibfnamefont
  {T.}~\bibnamefont {Autry}}, \bibinfo {author} {\bibfnamefont
  {T.}~\bibnamefont {Taniguchi}}, \bibinfo {author} {\bibfnamefont
  {K.}~\bibnamefont {Watanabe}}, \bibinfo {author} {\bibfnamefont
  {N.}~\bibnamefont {Lu}}, \bibinfo {author} {\bibfnamefont {S.~K.}\
  \bibnamefont {Banerjee}}, \bibinfo {author} {\bibfnamefont {K.~L.}\
  \bibnamefont {Silverman}}, \bibinfo {author} {\bibfnamefont {S.}~\bibnamefont
  {Kim}}, \bibinfo {author} {\bibfnamefont {E.}~\bibnamefont {Tutuc}}, \bibinfo
  {author} {\bibfnamefont {L.}~\bibnamefont {Yang}}, \bibinfo {author}
  {\bibfnamefont {A.~H.}\ \bibnamefont {MacDonald}},\ and\ \bibinfo {author}
  {\bibfnamefont {X.}~\bibnamefont {Li}},\ }\bibfield  {title} {\bibinfo
  {title} {Evidence for moiré excitons in van der {W}aals heterostructures},\
  }\href {https://doi.org/10.1038/s41586-019-0975-z} {\bibfield  {journal}
  {\bibinfo  {journal} {Nature}\ }\textbf {\bibinfo {volume} {567}},\ \bibinfo
  {pages} {71} (\bibinfo {year} {2019})}\BibitemShut {NoStop}%
\bibitem [{\citenamefont {Jin}\ \emph {et~al.}(2019)\citenamefont {Jin},
  \citenamefont {Regan}, \citenamefont {Yan}, \citenamefont {Iqbal
  Bakti~Utama}, \citenamefont {Wang}, \citenamefont {Zhao}, \citenamefont
  {Qin}, \citenamefont {Yang}, \citenamefont {Zheng}, \citenamefont {Shi},
  \citenamefont {Watanabe}, \citenamefont {Taniguchi}, \citenamefont {Tongay},
  \citenamefont {Zettl},\ and\ \citenamefont {Wang}}]{Jin2019}%
  \BibitemOpen
  \bibfield  {author} {\bibinfo {author} {\bibfnamefont {C.}~\bibnamefont
  {Jin}}, \bibinfo {author} {\bibfnamefont {E.~C.}\ \bibnamefont {Regan}},
  \bibinfo {author} {\bibfnamefont {A.}~\bibnamefont {Yan}}, \bibinfo {author}
  {\bibfnamefont {M.}~\bibnamefont {Iqbal Bakti~Utama}}, \bibinfo {author}
  {\bibfnamefont {D.}~\bibnamefont {Wang}}, \bibinfo {author} {\bibfnamefont
  {S.}~\bibnamefont {Zhao}}, \bibinfo {author} {\bibfnamefont {Y.}~\bibnamefont
  {Qin}}, \bibinfo {author} {\bibfnamefont {S.}~\bibnamefont {Yang}}, \bibinfo
  {author} {\bibfnamefont {Z.}~\bibnamefont {Zheng}}, \bibinfo {author}
  {\bibfnamefont {S.}~\bibnamefont {Shi}}, \bibinfo {author} {\bibfnamefont
  {K.}~\bibnamefont {Watanabe}}, \bibinfo {author} {\bibfnamefont
  {T.}~\bibnamefont {Taniguchi}}, \bibinfo {author} {\bibfnamefont
  {S.}~\bibnamefont {Tongay}}, \bibinfo {author} {\bibfnamefont
  {A.}~\bibnamefont {Zettl}},\ and\ \bibinfo {author} {\bibfnamefont
  {F.}~\bibnamefont {Wang}},\ }\bibfield  {title} {\bibinfo {title}
  {Observation of moiré excitons in {WSe}$_2$/{WS}$_2$ heterostructure
  superlattices},\ }\href {https://doi.org/10.1038/s41586-019-0976-y}
  {\bibfield  {journal} {\bibinfo  {journal} {Nature}\ }\textbf {\bibinfo
  {volume} {567}},\ \bibinfo {pages} {76} (\bibinfo {year} {2019})}\BibitemShut
  {NoStop}%
\bibitem [{\citenamefont {Alexeev}\ \emph {et~al.}(2019)\citenamefont
  {Alexeev}, \citenamefont {Ruiz-Tijerina}, \citenamefont {Danovich},
  \citenamefont {Hamer}, \citenamefont {Terry}, \citenamefont {Nayak},
  \citenamefont {Ahn}, \citenamefont {Pak}, \citenamefont {Lee}, \citenamefont
  {Sohn}, \citenamefont {Molas}, \citenamefont {Koperski}, \citenamefont
  {Watanabe}, \citenamefont {Taniguchi}, \citenamefont {Novoselov},
  \citenamefont {Gorbachev}, \citenamefont {Shin}, \citenamefont {Fal’ko},\
  and\ \citenamefont {Tartakovskii}}]{Alexeev2019}%
  \BibitemOpen
  \bibfield  {author} {\bibinfo {author} {\bibfnamefont {E.~M.}\ \bibnamefont
  {Alexeev}}, \bibinfo {author} {\bibfnamefont {D.~A.}\ \bibnamefont
  {Ruiz-Tijerina}}, \bibinfo {author} {\bibfnamefont {M.}~\bibnamefont
  {Danovich}}, \bibinfo {author} {\bibfnamefont {M.~J.}\ \bibnamefont {Hamer}},
  \bibinfo {author} {\bibfnamefont {D.~J.}\ \bibnamefont {Terry}}, \bibinfo
  {author} {\bibfnamefont {P.~K.}\ \bibnamefont {Nayak}}, \bibinfo {author}
  {\bibfnamefont {S.}~\bibnamefont {Ahn}}, \bibinfo {author} {\bibfnamefont
  {S.}~\bibnamefont {Pak}}, \bibinfo {author} {\bibfnamefont {J.}~\bibnamefont
  {Lee}}, \bibinfo {author} {\bibfnamefont {J.~I.}\ \bibnamefont {Sohn}},
  \bibinfo {author} {\bibfnamefont {M.~R.}\ \bibnamefont {Molas}}, \bibinfo
  {author} {\bibfnamefont {M.}~\bibnamefont {Koperski}}, \bibinfo {author}
  {\bibfnamefont {K.}~\bibnamefont {Watanabe}}, \bibinfo {author}
  {\bibfnamefont {T.}~\bibnamefont {Taniguchi}}, \bibinfo {author}
  {\bibfnamefont {K.~S.}\ \bibnamefont {Novoselov}}, \bibinfo {author}
  {\bibfnamefont {R.~V.}\ \bibnamefont {Gorbachev}}, \bibinfo {author}
  {\bibfnamefont {H.~S.}\ \bibnamefont {Shin}}, \bibinfo {author}
  {\bibfnamefont {V.~I.}\ \bibnamefont {Fal’ko}},\ and\ \bibinfo {author}
  {\bibfnamefont {A.~I.}\ \bibnamefont {Tartakovskii}},\ }\bibfield  {title}
  {\bibinfo {title} {Resonantly hybridized excitons in moiré superlattices in
  van der {W}aals heterostructures},\ }\href
  {https://doi.org/10.1038/s41586-019-0986-9} {\bibfield  {journal} {\bibinfo
  {journal} {Nature}\ }\textbf {\bibinfo {volume} {567}},\ \bibinfo {pages}
  {81} (\bibinfo {year} {2019})}\BibitemShut {NoStop}%
\bibitem [{\citenamefont {Bai}\ \emph {et~al.}(2020)\citenamefont {Bai},
  \citenamefont {Zhou}, \citenamefont {Wang}, \citenamefont {Wu}, \citenamefont
  {McGilly}, \citenamefont {Halbertal}, \citenamefont {Lo}, \citenamefont
  {Liu}, \citenamefont {Ardelean}, \citenamefont {Rivera}, \citenamefont
  {Finney}, \citenamefont {Yang}, \citenamefont {Basov}, \citenamefont {Yao},
  \citenamefont {Xu}, \citenamefont {Hone}, \citenamefont {Pasupathy},\ and\
  \citenamefont {Zhu}}]{Bai2020}%
  \BibitemOpen
  \bibfield  {author} {\bibinfo {author} {\bibfnamefont {Y.}~\bibnamefont
  {Bai}}, \bibinfo {author} {\bibfnamefont {L.}~\bibnamefont {Zhou}}, \bibinfo
  {author} {\bibfnamefont {J.}~\bibnamefont {Wang}}, \bibinfo {author}
  {\bibfnamefont {W.}~\bibnamefont {Wu}}, \bibinfo {author} {\bibfnamefont
  {L.~J.}\ \bibnamefont {McGilly}}, \bibinfo {author} {\bibfnamefont
  {D.}~\bibnamefont {Halbertal}}, \bibinfo {author} {\bibfnamefont {C.~F.~B.}\
  \bibnamefont {Lo}}, \bibinfo {author} {\bibfnamefont {F.}~\bibnamefont
  {Liu}}, \bibinfo {author} {\bibfnamefont {J.}~\bibnamefont {Ardelean}},
  \bibinfo {author} {\bibfnamefont {P.}~\bibnamefont {Rivera}}, \bibinfo
  {author} {\bibfnamefont {N.~R.}\ \bibnamefont {Finney}}, \bibinfo {author}
  {\bibfnamefont {X.-C.}\ \bibnamefont {Yang}}, \bibinfo {author}
  {\bibfnamefont {D.~N.}\ \bibnamefont {Basov}}, \bibinfo {author}
  {\bibfnamefont {W.}~\bibnamefont {Yao}}, \bibinfo {author} {\bibfnamefont
  {X.}~\bibnamefont {Xu}}, \bibinfo {author} {\bibfnamefont {J.}~\bibnamefont
  {Hone}}, \bibinfo {author} {\bibfnamefont {A.~N.}\ \bibnamefont
  {Pasupathy}},\ and\ \bibinfo {author} {\bibfnamefont {X.-Y.}\ \bibnamefont
  {Zhu}},\ }\bibfield  {title} {\bibinfo {title} {Excitons in strain-induced
  one-dimensional moiré potentials at transition metal dichalcogenide
  heterojunctions},\ }\href {https://doi.org/10.1038/s41563-020-0730-8}
  {\bibfield  {journal} {\bibinfo  {journal} {Nat. Mater.}\ }\textbf {\bibinfo
  {volume} {19}},\ \bibinfo {pages} {1068} (\bibinfo {year}
  {2020})}\BibitemShut {NoStop}%
\bibitem [{\citenamefont {Tartakovskii}(2020)}]{Tartakovskii2020}%
  \BibitemOpen
  \bibfield  {author} {\bibinfo {author} {\bibfnamefont {A.}~\bibnamefont
  {Tartakovskii}},\ }\bibfield  {title} {\bibinfo {title} {Moiré or not},\
  }\href {https://doi.org/10.1038/s41563-020-0693-9} {\bibfield  {journal}
  {\bibinfo  {journal} {Nat. Mater.}\ }\textbf {\bibinfo {volume} {19}},\
  \bibinfo {pages} {581} (\bibinfo {year} {2020})}\BibitemShut {NoStop}%
\bibitem [{\citenamefont {Alden}\ \emph {et~al.}(2013)\citenamefont {Alden},
  \citenamefont {Tsen}, \citenamefont {Huang}, \citenamefont {Hovden},
  \citenamefont {Brown}, \citenamefont {Park}, \citenamefont {Muller},\ and\
  \citenamefont {McEuen}}]{Alden2013}%
  \BibitemOpen
  \bibfield  {author} {\bibinfo {author} {\bibfnamefont {J.~S.}\ \bibnamefont
  {Alden}}, \bibinfo {author} {\bibfnamefont {A.~W.}\ \bibnamefont {Tsen}},
  \bibinfo {author} {\bibfnamefont {P.~Y.}\ \bibnamefont {Huang}}, \bibinfo
  {author} {\bibfnamefont {R.}~\bibnamefont {Hovden}}, \bibinfo {author}
  {\bibfnamefont {L.}~\bibnamefont {Brown}}, \bibinfo {author} {\bibfnamefont
  {J.}~\bibnamefont {Park}}, \bibinfo {author} {\bibfnamefont {D.~A.}\
  \bibnamefont {Muller}},\ and\ \bibinfo {author} {\bibfnamefont {P.~L.}\
  \bibnamefont {McEuen}},\ }\bibfield  {title} {\bibinfo {title} {Strain
  solitons and topological defects in bilayer graphene},\ }\href
  {https://doi.org/10.1073/pnas.1309394110} {\bibfield  {journal} {\bibinfo
  {journal} {Proc. Natl. Acad. Sci.}\ }\textbf {\bibinfo {volume} {110}},\
  \bibinfo {pages} {11256} (\bibinfo {year} {2013})}\BibitemShut {NoStop}%
\bibitem [{\citenamefont {Wu}\ \emph {et~al.}(2018{\natexlab{b}})\citenamefont
  {Wu}, \citenamefont {Lovorn}, \citenamefont {Tutuc},\ and\ \citenamefont
  {MacDonald}}]{WuHubbard2018}%
  \BibitemOpen
  \bibfield  {author} {\bibinfo {author} {\bibfnamefont {F.}~\bibnamefont
  {Wu}}, \bibinfo {author} {\bibfnamefont {T.}~\bibnamefont {Lovorn}}, \bibinfo
  {author} {\bibfnamefont {E.}~\bibnamefont {Tutuc}},\ and\ \bibinfo {author}
  {\bibfnamefont {A.~H.}\ \bibnamefont {MacDonald}},\ }\bibfield  {title}
  {\bibinfo {title} {Hubbard model physics in transition metal dichalcogenide
  moir\'e bands},\ }\href {https://doi.org/10.1103/PhysRevLett.121.026402}
  {\bibfield  {journal} {\bibinfo  {journal} {PRL}\ }\textbf {\bibinfo {volume}
  {121}},\ \bibinfo {pages} {026402} (\bibinfo {year}
  {2018}{\natexlab{b}})}\BibitemShut {NoStop}%
\bibitem [{\citenamefont {Tang}\ \emph {et~al.}(2020)\citenamefont {Tang},
  \citenamefont {Li}, \citenamefont {Li}, \citenamefont {Xu}, \citenamefont
  {Liu}, \citenamefont {Barmak}, \citenamefont {Watanabe}, \citenamefont
  {Taniguchi}, \citenamefont {MacDonald}, \citenamefont {Shan},\ and\
  \citenamefont {Mak}}]{Tang2020}%
  \BibitemOpen
  \bibfield  {author} {\bibinfo {author} {\bibfnamefont {Y.}~\bibnamefont
  {Tang}}, \bibinfo {author} {\bibfnamefont {L.}~\bibnamefont {Li}}, \bibinfo
  {author} {\bibfnamefont {T.}~\bibnamefont {Li}}, \bibinfo {author}
  {\bibfnamefont {Y.}~\bibnamefont {Xu}}, \bibinfo {author} {\bibfnamefont
  {S.}~\bibnamefont {Liu}}, \bibinfo {author} {\bibfnamefont {K.}~\bibnamefont
  {Barmak}}, \bibinfo {author} {\bibfnamefont {K.}~\bibnamefont {Watanabe}},
  \bibinfo {author} {\bibfnamefont {T.}~\bibnamefont {Taniguchi}}, \bibinfo
  {author} {\bibfnamefont {A.~H.}\ \bibnamefont {MacDonald}}, \bibinfo {author}
  {\bibfnamefont {J.}~\bibnamefont {Shan}},\ and\ \bibinfo {author}
  {\bibfnamefont {K.~F.}\ \bibnamefont {Mak}},\ }\bibfield  {title} {\bibinfo
  {title} {Simulation of {H}ubbard model physics in {WSe}$_2$/{WS}$_2$ moiré
  superlattices},\ }\href {https://doi.org/10.1038/s41586-020-2085-3}
  {\bibfield  {journal} {\bibinfo  {journal} {Nature}\ }\textbf {\bibinfo
  {volume} {579}},\ \bibinfo {pages} {353} (\bibinfo {year}
  {2020})}\BibitemShut {NoStop}%
\bibitem [{\citenamefont {Wang}\ \emph
  {et~al.}(2020{\natexlab{a}})\citenamefont {Wang}, \citenamefont {Shih},
  \citenamefont {Ghiotto}, \citenamefont {Xian}, \citenamefont {Rhodes},
  \citenamefont {Tan}, \citenamefont {Claassen}, \citenamefont {Kennes},
  \citenamefont {Bai}, \citenamefont {Kim}, \citenamefont {Watanabe},
  \citenamefont {Taniguchi}, \citenamefont {Zhu}, \citenamefont {Hone},
  \citenamefont {Rubio}, \citenamefont {Pasupathy},\ and\ \citenamefont
  {Dean}}]{Wang2020a}%
  \BibitemOpen
  \bibfield  {author} {\bibinfo {author} {\bibfnamefont {L.}~\bibnamefont
  {Wang}}, \bibinfo {author} {\bibfnamefont {E.-M.}\ \bibnamefont {Shih}},
  \bibinfo {author} {\bibfnamefont {A.}~\bibnamefont {Ghiotto}}, \bibinfo
  {author} {\bibfnamefont {L.}~\bibnamefont {Xian}}, \bibinfo {author}
  {\bibfnamefont {D.~A.}\ \bibnamefont {Rhodes}}, \bibinfo {author}
  {\bibfnamefont {C.}~\bibnamefont {Tan}}, \bibinfo {author} {\bibfnamefont
  {M.}~\bibnamefont {Claassen}}, \bibinfo {author} {\bibfnamefont {D.~M.}\
  \bibnamefont {Kennes}}, \bibinfo {author} {\bibfnamefont {Y.}~\bibnamefont
  {Bai}}, \bibinfo {author} {\bibfnamefont {B.}~\bibnamefont {Kim}}, \bibinfo
  {author} {\bibfnamefont {K.}~\bibnamefont {Watanabe}}, \bibinfo {author}
  {\bibfnamefont {T.}~\bibnamefont {Taniguchi}}, \bibinfo {author}
  {\bibfnamefont {X.}~\bibnamefont {Zhu}}, \bibinfo {author} {\bibfnamefont
  {J.}~\bibnamefont {Hone}}, \bibinfo {author} {\bibfnamefont {A.}~\bibnamefont
  {Rubio}}, \bibinfo {author} {\bibfnamefont {A.~N.}\ \bibnamefont
  {Pasupathy}},\ and\ \bibinfo {author} {\bibfnamefont {C.~R.}\ \bibnamefont
  {Dean}},\ }\bibfield  {title} {\bibinfo {title} {Correlated electronic phases
  in twisted bilayer transition metal dichalcogenides},\ }\href
  {https://doi.org/10.1038/s41563-020-0708-6} {\bibfield  {journal} {\bibinfo
  {journal} {Nat. Mater.}\ }\textbf {\bibinfo {volume} {19}},\ \bibinfo {pages}
  {861} (\bibinfo {year} {2020}{\natexlab{a}})}\BibitemShut {NoStop}%
\bibitem [{\citenamefont {Huang}\ \emph {et~al.}(2021)\citenamefont {Huang},
  \citenamefont {Wang}, \citenamefont {Miao}, \citenamefont {Wang},
  \citenamefont {Li}, \citenamefont {Lian}, \citenamefont {Taniguchi},
  \citenamefont {Watanabe}, \citenamefont {Okamoto}, \citenamefont {Xiao},
  \citenamefont {Shi},\ and\ \citenamefont {Cui}}]{Huang2021}%
  \BibitemOpen
  \bibfield  {author} {\bibinfo {author} {\bibfnamefont {X.}~\bibnamefont
  {Huang}}, \bibinfo {author} {\bibfnamefont {T.}~\bibnamefont {Wang}},
  \bibinfo {author} {\bibfnamefont {S.}~\bibnamefont {Miao}}, \bibinfo {author}
  {\bibfnamefont {C.}~\bibnamefont {Wang}}, \bibinfo {author} {\bibfnamefont
  {Z.}~\bibnamefont {Li}}, \bibinfo {author} {\bibfnamefont {Z.}~\bibnamefont
  {Lian}}, \bibinfo {author} {\bibfnamefont {T.}~\bibnamefont {Taniguchi}},
  \bibinfo {author} {\bibfnamefont {K.}~\bibnamefont {Watanabe}}, \bibinfo
  {author} {\bibfnamefont {S.}~\bibnamefont {Okamoto}}, \bibinfo {author}
  {\bibfnamefont {D.}~\bibnamefont {Xiao}}, \bibinfo {author} {\bibfnamefont
  {S.-F.}\ \bibnamefont {Shi}},\ and\ \bibinfo {author} {\bibfnamefont {Y.-T.}\
  \bibnamefont {Cui}},\ }\bibfield  {title} {\bibinfo {title} {{Correlated
  insulating states at fractional fillings of the WS$_2$/WSe$_2$ moir\'e
  lattice}},\ }\href {https://doi.org/10.1038/s41567-021-01171-w} {\bibfield
  {journal} {\bibinfo  {journal} {Nat. Phys.}\ }\textbf {\bibinfo {volume}
  {17}},\ \bibinfo {pages} {715} (\bibinfo {year} {2021})}\BibitemShut
  {NoStop}%
\bibitem [{\citenamefont {Ghiotto}\ \emph {et~al.}(2021)\citenamefont
  {Ghiotto}, \citenamefont {Shih}, \citenamefont {Pereira}, \citenamefont
  {Rhodes}, \citenamefont {Kim}, \citenamefont {Zang}, \citenamefont {Millis},
  \citenamefont {Watanabe}, \citenamefont {Taniguchi}, \citenamefont {Hone},
  \citenamefont {Wang}, \citenamefont {Dean},\ and\ \citenamefont
  {Pasupathy}}]{Ghiotto2021}%
  \BibitemOpen
  \bibfield  {author} {\bibinfo {author} {\bibfnamefont {A.}~\bibnamefont
  {Ghiotto}}, \bibinfo {author} {\bibfnamefont {E.-M.}\ \bibnamefont {Shih}},
  \bibinfo {author} {\bibfnamefont {G.~S. S.~G.}\ \bibnamefont {Pereira}},
  \bibinfo {author} {\bibfnamefont {D.~A.}\ \bibnamefont {Rhodes}}, \bibinfo
  {author} {\bibfnamefont {B.}~\bibnamefont {Kim}}, \bibinfo {author}
  {\bibfnamefont {J.}~\bibnamefont {Zang}}, \bibinfo {author} {\bibfnamefont
  {A.~J.}\ \bibnamefont {Millis}}, \bibinfo {author} {\bibfnamefont
  {K.}~\bibnamefont {Watanabe}}, \bibinfo {author} {\bibfnamefont
  {T.}~\bibnamefont {Taniguchi}}, \bibinfo {author} {\bibfnamefont {J.~C.}\
  \bibnamefont {Hone}}, \bibinfo {author} {\bibfnamefont {L.}~\bibnamefont
  {Wang}}, \bibinfo {author} {\bibfnamefont {C.~R.}\ \bibnamefont {Dean}},\
  and\ \bibinfo {author} {\bibfnamefont {A.~N.}\ \bibnamefont {Pasupathy}},\
  }\bibfield  {title} {\bibinfo {title} {Quantum criticality in twisted
  transition metal dichalcogenides},\ }\href
  {https://doi.org/10.1038/s41586-021-03815-6} {\bibfield  {journal} {\bibinfo
  {journal} {Nature}\ }\textbf {\bibinfo {volume} {597}},\ \bibinfo {pages}
  {345} (\bibinfo {year} {2021})}\BibitemShut {NoStop}%
\bibitem [{\citenamefont {Li}\ \emph {et~al.}(2021{\natexlab{a}})\citenamefont
  {Li}, \citenamefont {Jiang}, \citenamefont {Li}, \citenamefont {Zhang},
  \citenamefont {Kang}, \citenamefont {Zhu}, \citenamefont {Watanabe},
  \citenamefont {Taniguchi}, \citenamefont {Chowdhury}, \citenamefont {Fu},
  \citenamefont {Shan},\ and\ \citenamefont {Mak}}]{Li-Mak2021}%
  \BibitemOpen
  \bibfield  {author} {\bibinfo {author} {\bibfnamefont {T.}~\bibnamefont
  {Li}}, \bibinfo {author} {\bibfnamefont {S.}~\bibnamefont {Jiang}}, \bibinfo
  {author} {\bibfnamefont {L.}~\bibnamefont {Li}}, \bibinfo {author}
  {\bibfnamefont {Y.}~\bibnamefont {Zhang}}, \bibinfo {author} {\bibfnamefont
  {K.}~\bibnamefont {Kang}}, \bibinfo {author} {\bibfnamefont {J.}~\bibnamefont
  {Zhu}}, \bibinfo {author} {\bibfnamefont {K.}~\bibnamefont {Watanabe}},
  \bibinfo {author} {\bibfnamefont {T.}~\bibnamefont {Taniguchi}}, \bibinfo
  {author} {\bibfnamefont {D.}~\bibnamefont {Chowdhury}}, \bibinfo {author}
  {\bibfnamefont {L.}~\bibnamefont {Fu}}, \bibinfo {author} {\bibfnamefont
  {J.}~\bibnamefont {Shan}},\ and\ \bibinfo {author} {\bibfnamefont {K.~F.}\
  \bibnamefont {Mak}},\ }\bibfield  {title} {\bibinfo {title} {Continuous mott
  transition in semiconductor moiré superlattices},\ }\href
  {https://doi.org/10.1038/s41586-021-03853-0} {\bibfield  {journal} {\bibinfo
  {journal} {Nature}\ }\textbf {\bibinfo {volume} {597}},\ \bibinfo {pages}
  {350} (\bibinfo {year} {2021}{\natexlab{a}})}\BibitemShut {NoStop}%
\bibitem [{\citenamefont {Regan}\ \emph {et~al.}(2020)\citenamefont {Regan},
  \citenamefont {Wang}, \citenamefont {Jin}, \citenamefont {Bakti~Utama},
  \citenamefont {Gao}, \citenamefont {Wei}, \citenamefont {Zhao}, \citenamefont
  {Zhao}, \citenamefont {Zhang}, \citenamefont {Yumigeta}, \citenamefont
  {Blei}, \citenamefont {Carlström}, \citenamefont {Watanabe}, \citenamefont
  {Taniguchi}, \citenamefont {Tongay}, \citenamefont {Crommie}, \citenamefont
  {Zettl},\ and\ \citenamefont {Wang}}]{Regan2020}%
  \BibitemOpen
  \bibfield  {author} {\bibinfo {author} {\bibfnamefont {E.~C.}\ \bibnamefont
  {Regan}}, \bibinfo {author} {\bibfnamefont {D.}~\bibnamefont {Wang}},
  \bibinfo {author} {\bibfnamefont {C.}~\bibnamefont {Jin}}, \bibinfo {author}
  {\bibfnamefont {M.~I.}\ \bibnamefont {Bakti~Utama}}, \bibinfo {author}
  {\bibfnamefont {B.}~\bibnamefont {Gao}}, \bibinfo {author} {\bibfnamefont
  {X.}~\bibnamefont {Wei}}, \bibinfo {author} {\bibfnamefont {S.}~\bibnamefont
  {Zhao}}, \bibinfo {author} {\bibfnamefont {W.}~\bibnamefont {Zhao}}, \bibinfo
  {author} {\bibfnamefont {Z.}~\bibnamefont {Zhang}}, \bibinfo {author}
  {\bibfnamefont {K.}~\bibnamefont {Yumigeta}}, \bibinfo {author}
  {\bibfnamefont {M.}~\bibnamefont {Blei}}, \bibinfo {author} {\bibfnamefont
  {J.~D.}\ \bibnamefont {Carlström}}, \bibinfo {author} {\bibfnamefont
  {K.}~\bibnamefont {Watanabe}}, \bibinfo {author} {\bibfnamefont
  {T.}~\bibnamefont {Taniguchi}}, \bibinfo {author} {\bibfnamefont
  {S.}~\bibnamefont {Tongay}}, \bibinfo {author} {\bibfnamefont
  {M.}~\bibnamefont {Crommie}}, \bibinfo {author} {\bibfnamefont
  {A.}~\bibnamefont {Zettl}},\ and\ \bibinfo {author} {\bibfnamefont
  {F.}~\bibnamefont {Wang}},\ }\bibfield  {title} {\bibinfo {title} {Mott and
  generalized {W}igner crystal states in {WSe}$_2$/{WS}$_2$ moiré
  superlattices},\ }\href {https://doi.org/10.1038/s41586-020-2092-4}
  {\bibfield  {journal} {\bibinfo  {journal} {Nature}\ }\textbf {\bibinfo
  {volume} {579}},\ \bibinfo {pages} {359} (\bibinfo {year}
  {2020})}\BibitemShut {NoStop}%
\bibitem [{\citenamefont {Shimazaki}\ \emph {et~al.}(2020)\citenamefont
  {Shimazaki}, \citenamefont {Schwartz}, \citenamefont {Watanabe},
  \citenamefont {Taniguchi}, \citenamefont {Kroner},\ and\ \citenamefont
  {Imamoğlu}}]{Shimazaki2020}%
  \BibitemOpen
  \bibfield  {author} {\bibinfo {author} {\bibfnamefont {Y.}~\bibnamefont
  {Shimazaki}}, \bibinfo {author} {\bibfnamefont {I.}~\bibnamefont {Schwartz}},
  \bibinfo {author} {\bibfnamefont {K.}~\bibnamefont {Watanabe}}, \bibinfo
  {author} {\bibfnamefont {T.}~\bibnamefont {Taniguchi}}, \bibinfo {author}
  {\bibfnamefont {M.}~\bibnamefont {Kroner}},\ and\ \bibinfo {author}
  {\bibfnamefont {A.}~\bibnamefont {Imamoğlu}},\ }\bibfield  {title} {\bibinfo
  {title} {Strongly correlated electrons and hybrid excitons in a moiré
  heterostructure},\ }\href {https://doi.org/10.1038/s41586-020-2191-2}
  {\bibfield  {journal} {\bibinfo  {journal} {Nature}\ }\textbf {\bibinfo
  {volume} {580}},\ \bibinfo {pages} {472} (\bibinfo {year}
  {2020})}\BibitemShut {NoStop}%
\bibitem [{\citenamefont {Xu}\ \emph {et~al.}(2020)\citenamefont {Xu},
  \citenamefont {Liu}, \citenamefont {Rhodes}, \citenamefont {Watanabe},
  \citenamefont {Taniguchi}, \citenamefont {Hone}, \citenamefont {Elser},
  \citenamefont {Mak},\ and\ \citenamefont {Shan}}]{Xu2020}%
  \BibitemOpen
  \bibfield  {author} {\bibinfo {author} {\bibfnamefont {Y.}~\bibnamefont
  {Xu}}, \bibinfo {author} {\bibfnamefont {S.}~\bibnamefont {Liu}}, \bibinfo
  {author} {\bibfnamefont {D.~A.}\ \bibnamefont {Rhodes}}, \bibinfo {author}
  {\bibfnamefont {K.}~\bibnamefont {Watanabe}}, \bibinfo {author}
  {\bibfnamefont {T.}~\bibnamefont {Taniguchi}}, \bibinfo {author}
  {\bibfnamefont {J.}~\bibnamefont {Hone}}, \bibinfo {author} {\bibfnamefont
  {V.}~\bibnamefont {Elser}}, \bibinfo {author} {\bibfnamefont {K.~F.}\
  \bibnamefont {Mak}},\ and\ \bibinfo {author} {\bibfnamefont {J.}~\bibnamefont
  {Shan}},\ }\bibfield  {title} {\bibinfo {title} {Correlated insulating states
  at fractional fillings of moiré superlattices},\ }\href
  {https://doi.org/10.1038/s41586-020-2868-6} {\bibfield  {journal} {\bibinfo
  {journal} {Nature}\ }\textbf {\bibinfo {volume} {587}},\ \bibinfo {pages}
  {214} (\bibinfo {year} {2020})}\BibitemShut {NoStop}%
\bibitem [{\citenamefont {Li}\ \emph {et~al.}(2021{\natexlab{b}})\citenamefont
  {Li}, \citenamefont {Li}, \citenamefont {Regan}, \citenamefont {Wang},
  \citenamefont {Zhao}, \citenamefont {Kahn}, \citenamefont {Yumigeta},
  \citenamefont {Blei}, \citenamefont {Taniguchi}, \citenamefont {Watanabe},
  \citenamefont {Tongay}, \citenamefont {Zettl}, \citenamefont {Crommie},\ and\
  \citenamefont {Wang}}]{Li-Wang2021}%
  \BibitemOpen
  \bibfield  {author} {\bibinfo {author} {\bibfnamefont {H.}~\bibnamefont
  {Li}}, \bibinfo {author} {\bibfnamefont {S.}~\bibnamefont {Li}}, \bibinfo
  {author} {\bibfnamefont {E.~C.}\ \bibnamefont {Regan}}, \bibinfo {author}
  {\bibfnamefont {D.}~\bibnamefont {Wang}}, \bibinfo {author} {\bibfnamefont
  {W.}~\bibnamefont {Zhao}}, \bibinfo {author} {\bibfnamefont {S.}~\bibnamefont
  {Kahn}}, \bibinfo {author} {\bibfnamefont {K.}~\bibnamefont {Yumigeta}},
  \bibinfo {author} {\bibfnamefont {M.}~\bibnamefont {Blei}}, \bibinfo {author}
  {\bibfnamefont {T.}~\bibnamefont {Taniguchi}}, \bibinfo {author}
  {\bibfnamefont {K.}~\bibnamefont {Watanabe}}, \bibinfo {author}
  {\bibfnamefont {S.}~\bibnamefont {Tongay}}, \bibinfo {author} {\bibfnamefont
  {A.}~\bibnamefont {Zettl}}, \bibinfo {author} {\bibfnamefont {M.~F.}\
  \bibnamefont {Crommie}},\ and\ \bibinfo {author} {\bibfnamefont
  {F.}~\bibnamefont {Wang}},\ }\bibfield  {title} {\bibinfo {title} {Imaging
  two-dimensional generalized wigner crystals},\ }\href
  {https://doi.org/10.1038/s41586-021-03874-9} {\bibfield  {journal} {\bibinfo
  {journal} {Nature}\ }\textbf {\bibinfo {volume} {597}},\ \bibinfo {pages}
  {650} (\bibinfo {year} {2021}{\natexlab{b}})}\BibitemShut {NoStop}%
\bibitem [{\citenamefont {Zhang}\ \emph {et~al.}(2018)\citenamefont {Zhang},
  \citenamefont {Surrente}, \citenamefont {Baranowski}, \citenamefont {Maude},
  \citenamefont {Gant}, \citenamefont {{Castellanos-Gomez}},\ and\
  \citenamefont {Plochocka}}]{ZhangPlochocka2018}%
  \BibitemOpen
  \bibfield  {author} {\bibinfo {author} {\bibfnamefont {N.}~\bibnamefont
  {Zhang}}, \bibinfo {author} {\bibfnamefont {A.}~\bibnamefont {Surrente}},
  \bibinfo {author} {\bibfnamefont {M.}~\bibnamefont {Baranowski}}, \bibinfo
  {author} {\bibfnamefont {D.~K.}\ \bibnamefont {Maude}}, \bibinfo {author}
  {\bibfnamefont {P.}~\bibnamefont {Gant}}, \bibinfo {author} {\bibfnamefont
  {A.}~\bibnamefont {{Castellanos-Gomez}}},\ and\ \bibinfo {author}
  {\bibfnamefont {P.}~\bibnamefont {Plochocka}},\ }\bibfield  {title} {\bibinfo
  {title} {Moir{\'e} {{Intralayer Excitons}} in a {{MoSe$_2$}}/{{MoS$_2$
  Heterostructure}}},\ }\href {https://doi.org/10.1021/acs.nanolett.8b03266}
  {\bibfield  {journal} {\bibinfo  {journal} {Nano Lett.}\ }\textbf {\bibinfo
  {volume} {18}},\ \bibinfo {pages} {7651} (\bibinfo {year}
  {2018})}\BibitemShut {NoStop}%
\bibitem [{\citenamefont {Gong}\ \emph {et~al.}(2013)\citenamefont {Gong},
  \citenamefont {Liu}, \citenamefont {Yu}, \citenamefont {Xiao}, \citenamefont
  {Cui}, \citenamefont {Xu},\ and\ \citenamefont {Yao}}]{Gong2013}%
  \BibitemOpen
  \bibfield  {author} {\bibinfo {author} {\bibfnamefont {Z.}~\bibnamefont
  {Gong}}, \bibinfo {author} {\bibfnamefont {G.-B.}\ \bibnamefont {Liu}},
  \bibinfo {author} {\bibfnamefont {H.}~\bibnamefont {Yu}}, \bibinfo {author}
  {\bibfnamefont {D.}~\bibnamefont {Xiao}}, \bibinfo {author} {\bibfnamefont
  {X.}~\bibnamefont {Cui}}, \bibinfo {author} {\bibfnamefont {X.}~\bibnamefont
  {Xu}},\ and\ \bibinfo {author} {\bibfnamefont {W.}~\bibnamefont {Yao}},\
  }\bibfield  {title} {\bibinfo {title} {Magnetoelectric effects and
  valley-controlled spin quantum gates in transition metal dichalcogenide
  bilayers},\ }\href {https://www.nature.com/articles/ncomms3053} {\bibfield
  {journal} {\bibinfo  {journal} {Nat. Commun.}\ }\textbf {\bibinfo {volume}
  {4}},\ \bibinfo {pages} {1} (\bibinfo {year} {2013})}\BibitemShut {NoStop}%
\bibitem [{\citenamefont {Scuri}\ \emph {et~al.}(2020)\citenamefont {Scuri},
  \citenamefont {Andersen}, \citenamefont {Zhou}, \citenamefont {Wild},
  \citenamefont {Sung}, \citenamefont {Gelly}, \citenamefont
  {B{\ifmmode\acute{e}\else\'{e}\fi}rub{\ifmmode\acute{e}\else\'{e}\fi}},
  \citenamefont {Heo}, \citenamefont {Shao}, \citenamefont {Joe}, \citenamefont
  {Mier~Valdivia}, \citenamefont {Taniguchi}, \citenamefont {Watanabe},
  \citenamefont {Lon{\ifmmode\check{c}\else\v{c}\fi}ar}, \citenamefont {Kim},
  \citenamefont {Lukin},\ and\ \citenamefont {Park}}]{Scuri2020}%
  \BibitemOpen
  \bibfield  {author} {\bibinfo {author} {\bibfnamefont {G.}~\bibnamefont
  {Scuri}}, \bibinfo {author} {\bibfnamefont {T.~I.}\ \bibnamefont {Andersen}},
  \bibinfo {author} {\bibfnamefont {Y.}~\bibnamefont {Zhou}}, \bibinfo {author}
  {\bibfnamefont {D.~S.}\ \bibnamefont {Wild}}, \bibinfo {author}
  {\bibfnamefont {J.}~\bibnamefont {Sung}}, \bibinfo {author} {\bibfnamefont
  {R.~J.}\ \bibnamefont {Gelly}}, \bibinfo {author} {\bibfnamefont
  {D.}~\bibnamefont
  {B{\ifmmode\acute{e}\else\'{e}\fi}rub{\ifmmode\acute{e}\else\'{e}\fi}}},
  \bibinfo {author} {\bibfnamefont {H.}~\bibnamefont {Heo}}, \bibinfo {author}
  {\bibfnamefont {L.}~\bibnamefont {Shao}}, \bibinfo {author} {\bibfnamefont
  {A.~Y.}\ \bibnamefont {Joe}}, \bibinfo {author} {\bibfnamefont {A.~M.}\
  \bibnamefont {Mier~Valdivia}}, \bibinfo {author} {\bibfnamefont
  {T.}~\bibnamefont {Taniguchi}}, \bibinfo {author} {\bibfnamefont
  {K.}~\bibnamefont {Watanabe}}, \bibinfo {author} {\bibfnamefont
  {M.}~\bibnamefont {Lon{\ifmmode\check{c}\else\v{c}\fi}ar}}, \bibinfo {author}
  {\bibfnamefont {P.}~\bibnamefont {Kim}}, \bibinfo {author} {\bibfnamefont
  {M.~D.}\ \bibnamefont {Lukin}},\ and\ \bibinfo {author} {\bibfnamefont
  {H.}~\bibnamefont {Park}},\ }\bibfield  {title} {\bibinfo {title}
  {{Electrically Tunable Valley Dynamics in Twisted
  ${\mathrm{WSe}}_{2}/{\mathrm{WSe}}_{2}$ Bilayers}},\ }\href
  {https://doi.org/10.1103/PhysRevLett.124.217403} {\bibfield  {journal}
  {\bibinfo  {journal} {Phys. Rev. Lett.}\ }\textbf {\bibinfo {volume} {124}},\
  \bibinfo {pages} {217403} (\bibinfo {year} {2020})}\BibitemShut {NoStop}%
\bibitem [{\citenamefont {Lin}\ \emph {et~al.}(2021)\citenamefont {Lin},
  \citenamefont {Faria~Junior}, \citenamefont {Bauer}, \citenamefont {Peng},
  \citenamefont {Monserrat}, \citenamefont {Gmitra}, \citenamefont {Fabian},
  \citenamefont {Bange},\ and\ \citenamefont {Lupton}}]{LinNonlinear2021}%
  \BibitemOpen
  \bibfield  {author} {\bibinfo {author} {\bibfnamefont {K.-Q.}\ \bibnamefont
  {Lin}}, \bibinfo {author} {\bibfnamefont {P.~E.}\ \bibnamefont
  {Faria~Junior}}, \bibinfo {author} {\bibfnamefont {J.~M.}\ \bibnamefont
  {Bauer}}, \bibinfo {author} {\bibfnamefont {B.}~\bibnamefont {Peng}},
  \bibinfo {author} {\bibfnamefont {B.}~\bibnamefont {Monserrat}}, \bibinfo
  {author} {\bibfnamefont {M.}~\bibnamefont {Gmitra}}, \bibinfo {author}
  {\bibfnamefont {J.}~\bibnamefont {Fabian}}, \bibinfo {author} {\bibfnamefont
  {S.}~\bibnamefont {Bange}},\ and\ \bibinfo {author} {\bibfnamefont {J.~M.}\
  \bibnamefont {Lupton}},\ }\bibfield  {title} {\bibinfo {title} {{Twist-angle
  engineering of excitonic quantum interference and optical nonlinearities in
  stacked 2D semiconductors}},\ }\href
  {https://doi.org/10.1038/s41467-021-21547-z} {\bibfield  {journal} {\bibinfo
  {journal} {Nat. Commun.}\ }\textbf {\bibinfo {volume} {12}},\ \bibinfo
  {pages} {1} (\bibinfo {year} {2021})}\BibitemShut {NoStop}%
\bibitem [{\citenamefont {Ma}\ \emph {et~al.}(2021)\citenamefont {Ma},
  \citenamefont {Nguyen}, \citenamefont {Wang}, \citenamefont {Zeng},
  \citenamefont {Watanabe}, \citenamefont {Taniguchi}, \citenamefont
  {MacDonald}, \citenamefont {Mak},\ and\ \citenamefont {Shan}}]{Ma2021}%
  \BibitemOpen
  \bibfield  {author} {\bibinfo {author} {\bibfnamefont {L.}~\bibnamefont
  {Ma}}, \bibinfo {author} {\bibfnamefont {P.~X.}\ \bibnamefont {Nguyen}},
  \bibinfo {author} {\bibfnamefont {Z.}~\bibnamefont {Wang}}, \bibinfo {author}
  {\bibfnamefont {Y.}~\bibnamefont {Zeng}}, \bibinfo {author} {\bibfnamefont
  {K.}~\bibnamefont {Watanabe}}, \bibinfo {author} {\bibfnamefont
  {T.}~\bibnamefont {Taniguchi}}, \bibinfo {author} {\bibfnamefont {A.~H.}\
  \bibnamefont {MacDonald}}, \bibinfo {author} {\bibfnamefont {K.~F.}\
  \bibnamefont {Mak}},\ and\ \bibinfo {author} {\bibfnamefont {J.}~\bibnamefont
  {Shan}},\ }\bibfield  {title} {\bibinfo {title} {Strongly correlated
  excitonic insulator in atomic double layers},\ }\href
  {https://doi.org/10.1038/s41586-021-03947-9} {\bibfield  {journal} {\bibinfo
  {journal} {Nature}\ }\textbf {\bibinfo {volume} {598}},\ \bibinfo {pages}
  {585} (\bibinfo {year} {2021})}\BibitemShut {NoStop}%
\bibitem [{\citenamefont {Wilson}\ \emph {et~al.}(2021)\citenamefont {Wilson},
  \citenamefont {Yao}, \citenamefont {Shan},\ and\ \citenamefont
  {Xu}}]{Wilson2021}%
  \BibitemOpen
  \bibfield  {author} {\bibinfo {author} {\bibfnamefont {N.~P.}\ \bibnamefont
  {Wilson}}, \bibinfo {author} {\bibfnamefont {W.}~\bibnamefont {Yao}},
  \bibinfo {author} {\bibfnamefont {J.}~\bibnamefont {Shan}},\ and\ \bibinfo
  {author} {\bibfnamefont {X.}~\bibnamefont {Xu}},\ }\bibfield  {title}
  {\bibinfo {title} {Excitons and emergent quantum phenomena in stacked 2d
  semiconductors},\ }\href {https://doi.org/10.1038/s41586-021-03979-1}
  {\bibfield  {journal} {\bibinfo  {journal} {Nature}\ }\textbf {\bibinfo
  {volume} {599}},\ \bibinfo {pages} {383} (\bibinfo {year}
  {2021})}\BibitemShut {NoStop}%
\bibitem [{\citenamefont {Joe}\ \emph {et~al.}(2021)\citenamefont {Joe},
  \citenamefont {Jauregui}, \citenamefont {Pistunova}, \citenamefont
  {Mier~Valdivia}, \citenamefont {Lu}, \citenamefont {Wild}, \citenamefont
  {Scuri}, \citenamefont {De~Greve}, \citenamefont {Gelly}, \citenamefont
  {Zhou}, \citenamefont {Sung}, \citenamefont {Sushko}, \citenamefont
  {Taniguchi}, \citenamefont {Watanabe}, \citenamefont {Smirnov}, \citenamefont
  {Lukin}, \citenamefont {Park},\ and\ \citenamefont {Kim}}]{Joe2021}%
  \BibitemOpen
  \bibfield  {author} {\bibinfo {author} {\bibfnamefont {A.~Y.}\ \bibnamefont
  {Joe}}, \bibinfo {author} {\bibfnamefont {L.~A.}\ \bibnamefont {Jauregui}},
  \bibinfo {author} {\bibfnamefont {K.}~\bibnamefont {Pistunova}}, \bibinfo
  {author} {\bibfnamefont {A.~M.}\ \bibnamefont {Mier~Valdivia}}, \bibinfo
  {author} {\bibfnamefont {Z.}~\bibnamefont {Lu}}, \bibinfo {author}
  {\bibfnamefont {D.~S.}\ \bibnamefont {Wild}}, \bibinfo {author}
  {\bibfnamefont {G.}~\bibnamefont {Scuri}}, \bibinfo {author} {\bibfnamefont
  {K.}~\bibnamefont {De~Greve}}, \bibinfo {author} {\bibfnamefont {R.~J.}\
  \bibnamefont {Gelly}}, \bibinfo {author} {\bibfnamefont {Y.}~\bibnamefont
  {Zhou}}, \bibinfo {author} {\bibfnamefont {J.}~\bibnamefont {Sung}}, \bibinfo
  {author} {\bibfnamefont {A.}~\bibnamefont {Sushko}}, \bibinfo {author}
  {\bibfnamefont {T.}~\bibnamefont {Taniguchi}}, \bibinfo {author}
  {\bibfnamefont {K.}~\bibnamefont {Watanabe}}, \bibinfo {author}
  {\bibfnamefont {D.}~\bibnamefont {Smirnov}}, \bibinfo {author} {\bibfnamefont
  {M.~D.}\ \bibnamefont {Lukin}}, \bibinfo {author} {\bibfnamefont
  {H.}~\bibnamefont {Park}},\ and\ \bibinfo {author} {\bibfnamefont
  {P.}~\bibnamefont {Kim}},\ }\bibfield  {title} {\bibinfo {title}
  {Electrically controlled emission from singlet and triplet exciton species in
  atomically thin light-emitting diodes},\ }\href
  {https://doi.org/10.1103/PhysRevB.103.L161411} {\bibfield  {journal}
  {\bibinfo  {journal} {Phys. Rev. B}\ }\textbf {\bibinfo {volume} {103}},\
  \bibinfo {pages} {L161411} (\bibinfo {year} {2021})}\BibitemShut {NoStop}%
\bibitem [{\citenamefont {Ciarrocchi}\ \emph {et~al.}(2019)\citenamefont
  {Ciarrocchi}, \citenamefont {Unuchek}, \citenamefont {Avsar}, \citenamefont
  {Watanabe}, \citenamefont {Taniguchi},\ and\ \citenamefont
  {Kis}}]{Ciarrocchi2019}%
  \BibitemOpen
  \bibfield  {author} {\bibinfo {author} {\bibfnamefont {A.}~\bibnamefont
  {Ciarrocchi}}, \bibinfo {author} {\bibfnamefont {D.}~\bibnamefont {Unuchek}},
  \bibinfo {author} {\bibfnamefont {A.}~\bibnamefont {Avsar}}, \bibinfo
  {author} {\bibfnamefont {K.}~\bibnamefont {Watanabe}}, \bibinfo {author}
  {\bibfnamefont {T.}~\bibnamefont {Taniguchi}},\ and\ \bibinfo {author}
  {\bibfnamefont {A.}~\bibnamefont {Kis}},\ }\bibfield  {title} {\bibinfo
  {title} {Polarization switching and electrical control of interlayer excitons
  in two-dimensional van der {W}aals heterostructures},\ }\href
  {https://doi.org/10.1038/s41566-018-0325-y} {\bibfield  {journal} {\bibinfo
  {journal} {Nat. Photon.}\ }\textbf {\bibinfo {volume} {13}},\ \bibinfo
  {pages} {131} (\bibinfo {year} {2019})}\BibitemShut {NoStop}%
\bibitem [{\citenamefont {Brotons-Gisbert}\ \emph {et~al.}(2020)\citenamefont
  {Brotons-Gisbert}, \citenamefont {Baek}, \citenamefont {Molina-Sánchez},
  \citenamefont {Campbell}, \citenamefont {Scerri}, \citenamefont {White},
  \citenamefont {Watanabe}, \citenamefont {Taniguchi}, \citenamefont {Bonato},\
  and\ \citenamefont {Gerardot}}]{Brotons-Gisbert2020}%
  \BibitemOpen
  \bibfield  {author} {\bibinfo {author} {\bibfnamefont {M.}~\bibnamefont
  {Brotons-Gisbert}}, \bibinfo {author} {\bibfnamefont {H.}~\bibnamefont
  {Baek}}, \bibinfo {author} {\bibfnamefont {A.}~\bibnamefont
  {Molina-Sánchez}}, \bibinfo {author} {\bibfnamefont {A.}~\bibnamefont
  {Campbell}}, \bibinfo {author} {\bibfnamefont {E.}~\bibnamefont {Scerri}},
  \bibinfo {author} {\bibfnamefont {D.}~\bibnamefont {White}}, \bibinfo
  {author} {\bibfnamefont {K.}~\bibnamefont {Watanabe}}, \bibinfo {author}
  {\bibfnamefont {T.}~\bibnamefont {Taniguchi}}, \bibinfo {author}
  {\bibfnamefont {C.}~\bibnamefont {Bonato}},\ and\ \bibinfo {author}
  {\bibfnamefont {B.~D.}\ \bibnamefont {Gerardot}},\ }\bibfield  {title}
  {\bibinfo {title} {Spin-layer locking of interlayer excitons trapped in
  moiré potentials},\ }\href {https://doi.org/10.1038/s41563-020-0687-7}
  {\bibfield  {journal} {\bibinfo  {journal} {Nat. Mater.}\ }\textbf {\bibinfo
  {volume} {19}},\ \bibinfo {pages} {630} (\bibinfo {year} {2020})}\BibitemShut
  {NoStop}%
\bibitem [{\citenamefont {Gillen}\ and\ \citenamefont
  {Maultzsch}(2018)}]{Gillen2018}%
  \BibitemOpen
  \bibfield  {author} {\bibinfo {author} {\bibfnamefont {R.}~\bibnamefont
  {Gillen}}\ and\ \bibinfo {author} {\bibfnamefont {J.}~\bibnamefont
  {Maultzsch}},\ }\bibfield  {title} {\bibinfo {title} {Interlayer excitons in
  {MoSe}$_2$/{WSe}$_2$ heterostructures from first principles},\ }\href
  {https://doi.org/10.1103/PhysRevB.97.165306} {\bibfield  {journal} {\bibinfo
  {journal} {Phys. Rev. B}\ }\textbf {\bibinfo {volume} {97}},\ \bibinfo
  {pages} {165306} (\bibinfo {year} {2018})}\BibitemShut {NoStop}%
\bibitem [{\citenamefont {F{\ifmmode\ddot{o}\else\"{o}\fi}rg}\ \emph
  {et~al.}(2021)\citenamefont {F{\ifmmode\ddot{o}\else\"{o}\fi}rg},
  \citenamefont {Baimuratov}, \citenamefont {Kruchinin}, \citenamefont {Vovk},
  \citenamefont {Scherzer}, \citenamefont
  {F{\ifmmode\ddot{o}\else\"{o}\fi}rste}, \citenamefont {Funk}, \citenamefont
  {Watanabe}, \citenamefont {Taniguchi},\ and\ \citenamefont
  {H{\ifmmode\ddot{o}\else\"{o}\fi}gele}}]{Forg2021}%
  \BibitemOpen
  \bibfield  {author} {\bibinfo {author} {\bibfnamefont {M.}~\bibnamefont
  {F{\ifmmode\ddot{o}\else\"{o}\fi}rg}}, \bibinfo {author} {\bibfnamefont
  {A.~S.}\ \bibnamefont {Baimuratov}}, \bibinfo {author} {\bibfnamefont
  {S.~{\relax Yu}.}\ \bibnamefont {Kruchinin}}, \bibinfo {author}
  {\bibfnamefont {I.~A.}\ \bibnamefont {Vovk}}, \bibinfo {author}
  {\bibfnamefont {J.}~\bibnamefont {Scherzer}}, \bibinfo {author}
  {\bibfnamefont {J.}~\bibnamefont {F{\ifmmode\ddot{o}\else\"{o}\fi}rste}},
  \bibinfo {author} {\bibfnamefont {V.}~\bibnamefont {Funk}}, \bibinfo {author}
  {\bibfnamefont {K.}~\bibnamefont {Watanabe}}, \bibinfo {author}
  {\bibfnamefont {T.}~\bibnamefont {Taniguchi}},\ and\ \bibinfo {author}
  {\bibfnamefont {A.}~\bibnamefont {H{\ifmmode\ddot{o}\else\"{o}\fi}gele}},\
  }\bibfield  {title} {\bibinfo {title} {{Moir{\ifmmode\acute{e}\else\'{e}\fi}
  excitons in {MoSe}$_2$-{WSe}$_2$ heterobilayers and heterotrilayers}},\
  }\href {https://doi.org/10.1038/s41467-021-21822-z} {\bibfield  {journal}
  {\bibinfo  {journal} {Nat. Commun.}\ }\textbf {\bibinfo {volume} {12}},\
  \bibinfo {pages} {1} (\bibinfo {year} {2021})}\BibitemShut {NoStop}%
\bibitem [{\citenamefont {F{\ifmmode\ddot{o}\else\"{o}\fi}rg}\ \emph
  {et~al.}(2019)\citenamefont {F{\ifmmode\ddot{o}\else\"{o}\fi}rg},
  \citenamefont {Colombier}, \citenamefont {Patel}, \citenamefont {Lindlau},
  \citenamefont {Mohite}, \citenamefont {Yamaguchi}, \citenamefont {Glazov},
  \citenamefont {Hunger},\ and\ \citenamefont
  {H{\ifmmode\ddot{o}\else\"{o}\fi}gele}}]{Forg2019}%
  \BibitemOpen
  \bibfield  {author} {\bibinfo {author} {\bibfnamefont {M.}~\bibnamefont
  {F{\ifmmode\ddot{o}\else\"{o}\fi}rg}}, \bibinfo {author} {\bibfnamefont
  {L.}~\bibnamefont {Colombier}}, \bibinfo {author} {\bibfnamefont {R.~K.}\
  \bibnamefont {Patel}}, \bibinfo {author} {\bibfnamefont {J.}~\bibnamefont
  {Lindlau}}, \bibinfo {author} {\bibfnamefont {A.~D.}\ \bibnamefont {Mohite}},
  \bibinfo {author} {\bibfnamefont {H.}~\bibnamefont {Yamaguchi}}, \bibinfo
  {author} {\bibfnamefont {M.~M.}\ \bibnamefont {Glazov}}, \bibinfo {author}
  {\bibfnamefont {D.}~\bibnamefont {Hunger}},\ and\ \bibinfo {author}
  {\bibfnamefont {A.}~\bibnamefont {H{\ifmmode\ddot{o}\else\"{o}\fi}gele}},\
  }\bibfield  {title} {\bibinfo {title} {{Cavity-control of interlayer excitons
  in van der Waals heterostructures}},\ }\href
  {https://doi.org/10.1038/s41467-019-11620-z} {\bibfield  {journal} {\bibinfo
  {journal} {Nat. Commun.}\ }\textbf {\bibinfo {volume} {10}},\ \bibinfo
  {pages} {1} (\bibinfo {year} {2019})}\BibitemShut {NoStop}%
\bibitem [{\citenamefont {Wo\ifmmode~\acute{z}\else \'{z}\fi{}niak}\ \emph
  {et~al.}(2020)\citenamefont {Wo\ifmmode~\acute{z}\else \'{z}\fi{}niak},
  \citenamefont {Faria~Junior}, \citenamefont {Seifert}, \citenamefont
  {Chaves},\ and\ \citenamefont {Kunstmann}}]{Wozniak2020}%
  \BibitemOpen
  \bibfield  {author} {\bibinfo {author} {\bibfnamefont {T.}~\bibnamefont
  {Wo\ifmmode~\acute{z}\else \'{z}\fi{}niak}}, \bibinfo {author} {\bibfnamefont
  {P.~E.}\ \bibnamefont {Faria~Junior}}, \bibinfo {author} {\bibfnamefont
  {G.}~\bibnamefont {Seifert}}, \bibinfo {author} {\bibfnamefont
  {A.}~\bibnamefont {Chaves}},\ and\ \bibinfo {author} {\bibfnamefont
  {J.}~\bibnamefont {Kunstmann}},\ }\bibfield  {title} {\bibinfo {title}
  {{Exciton $g$ factors of van der Waals heterostructures from first-principles
  calculations}},\ }\href {https://doi.org/10.1103/PhysRevB.101.235408}
  {\bibfield  {journal} {\bibinfo  {journal} {Phys. Rev. B}\ }\textbf {\bibinfo
  {volume} {101}},\ \bibinfo {pages} {235408} (\bibinfo {year}
  {2020})}\BibitemShut {NoStop}%
\bibitem [{\citenamefont {Xuan}\ and\ \citenamefont {Quek}(2020)}]{Xuan2020}%
  \BibitemOpen
  \bibfield  {author} {\bibinfo {author} {\bibfnamefont {F.}~\bibnamefont
  {Xuan}}\ and\ \bibinfo {author} {\bibfnamefont {S.~Y.}\ \bibnamefont
  {Quek}},\ }\bibfield  {title} {\bibinfo {title} {Valley zeeman effect and
  landau levels in two-dimensional transition metal dichalcogenides},\ }\href
  {https://doi.org/10.1103/PhysRevResearch.2.033256} {\bibfield  {journal}
  {\bibinfo  {journal} {Phys. Rev. Research}\ }\textbf {\bibinfo {volume}
  {2}},\ \bibinfo {pages} {033256} (\bibinfo {year} {2020})}\BibitemShut
  {NoStop}%
\bibitem [{\citenamefont {Pizzocchero}\ \emph {et~al.}(2016)\citenamefont
  {Pizzocchero}, \citenamefont {Gammelgaard}, \citenamefont {Jessen},
  \citenamefont {Caridad}, \citenamefont {Wang}, \citenamefont {Hone},
  \citenamefont {Bøggild},\ and\ \citenamefont {Booth}}]{Pizzocchero2016}%
  \BibitemOpen
  \bibfield  {author} {\bibinfo {author} {\bibfnamefont {F.}~\bibnamefont
  {Pizzocchero}}, \bibinfo {author} {\bibfnamefont {L.}~\bibnamefont
  {Gammelgaard}}, \bibinfo {author} {\bibfnamefont {B.~S.}\ \bibnamefont
  {Jessen}}, \bibinfo {author} {\bibfnamefont {J.~M.}\ \bibnamefont {Caridad}},
  \bibinfo {author} {\bibfnamefont {L.}~\bibnamefont {Wang}}, \bibinfo {author}
  {\bibfnamefont {J.}~\bibnamefont {Hone}}, \bibinfo {author} {\bibfnamefont
  {P.}~\bibnamefont {Bøggild}},\ and\ \bibinfo {author} {\bibfnamefont
  {T.~J.}\ \bibnamefont {Booth}},\ }\bibfield  {title} {\bibinfo {title} {The
  hot pick-up technique for batch assembly of van der {W}aals
  heterostructures},\ }\href {https://doi.org/10.1038/ncomms11894} {\bibfield
  {journal} {\bibinfo  {journal} {Nat. Commun.}\ }\textbf {\bibinfo {volume}
  {7}},\ \bibinfo {pages} {11894} (\bibinfo {year} {2016})}\BibitemShut
  {NoStop}%
\bibitem [{\citenamefont {Cadiz}\ \emph {et~al.}(2017)\citenamefont {Cadiz},
  \citenamefont {Courtade}, \citenamefont {Robert}, \citenamefont {Wang},
  \citenamefont {Shen}, \citenamefont {Cai}, \citenamefont {Taniguchi},
  \citenamefont {Watanabe}, \citenamefont {Carrere}, \citenamefont {Lagarde},
  \citenamefont {Manca}, \citenamefont {Amand}, \citenamefont {Renucci},
  \citenamefont {Tongay}, \citenamefont {Marie},\ and\ \citenamefont
  {Urbaszek}}]{Cadiz2017}%
  \BibitemOpen
  \bibfield  {author} {\bibinfo {author} {\bibfnamefont {F.}~\bibnamefont
  {Cadiz}}, \bibinfo {author} {\bibfnamefont {E.}~\bibnamefont {Courtade}},
  \bibinfo {author} {\bibfnamefont {C.}~\bibnamefont {Robert}}, \bibinfo
  {author} {\bibfnamefont {G.}~\bibnamefont {Wang}}, \bibinfo {author}
  {\bibfnamefont {Y.}~\bibnamefont {Shen}}, \bibinfo {author} {\bibfnamefont
  {H.}~\bibnamefont {Cai}}, \bibinfo {author} {\bibfnamefont {T.}~\bibnamefont
  {Taniguchi}}, \bibinfo {author} {\bibfnamefont {K.}~\bibnamefont {Watanabe}},
  \bibinfo {author} {\bibfnamefont {H.}~\bibnamefont {Carrere}}, \bibinfo
  {author} {\bibfnamefont {D.}~\bibnamefont {Lagarde}}, \bibinfo {author}
  {\bibfnamefont {M.}~\bibnamefont {Manca}}, \bibinfo {author} {\bibfnamefont
  {T.}~\bibnamefont {Amand}}, \bibinfo {author} {\bibfnamefont
  {P.}~\bibnamefont {Renucci}}, \bibinfo {author} {\bibfnamefont
  {S.}~\bibnamefont {Tongay}}, \bibinfo {author} {\bibfnamefont
  {X.}~\bibnamefont {Marie}},\ and\ \bibinfo {author} {\bibfnamefont
  {B.}~\bibnamefont {Urbaszek}},\ }\bibfield  {title} {\bibinfo {title}
  {{Excitonic linewidth approaching the homogeneous limit in MoS$_2$-based van
  der Waals heterostructures}},\ }\href
  {https://doi.org/10.1103/PhysRevX.7.021026} {\bibfield  {journal} {\bibinfo
  {journal} {Phys. Rev. X}\ }\textbf {\bibinfo {volume} {7}},\ \bibinfo {pages}
  {021026} (\bibinfo {year} {2017})}\BibitemShut {NoStop}%
\bibitem [{\citenamefont {Neumann}\ \emph {et~al.}(2017)\citenamefont
  {Neumann}, \citenamefont {Lindlau}, \citenamefont {Colombier}, \citenamefont
  {Nutz}, \citenamefont {Najmaei}, \citenamefont {Lou}, \citenamefont {Mohite},
  \citenamefont {Yamaguchi},\ and\ \citenamefont
  {H{\ifmmode\ddot{o}\else\"{o}\fi}gele}}]{Neumann2017}%
  \BibitemOpen
  \bibfield  {author} {\bibinfo {author} {\bibfnamefont {A.}~\bibnamefont
  {Neumann}}, \bibinfo {author} {\bibfnamefont {J.}~\bibnamefont {Lindlau}},
  \bibinfo {author} {\bibfnamefont {L.}~\bibnamefont {Colombier}}, \bibinfo
  {author} {\bibfnamefont {M.}~\bibnamefont {Nutz}}, \bibinfo {author}
  {\bibfnamefont {S.}~\bibnamefont {Najmaei}}, \bibinfo {author} {\bibfnamefont
  {J.}~\bibnamefont {Lou}}, \bibinfo {author} {\bibfnamefont {A.~D.}\
  \bibnamefont {Mohite}}, \bibinfo {author} {\bibfnamefont {H.}~\bibnamefont
  {Yamaguchi}},\ and\ \bibinfo {author} {\bibfnamefont {A.}~\bibnamefont
  {H{\ifmmode\ddot{o}\else\"{o}\fi}gele}},\ }\bibfield  {title} {\bibinfo
  {title} {{Opto-valleytronic imaging of atomically thin semiconductors}},\
  }\href {https://doi.org/10.1038/nnano.2016.282} {\bibfield  {journal}
  {\bibinfo  {journal} {Nat. Nanotechnol.}\ }\textbf {\bibinfo {volume} {12}},\
  \bibinfo {pages} {329} (\bibinfo {year} {2017})}\BibitemShut {NoStop}%
\bibitem [{\citenamefont {Alexeev}\ \emph {et~al.}(2020)\citenamefont
  {Alexeev}, \citenamefont {Mullin}, \citenamefont {Ares}, \citenamefont
  {Nevison-Andrews}, \citenamefont {Skrypka}, \citenamefont {Godde},
  \citenamefont {Kozikov}, \citenamefont {Hague}, \citenamefont {Wang},
  \citenamefont {Novoselov}, \citenamefont {Fumagalli}, \citenamefont {Hobbs},\
  and\ \citenamefont {Tartakovskii}}]{Alexeev2020}%
  \BibitemOpen
  \bibfield  {author} {\bibinfo {author} {\bibfnamefont {E.~M.}\ \bibnamefont
  {Alexeev}}, \bibinfo {author} {\bibfnamefont {N.}~\bibnamefont {Mullin}},
  \bibinfo {author} {\bibfnamefont {P.}~\bibnamefont {Ares}}, \bibinfo {author}
  {\bibfnamefont {H.}~\bibnamefont {Nevison-Andrews}}, \bibinfo {author}
  {\bibfnamefont {O.}~\bibnamefont {Skrypka}}, \bibinfo {author} {\bibfnamefont
  {T.}~\bibnamefont {Godde}}, \bibinfo {author} {\bibfnamefont
  {A.}~\bibnamefont {Kozikov}}, \bibinfo {author} {\bibfnamefont
  {L.}~\bibnamefont {Hague}}, \bibinfo {author} {\bibfnamefont
  {Y.}~\bibnamefont {Wang}}, \bibinfo {author} {\bibfnamefont {K.~S.}\
  \bibnamefont {Novoselov}}, \bibinfo {author} {\bibfnamefont {L.}~\bibnamefont
  {Fumagalli}}, \bibinfo {author} {\bibfnamefont {J.~K.}\ \bibnamefont
  {Hobbs}},\ and\ \bibinfo {author} {\bibfnamefont {A.~I.}\ \bibnamefont
  {Tartakovskii}},\ }\bibfield  {title} {\bibinfo {title} {Emergence of highly
  linearly polarized interlayer exciton emission in {MoSe}$_2$/{WSe}$_2$
  heterobilayers with transfer-induced layer corrugation},\ }\href
  {https://doi.org/10.1021/acsnano.0c01146} {\bibfield  {journal} {\bibinfo
  {journal} {ACS Nano}\ }\textbf {\bibinfo {volume} {14}},\ \bibinfo {pages}
  {11110} (\bibinfo {year} {2020})}\BibitemShut {NoStop}%
\bibitem [{\citenamefont {Holler}\ \emph {et~al.}(2020)\citenamefont {Holler},
  \citenamefont {Meier}, \citenamefont {Kempf}, \citenamefont {Nagler},
  \citenamefont {Watanabe}, \citenamefont {Taniguchi}, \citenamefont {Korn},\
  and\ \citenamefont {Schüller}}]{Holler2020}%
  \BibitemOpen
  \bibfield  {author} {\bibinfo {author} {\bibfnamefont {J.}~\bibnamefont
  {Holler}}, \bibinfo {author} {\bibfnamefont {S.}~\bibnamefont {Meier}},
  \bibinfo {author} {\bibfnamefont {M.}~\bibnamefont {Kempf}}, \bibinfo
  {author} {\bibfnamefont {P.}~\bibnamefont {Nagler}}, \bibinfo {author}
  {\bibfnamefont {K.}~\bibnamefont {Watanabe}}, \bibinfo {author}
  {\bibfnamefont {T.}~\bibnamefont {Taniguchi}}, \bibinfo {author}
  {\bibfnamefont {T.}~\bibnamefont {Korn}},\ and\ \bibinfo {author}
  {\bibfnamefont {C.}~\bibnamefont {Schüller}},\ }\bibfield  {title} {\bibinfo
  {title} {{Low-frequency Raman scattering in WSe$_2$-MoSe$_2$ heterobilayers:
  Evidence for atomic reconstruction}},\ }\href
  {https://doi.org/10.1063/5.0012249} {\bibfield  {journal} {\bibinfo
  {journal} {Appl. Phys. Lett.}\ }\textbf {\bibinfo {volume} {117}},\ \bibinfo
  {pages} {013104} (\bibinfo {year} {2020})}\BibitemShut {NoStop}%
\bibitem [{\citenamefont {Zhang}\ \emph {et~al.}(2019)\citenamefont {Zhang},
  \citenamefont {Gogna}, \citenamefont {Burg}, \citenamefont {Horng},
  \citenamefont {Paik}, \citenamefont {Chou}, \citenamefont {Kim},
  \citenamefont {Tutuc},\ and\ \citenamefont {Deng}}]{Zhang2019}%
  \BibitemOpen
  \bibfield  {author} {\bibinfo {author} {\bibfnamefont {L.}~\bibnamefont
  {Zhang}}, \bibinfo {author} {\bibfnamefont {R.}~\bibnamefont {Gogna}},
  \bibinfo {author} {\bibfnamefont {G.~W.}\ \bibnamefont {Burg}}, \bibinfo
  {author} {\bibfnamefont {J.}~\bibnamefont {Horng}}, \bibinfo {author}
  {\bibfnamefont {E.}~\bibnamefont {Paik}}, \bibinfo {author} {\bibfnamefont
  {Y.-H.}\ \bibnamefont {Chou}}, \bibinfo {author} {\bibfnamefont
  {K.}~\bibnamefont {Kim}}, \bibinfo {author} {\bibfnamefont {E.}~\bibnamefont
  {Tutuc}},\ and\ \bibinfo {author} {\bibfnamefont {H.}~\bibnamefont {Deng}},\
  }\bibfield  {title} {\bibinfo {title} {Highly valley-polarized singlet and
  triplet interlayer excitons in van der waals heterostructure},\ }\href
  {https://doi.org/10.1103/PhysRevB.100.041402} {\bibfield  {journal} {\bibinfo
   {journal} {Phys. Rev. B}\ }\textbf {\bibinfo {volume} {100}},\ \bibinfo
  {pages} {041402} (\bibinfo {year} {2019})}\BibitemShut {NoStop}%
\bibitem [{\citenamefont {Wang}\ \emph
  {et~al.}(2020{\natexlab{b}})\citenamefont {Wang}, \citenamefont {Miao},
  \citenamefont {Li}, \citenamefont {Meng}, \citenamefont {Lu}, \citenamefont
  {Lian}, \citenamefont {Blei}, \citenamefont {Taniguchi}, \citenamefont
  {Watanabe}, \citenamefont {Tongay}, \citenamefont {Smirnov},\ and\
  \citenamefont {Shi}}]{Wang2020}%
  \BibitemOpen
  \bibfield  {author} {\bibinfo {author} {\bibfnamefont {T.}~\bibnamefont
  {Wang}}, \bibinfo {author} {\bibfnamefont {S.}~\bibnamefont {Miao}}, \bibinfo
  {author} {\bibfnamefont {Z.}~\bibnamefont {Li}}, \bibinfo {author}
  {\bibfnamefont {Y.}~\bibnamefont {Meng}}, \bibinfo {author} {\bibfnamefont
  {Z.}~\bibnamefont {Lu}}, \bibinfo {author} {\bibfnamefont {Z.}~\bibnamefont
  {Lian}}, \bibinfo {author} {\bibfnamefont {M.}~\bibnamefont {Blei}}, \bibinfo
  {author} {\bibfnamefont {T.}~\bibnamefont {Taniguchi}}, \bibinfo {author}
  {\bibfnamefont {K.}~\bibnamefont {Watanabe}}, \bibinfo {author}
  {\bibfnamefont {S.}~\bibnamefont {Tongay}}, \bibinfo {author} {\bibfnamefont
  {D.}~\bibnamefont {Smirnov}},\ and\ \bibinfo {author} {\bibfnamefont {S.-F.}\
  \bibnamefont {Shi}},\ }\bibfield  {title} {\bibinfo {title} {Giant
  {{Valley}}-{{Zeeman Splitting}} from {{Spin}}-{{Singlet}} and
  {{Spin}}-{{Triplet Interlayer Excitons}} in {{WSe}}{$_2$}/{{MoSe}}{$_2$}
  {{Heterostructure}}},\ }\href {https://doi.org/10.1021/acs.nanolett.9b04528}
  {\bibfield  {journal} {\bibinfo  {journal} {Nano Lett.}\ }\textbf {\bibinfo
  {volume} {20}},\ \bibinfo {pages} {694} (\bibinfo {year}
  {2020}{\natexlab{b}})}\BibitemShut {NoStop}%
\bibitem [{\citenamefont {Semina}\ \emph {et~al.}(2020)\citenamefont {Semina},
  \citenamefont {Glazov},\ and\ \citenamefont {Sherman}}]{Semina2020}%
  \BibitemOpen
  \bibfield  {author} {\bibinfo {author} {\bibfnamefont {M.~A.}\ \bibnamefont
  {Semina}}, \bibinfo {author} {\bibfnamefont {M.~M.}\ \bibnamefont {Glazov}},\
  and\ \bibinfo {author} {\bibfnamefont {E.}~\bibnamefont {Sherman}},\
  }\bibfield  {title} {\bibinfo {title} {Interlayer exciton–polaron in
  atomically thin semiconductors},\ }\href
  {https://doi.org/https://doi.org/10.1002/andp.202000339} {\bibfield
  {journal} {\bibinfo  {journal} {Ann. Phys.}\ }\textbf {\bibinfo {volume}
  {532}},\ \bibinfo {pages} {2000339} (\bibinfo {year} {2020})}\BibitemShut
  {NoStop}%
\bibitem [{\citenamefont {Vizner~Stern}\ \emph {et~al.}(2021)\citenamefont
  {Vizner~Stern}, \citenamefont {Waschitz}, \citenamefont {Cao}, \citenamefont
  {Nevo}, \citenamefont {Watanabe}, \citenamefont {Taniguchi}, \citenamefont
  {Sela}, \citenamefont {Urbakh}, \citenamefont {Hod},\ and\ \citenamefont
  {Ben~Shalom}}]{Vizner2021}%
  \BibitemOpen
  \bibfield  {author} {\bibinfo {author} {\bibfnamefont {M.}~\bibnamefont
  {Vizner~Stern}}, \bibinfo {author} {\bibfnamefont {Y.}~\bibnamefont
  {Waschitz}}, \bibinfo {author} {\bibfnamefont {W.}~\bibnamefont {Cao}},
  \bibinfo {author} {\bibfnamefont {I.}~\bibnamefont {Nevo}}, \bibinfo {author}
  {\bibfnamefont {K.}~\bibnamefont {Watanabe}}, \bibinfo {author}
  {\bibfnamefont {T.}~\bibnamefont {Taniguchi}}, \bibinfo {author}
  {\bibfnamefont {E.}~\bibnamefont {Sela}}, \bibinfo {author} {\bibfnamefont
  {M.}~\bibnamefont {Urbakh}}, \bibinfo {author} {\bibfnamefont
  {O.}~\bibnamefont {Hod}},\ and\ \bibinfo {author} {\bibfnamefont
  {M.}~\bibnamefont {Ben~Shalom}},\ }\bibfield  {title} {\bibinfo {title}
  {Interfacial ferroelectricity by van der {Waals} sliding},\ }\href
  {https://www.science.org/doi/10.1126/science.abe8177} {\bibfield  {journal}
  {\bibinfo  {journal} {Science}\ }\textbf {\bibinfo {volume} {372}},\ \bibinfo
  {pages} {1462} (\bibinfo {year} {2021})}\BibitemShut {NoStop}%
\bibitem [{\citenamefont {Woods}\ \emph {et~al.}(2021)\citenamefont {Woods},
  \citenamefont {Ares}, \citenamefont {Nevison-Andrews}, \citenamefont
  {Holwill}, \citenamefont {Fabregas}, \citenamefont {Guinea}, \citenamefont
  {Geim}, \citenamefont {Novoselov}, \citenamefont {Walet},\ and\ \citenamefont
  {Fumagalli}}]{Woods2021}%
  \BibitemOpen
  \bibfield  {author} {\bibinfo {author} {\bibfnamefont {C.}~\bibnamefont
  {Woods}}, \bibinfo {author} {\bibfnamefont {P.}~\bibnamefont {Ares}},
  \bibinfo {author} {\bibfnamefont {H.}~\bibnamefont {Nevison-Andrews}},
  \bibinfo {author} {\bibfnamefont {M.}~\bibnamefont {Holwill}}, \bibinfo
  {author} {\bibfnamefont {R.}~\bibnamefont {Fabregas}}, \bibinfo {author}
  {\bibfnamefont {F.}~\bibnamefont {Guinea}}, \bibinfo {author} {\bibfnamefont
  {A.}~\bibnamefont {Geim}}, \bibinfo {author} {\bibfnamefont {K.}~\bibnamefont
  {Novoselov}}, \bibinfo {author} {\bibfnamefont {N.}~\bibnamefont {Walet}},\
  and\ \bibinfo {author} {\bibfnamefont {L.}~\bibnamefont {Fumagalli}},\
  }\bibfield  {title} {\bibinfo {title} {Charge-polarized interfacial
  superlattices in marginally twisted hexagonal boron nitride},\ }\href
  {https://www.nature.com/articles/s41467-020-20667-2?proof=t%C2%A0} {\bibfield
   {journal} {\bibinfo  {journal} {Nat. Commun.}\ }\textbf {\bibinfo {volume}
  {12}},\ \bibinfo {pages} {1} (\bibinfo {year} {2021})}\BibitemShut {NoStop}%
\bibitem [{\citenamefont {Tiancheng}\ \emph {et~al.}(2021)\citenamefont
  {Tiancheng}, \citenamefont {Qi-Chao}, \citenamefont {Eric}, \citenamefont
  {Chong}, \citenamefont {Jimin}, \citenamefont {Takashi}, \citenamefont
  {Kenji}, \citenamefont {McGuire~Michael}, \citenamefont {Rainer},
  \citenamefont {Di}, \citenamefont {Ting}, \citenamefont {Jörg},\ and\
  \citenamefont {Xiaodong}}]{Tiancheng2021}%
  \BibitemOpen
  \bibfield  {author} {\bibinfo {author} {\bibfnamefont {S.}~\bibnamefont
  {Tiancheng}}, \bibinfo {author} {\bibfnamefont {S.}~\bibnamefont {Qi-Chao}},
  \bibinfo {author} {\bibfnamefont {A.}~\bibnamefont {Eric}}, \bibinfo {author}
  {\bibfnamefont {W.}~\bibnamefont {Chong}}, \bibinfo {author} {\bibfnamefont
  {Q.}~\bibnamefont {Jimin}}, \bibinfo {author} {\bibfnamefont
  {T.}~\bibnamefont {Takashi}}, \bibinfo {author} {\bibfnamefont
  {W.}~\bibnamefont {Kenji}}, \bibinfo {author} {\bibfnamefont
  {A.}~\bibnamefont {McGuire~Michael}}, \bibinfo {author} {\bibfnamefont
  {S.}~\bibnamefont {Rainer}}, \bibinfo {author} {\bibfnamefont
  {X.}~\bibnamefont {Di}}, \bibinfo {author} {\bibfnamefont {C.}~\bibnamefont
  {Ting}}, \bibinfo {author} {\bibfnamefont {W.}~\bibnamefont {Jörg}},\ and\
  \bibinfo {author} {\bibfnamefont {X.}~\bibnamefont {Xiaodong}},\ }\bibfield
  {title} {\bibinfo {title} {Direct visualization of magnetic domains and
  moiré magnetism in twisted 2d magnets},\ }\href
  {https://doi.org/10.1126/science.abj7478} {\bibfield  {journal} {\bibinfo
  {journal} {Science}\ }\textbf {\bibinfo {volume} {374}},\ \bibinfo {pages}
  {1140} (\bibinfo {year} {2021})}\BibitemShut {NoStop}%
\bibitem [{\citenamefont {Xie}\ \emph {et~al.}(2022)\citenamefont {Xie},
  \citenamefont {Luo}, \citenamefont {Ye}, \citenamefont {Ye}, \citenamefont
  {Ge}, \citenamefont {Sung}, \citenamefont {Rennich}, \citenamefont {Yan},
  \citenamefont {Fu}, \citenamefont {Tian}, \citenamefont {Lei}, \citenamefont
  {Hovden}, \citenamefont {Sun}, \citenamefont {He},\ and\ \citenamefont
  {Zhao}}]{Xie2022}%
  \BibitemOpen
  \bibfield  {author} {\bibinfo {author} {\bibfnamefont {H.}~\bibnamefont
  {Xie}}, \bibinfo {author} {\bibfnamefont {X.}~\bibnamefont {Luo}}, \bibinfo
  {author} {\bibfnamefont {G.}~\bibnamefont {Ye}}, \bibinfo {author}
  {\bibfnamefont {Z.}~\bibnamefont {Ye}}, \bibinfo {author} {\bibfnamefont
  {H.}~\bibnamefont {Ge}}, \bibinfo {author} {\bibfnamefont {S.~H.}\
  \bibnamefont {Sung}}, \bibinfo {author} {\bibfnamefont {E.}~\bibnamefont
  {Rennich}}, \bibinfo {author} {\bibfnamefont {S.}~\bibnamefont {Yan}},
  \bibinfo {author} {\bibfnamefont {Y.}~\bibnamefont {Fu}}, \bibinfo {author}
  {\bibfnamefont {S.}~\bibnamefont {Tian}}, \bibinfo {author} {\bibfnamefont
  {H.}~\bibnamefont {Lei}}, \bibinfo {author} {\bibfnamefont {R.}~\bibnamefont
  {Hovden}}, \bibinfo {author} {\bibfnamefont {K.}~\bibnamefont {Sun}},
  \bibinfo {author} {\bibfnamefont {R.}~\bibnamefont {He}},\ and\ \bibinfo
  {author} {\bibfnamefont {L.}~\bibnamefont {Zhao}},\ }\bibfield  {title}
  {\bibinfo {title} {Twist engineering of the two-dimensional magnetism in
  double bilayer chromium triiodide homostructures},\ }\href
  {https://doi.org/10.1038/s41567-021-01408-8} {\bibfield  {journal} {\bibinfo
  {journal} {Nat. Phys.}\ }\textbf {\bibinfo {volume} {18}},\ \bibinfo {pages}
  {30} (\bibinfo {year} {2022})}\BibitemShut {NoStop}%
\bibitem [{\citenamefont {Zhu}\ \emph {et~al.}(2017)\citenamefont {Zhu},
  \citenamefont {Shu}, \citenamefont {Jiang}, \citenamefont {Lv}, \citenamefont
  {Asokan}, \citenamefont {Omar}, \citenamefont {Yuan}, \citenamefont {Zhang},\
  and\ \citenamefont {Jin}}]{Zhu2017}%
  \BibitemOpen
  \bibfield  {author} {\bibinfo {author} {\bibfnamefont {D.}~\bibnamefont
  {Zhu}}, \bibinfo {author} {\bibfnamefont {H.}~\bibnamefont {Shu}}, \bibinfo
  {author} {\bibfnamefont {F.}~\bibnamefont {Jiang}}, \bibinfo {author}
  {\bibfnamefont {D.}~\bibnamefont {Lv}}, \bibinfo {author} {\bibfnamefont
  {V.}~\bibnamefont {Asokan}}, \bibinfo {author} {\bibfnamefont
  {O.}~\bibnamefont {Omar}}, \bibinfo {author} {\bibfnamefont {J.}~\bibnamefont
  {Yuan}}, \bibinfo {author} {\bibfnamefont {Z.}~\bibnamefont {Zhang}},\ and\
  \bibinfo {author} {\bibfnamefont {C.}~\bibnamefont {Jin}},\ }\bibfield
  {title} {\bibinfo {title} {{Capture the growth kinetics of {CVD} growth of
  two-dimensional {MoS$_2$}}},\ }\href
  {https://doi.org/10.1038/s41699-017-0010-x} {\bibfield  {journal} {\bibinfo
  {journal} {npj 2D Mater. Appl.}\ }\textbf {\bibinfo {volume} {1}},\ \bibinfo
  {pages} {1} (\bibinfo {year} {2017})}\BibitemShut {NoStop}%
\bibitem [{\citenamefont {Kim}\ \emph {et~al.}(2016)\citenamefont {Kim},
  \citenamefont {Yankowitz}, \citenamefont {Fallahazad}, \citenamefont {Kang},
  \citenamefont {Movva}, \citenamefont {Huang}, \citenamefont {Larentis},
  \citenamefont {Corbet}, \citenamefont {Taniguchi}, \citenamefont {Watanabe},
  \citenamefont {Banerjee}, \citenamefont {LeRoy},\ and\ \citenamefont
  {Tutuc}}]{Kim2016}%
  \BibitemOpen
  \bibfield  {author} {\bibinfo {author} {\bibfnamefont {K.}~\bibnamefont
  {Kim}}, \bibinfo {author} {\bibfnamefont {M.}~\bibnamefont {Yankowitz}},
  \bibinfo {author} {\bibfnamefont {B.}~\bibnamefont {Fallahazad}}, \bibinfo
  {author} {\bibfnamefont {S.}~\bibnamefont {Kang}}, \bibinfo {author}
  {\bibfnamefont {H.~C.~P.}\ \bibnamefont {Movva}}, \bibinfo {author}
  {\bibfnamefont {S.}~\bibnamefont {Huang}}, \bibinfo {author} {\bibfnamefont
  {S.}~\bibnamefont {Larentis}}, \bibinfo {author} {\bibfnamefont {C.~M.}\
  \bibnamefont {Corbet}}, \bibinfo {author} {\bibfnamefont {T.}~\bibnamefont
  {Taniguchi}}, \bibinfo {author} {\bibfnamefont {K.}~\bibnamefont {Watanabe}},
  \bibinfo {author} {\bibfnamefont {S.~K.}\ \bibnamefont {Banerjee}}, \bibinfo
  {author} {\bibfnamefont {B.~J.}\ \bibnamefont {LeRoy}},\ and\ \bibinfo
  {author} {\bibfnamefont {E.}~\bibnamefont {Tutuc}},\ }\bibfield  {title}
  {\bibinfo {title} {van der {W}aals heterostructures with high accuracy
  rotational alignment},\ }\href {https://doi.org/10.1021/acs.nanolett.5b05263}
  {\bibfield  {journal} {\bibinfo  {journal} {Nano Lett.}\ }\textbf {\bibinfo
  {volume} {16}},\ \bibinfo {pages} {1989} (\bibinfo {year}
  {2016})}\BibitemShut {NoStop}%
\end{thebibliography}
\end{document}


\title{SUPPLEMENTARY INFORMATION: \\
Excitons in mesoscopically reconstructed moir\'e heterostructures}

\author{Shen Zhao}
\affiliation{Fakult\"at f\"ur Physik, Munich Quantum Center, and
    Center for NanoScience (CeNS), Ludwig-Maximilians-Universit\"at
    M\"unchen, Geschwister-Scholl-Platz 1, 80539 M\"unchen, Germany}
\author{Zhijie Li}
\affiliation{Fakult\"at f\"ur Physik, Munich Quantum Center, and
    Center for NanoScience (CeNS), Ludwig-Maximilians-Universit\"at
    M\"unchen, Geschwister-Scholl-Platz 1, 80539 M\"unchen, Germany}
\author{Xin Huang}
\affiliation{Fakult\"at f\"ur Physik, Munich Quantum Center, and
    Center for NanoScience (CeNS), Ludwig-Maximilians-Universit\"at
    M\"unchen, Geschwister-Scholl-Platz 1, 80539 M\"unchen, Germany}
\affiliation{Present affiliations: Beijing National Laboratory for Condensed Matter Physics, Institute of Physics, Chinese Academy of Sciences, Beijing 100190, People's Republic of China \\ School of Physical Sciences, CAS Key Laboratory of Vacuum Physics, University of Chinese Academy of Sciences, Beijing 100190, People's Republic of China}
\author{Anna Rupp}
\affiliation{Fakult\"at f\"ur Physik, Munich Quantum Center, and
    Center for NanoScience (CeNS), Ludwig-Maximilians-Universit\"at
    M\"unchen, Geschwister-Scholl-Platz 1, 80539 M\"unchen, Germany}
\author{Jonas G{\"o}ser}
\affiliation{Fakult\"at f\"ur Physik, Munich Quantum Center, and
    Center for NanoScience (CeNS), Ludwig-Maximilians-Universit\"at
    M\"unchen, Geschwister-Scholl-Platz 1, 80539 M\"unchen, Germany}
\author{Ilia~A.~Vovk}
\affiliation{PhysNano Department, ITMO University, Saint Petersburg 197101, Russia}
\author{Stanislav~Yu.~Kruchinin}
\affiliation{Center for Computational Materials Sciences, Faculty of Physics, University of Vienna, Sensengasse 8/12, 1090 Vienna, Austria}
\affiliation{Nuance Communications Austria GmbH, Technologiestraße 8, 1120 Wien}
\author{Kenji Watanabe}
\affiliation{Research Center for Functional Materials, National Institute for Materials Science, 1-1 Namiki, Tsukuba 305-0044, Japan}
\author{Takashi Taniguchi}
\affiliation{International Center for Materials Nanoarchitectonics, 
    National Institute for Materials Science, 1-1 Namiki, Tsukuba 305-0044, Japan}
\author{Ismail Bilgin}
\affiliation{Fakult\"at f\"ur Physik, Munich Quantum Center, and
    Center for NanoScience (CeNS), Ludwig-Maximilians-Universit\"at
    M\"unchen, Geschwister-Scholl-Platz 1, 80539 M\"unchen, Germany}
\author{Anvar~S.~Baimuratov}
\affiliation{Fakult\"at f\"ur Physik, Munich Quantum Center, and
    Center for NanoScience (CeNS), Ludwig-Maximilians-Universit\"at
    M\"unchen, Geschwister-Scholl-Platz 1, 80539 M\"unchen, Germany}
\author{Alexander H{\"o}gele}
\affiliation{Fakult\"at f\"ur Physik, Munich Quantum Center, and
    Center for NanoScience (CeNS), Ludwig-Maximilians-Universit\"at
    M\"unchen, Geschwister-Scholl-Platz 1, 80539 M\"unchen, Germany}
\affiliation{Munich Center for Quantum Science and Technology (MCQST),
    Schellingtra\ss{}e 4, 80799 M\"unchen, Germany}

\date{}
\maketitle
\noindent \textbf{Supplementary Note $\mathbf{1}$: Mesoscopic reconstruction in SEM}

To image mesoscopic lattice reconstruction in MoSe$_2$-WSe$_2$ HBLs (as in Fig.~1f,g of the main text) we used the secondary electron (SE) imaging technique in low-energy scanning electron microscopy (SEM) initially pioneered on SiC crystals~\cite{Ashida2015} and recently adopted to twisted WSe$_2$ homobilayers~\cite{Andersen2021}. In brief, the technique is based on  out-lens detection of SEs emitted from atoms near the crystal surface by inelastic scattering of incoming primary electrons. In a TMD bilayer, the top and bottom hexagonal lattices form an effective cavity for electrons, rendering the scattering probability for the incoming beam in the cavity and thus the generation of SE dependent on the relative atomic positions in given stacking configurations. In optimized experimental geometries, the contrast of SE yield under channeling conditions allows to discriminate domains of different stacking configurations~\cite{Andersen2021}.
We adopted the technique to image R- and H-type MoSe$_2$-WSe$_2$ HBLs. The samples were prepared by the same method as for optical spectroscopy yet without hBN encapsulation. For both R- and H-type HBLs, the measurements were performed in a Raith eLine system (equipped with an Everhart-Thornley detector)at 1~keV electron beam energy and normal incidence for H-type and $\sim38^\circ$ tilt for R-type samples. Respective SEM images of H- and R-type MoSe$_2$-WSe$_2$ HBLs are shown in Supplementary Fig.~\ref{SEM_H} and \ref{SEM_R}. 

\begin{figure}[!ht]
\includegraphics[scale=1.0]{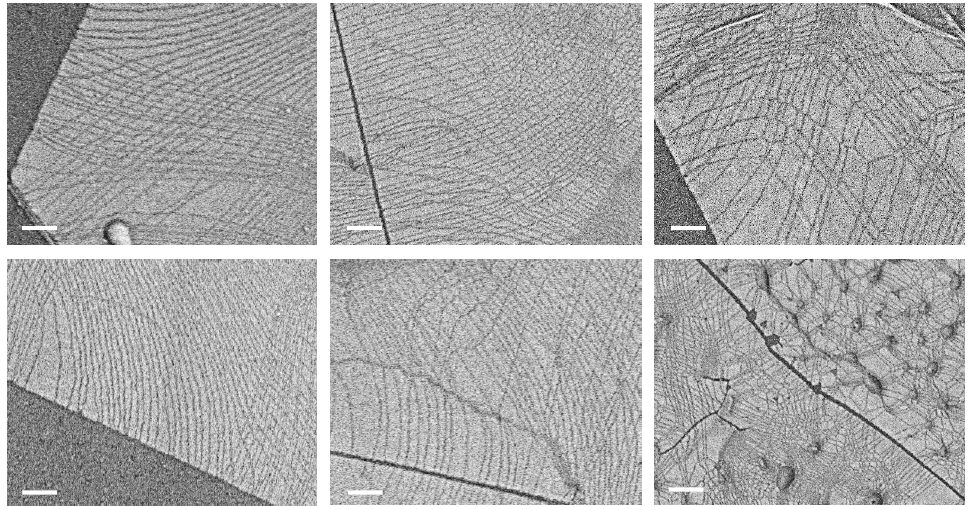}
\caption{SEM images of nearly aligned H-type MoSe$_2$-WSe$_2$ HBLs.
The scale bars are 600~nm.}
\label{SEM_H}
\end{figure}

\begin{figure}[!ht]
\includegraphics[scale=1.0]{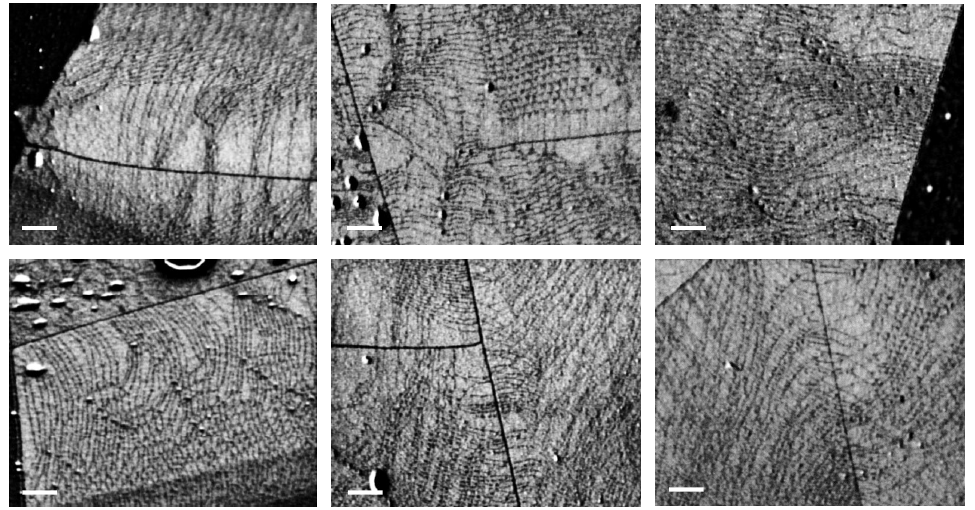}
\caption{SEM images of nearly aligned R-type MoSe$_2$-WSe$_2$ HBLs. The scale bars are 600~nm.}
\label{SEM_R}
\end{figure}

The images illustrate for nearly-aligned H- and R-type MoSe$_2$-WSe$_2$ HBLs clear departures from canonical moir\'e patterns or uniformly reconstructed periodic domains. They actually exhibit large variations in domain shape and size with the following common features: from the tips and edges to the cores, the reconstructed pattern evolves from large micron-sized two-dimensional (2D) domains to elongated one-dimensional (1D) stripes which merge into a network of zero-dimensional (0D) arrays with dimensions well below $100$~nm with variations in domain shape and size. In addition to line defects such as tips and edges, local strain such as interfacial bubbles in the bottom right panel of Fig.~\ref{SEM_H} can also give rise to mesoscopic reconstruction. The main difference between H- and R-type HBLs is that only one staking ($H_h^h$) is optimal in the former, whereas two stackings ($R_h^X$ and $R_h^M$) yield optimally reconstructed domains in the latter.

\clearpage
\noindent \textbf{Supplementary Note $\mathbf{2}$: Modelling of mesoscopic reconstruction}

Reconstruction of MoSe$_2$-WSe$_2$ HBL is driven by the interplay of interlayer adhesion energy and strain. The sum of the intralayer and interlayer energies of the lattice is given by the integral over the HBL area $S$ as \cite{Enaldiev2020}:
\begin{equation}
    \label{totalE}
    \mathcal{E}=\int_S \left[U(\mathbf{r})+W_s(\mathbf{r})\right]\mathrm{d}\mathbf{r},
\end{equation}
where $U(\mathbf{r})$ and $W_s(\mathbf{r})$ are the respective intralayer and interlayer strain energy densities which depend on the in-plane displacements in MoSe$_2$ and WSe$_2$ layers. Assuming equal elastic properties and lattice constants in both layers \cite{Carr2018}, the displacement fields are given by:
\begin{equation}
    \mathbf{u}_\mathrm{WSe_2}(\mathbf{r}) = 
    -\mathbf{u}_\mathrm{MoSe_2}(\mathbf{r}) = 
    \frac{\mathbf{u}(\mathbf{r})}{2} \equiv
    \left(
        \frac{u_x(\mathbf{r})}{2},
        \frac{u_y(\mathbf{r})}{2}
    \right).
\end{equation}

\noindent The intralayer strain energy density is given by:
\begin{equation}\label{eq_U}
    U(\mathbf{r}) = \frac{1}{2}
    \left[
        \frac{\lambda}{2}\left(\sum_{i}{u_{ii}(\mathbf{r})}\right)^2 +
        \mu\sum_{ij}{u_{ij}^2(\mathbf{r})}
    \right]
\end{equation}
with the strain tensor:
\begin{equation}\label{deriv}
    u_{ij}(\mathbf{r}) =
    \frac{1}{2}[\partial_ju_i(\mathbf{r})+\partial_iu_j(\mathbf{r})],
\end{equation}
where the first and second Lam\'{e} parameters $\lambda$ and $\mu$ are obtained by averaging over the respective parameters in MoSe$_2$ and WSe$_2$ MLs. 

\noindent The interlayer term $W_s(\mathbf{r}) = V_s(\mathbf{r})-\varepsilon Z_s^2(\mathbf{r})$ quantifies the adhesion energy density defined via the expressions~\cite{Enaldiev2021}:
\begin{align}
    V_s(\mathbf{r}) &=
    \sum_{n=1}^{3}\{A_1e^{-Qd_0}\cos{\phi_n(\mathbf{r})}+A_2e^{-Gd_0}\sin{[\phi_n(\mathbf{r})+\varphi_s]}\},\\
    Z_s(\mathbf{r}) &=
    \frac{1}{2\varepsilon}\sum_{n=1}^{3}\{QA_1e^{-Qd_0}\cos{\phi_n(\mathbf{r})}+GA_2e^{-Gd_0}\sin{[\phi_n(\mathbf{r})+\varphi_s]}\},
\end{align}
where the index $s = R,H$ denotes R-type and H-type stackings with the twist angle between the top and bottom layers $\theta$, and $\phi_n(\mathbf{r})=\mathbf{g}_n\cdot\mathbf{r}+\mathbf{G}_n\cdot\mathbf{u}(\mathbf{r})$ is defined through $\mathbf{g}_n=-\theta\,\mathbf{e}_z\times\mathbf{G}_n$ with the vectors of the first star of the reciprocal lattice $\pm\mathbf{G}_n$ $(n=1,2,3)$ and $G=|\mathbf{G}_n|=4\pi/(\sqrt{3}a)$ with the lattice constant $a$, $\varphi_\mathrm{R}=\pi/2$, $\varphi_\mathrm{H}=0$, and $Q=\sqrt{G^2+\rho^{-2}}$; $\varepsilon$,  $A_1$, $A_2$, $d_0$, and $\rho$ are fitting parameters. We use all material and fitting parameters for MoSe$_2$-WSe$_2$ HBL from Ref.~\cite{Enaldiev2020}: $a = 0.329$~nm, $d_0 = 0.69$~nm, $\rho = 0.052$~nm, $\lambda = 222$~eV/nm$^2$, $\mu = 306$~eV/nm$^2$, $A_1 = 77621500$~eV/nm$^2$, $A_2 = 84739$~eV/nm$^2$, and $\varepsilon = 189$~eV/nm$^4$.

\begin{figure}[!tb]
\includegraphics[scale=0.9]{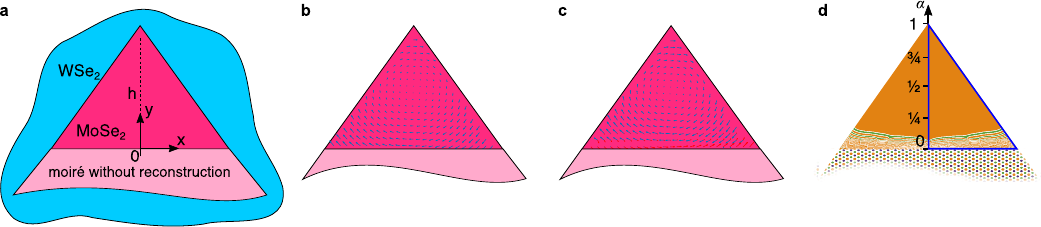}
\caption{\textbf{a}, Geometry of triangular MoSe$_2$-WSe$_2$ HBL tip used in simulations of mesoscopic reconstruction, with pink and blue colors denoting the top MoSe$_2$ ML and the bottom WSe$_2$ ML, respectively. \textbf{b} and \textbf{c}, Illustration of the initial and final displacement fields $\mathbf{u}^{\mathrm{i}}$ and $\mathbf{u}^{\mathrm{f}}$, respectively, for the reconstruction pattern of R-type HBL with $\alpha = 0.5$ shown in \textbf{d}.}
\label{triangle}
\end{figure}

In the next step, we assume an equilateral triangle for the shape of the HBL area $S$. Supplementary Fig.~\ref{triangle}a shows the triangle baseline placed on the $x$-axis as a borderline to the moir\'e region in the HBL with twist angle $\theta$ and without reconstruction. At the borderline, the displacement field is zero, $u_x(x,0) = u_y(x,0) = 0$, whereas the remaining two sides of the triangle are free from boundary conditions. We assume that the $y$-axis bisects the angle $\theta$ and divides the equilateral triangle into two right-angled triangles, yielding two constrains:
\begin{align}
    u_x(x,y) &= u_x(-x,y), \\
    u_y(x,y) &= -u_y(-x,y).
\end{align}
Finally, we discretize the displacement field $\mathbf{u}(\mathbf{r})$ with a square mesh, replace the integration in Eq.~\eqref{totalE} by the summation over the triangle area, and express the derivatives in Eq.~\eqref{deriv} through the finite differences method. The density of the square mesh (we choose $288$ moir\'e unit cells in all simulations of equilateral triangles) determines the number of discretization parameters (\textit{i.e.} if the mesh divides the height of triangle $h$ into $N$ equal sectors, the displacement field is defined by $\sim (N+1)^2/\sqrt 3$ parameters). To find the parameter set that minimizes the total energy in Eq.~\eqref{totalE}, we use the trust-region algorithm implemented in the Optimization Toolbox\textsuperscript{TM} of MATLAB\textsuperscript{\textregistered} to determine the local energy minimum of the final displacement field $\mathbf{u}^{\mathrm{f}}$ as a function of the initial displacement field $\mathbf{u}^{\mathrm{i}}$ set prior to reconstruction. In all calculations, we monitor the convergence of numerical results.

In Fig.~2a,d of the main text we used the following initial sets of displacement fields. The ideal moir\'e case was obtained with the initial set $\mathbf{u}^{\mathrm{i}} =\mathbf{0}$ (twisted HBL with $\theta = 0.4^\circ$) and without reconstruction. The periodic case illustrates the final set after the reconstruction with the initial field $\mathbf{u}^{\mathrm{i}}= \mathbf{0}$. All remaining cases show the final sets with the rotated initial displacement:
\begin{align}
    u^{\mathrm{i}}_{x} &=  \theta (y- \alpha h ), \\
    u^{\mathrm{i}}_{y} &= -\theta x,
\end{align}
which realizes the rotation of layers by $\pm\theta/2$ around the point $(0 ,\alpha h)$ and leads to untwisted HBL regions when the stacking expands from the rotation point to the entire area of the triangle. To denote these initial displacement fields we use the dimensionless $y$-coordinate of the rotation point, $\alpha$. 

Supplementary Fig.~\ref{triangle} illustrates the reconstruction in the tip of a triangular R-type HBL (Supplementary Fig.~\ref{triangle}a) for an initial field displacement $\mathbf{u}^{\mathrm{i}}$ with a rotation point at $\alpha = 0.5$ (Supplementary Fig.~\ref{triangle}b). The optimally reconstructed tip is described by the final displacement field $\mathbf{u}^{\mathrm{f}}$ (Supplementary Fig.~\ref{triangle}c) yielding the characteristic mesoscopic domain network (Supplementary Fig.~\ref{triangle}d). To complement the data of Fig.~2a,d of the main text, we show in Supplementary Figs.~\ref{fig_R_vector} and \ref{fig_H_vector} the vector fields of the initial and final displacements, $\mathbf{u}^{\mathrm{i}}$ and $\mathbf{u}^{\mathrm{f}}$, and by red arrows their difference $\mathbf{u}^{\mathrm{i}}-\mathbf{u}^{\mathrm{f}}$ after reconstruction. Additionally, we show the corresponding maps of intralayer and interlayer energies.

To illustrate the generalization of our results for more than one rotation point we show in Supplementary Fig.~\ref{tworotation} reconstruction maps for R- and H-type HBLs with a twist angle of $3^\circ$. The top panel shows periodic reconstruction, and the central panel illustrates reconstruction with one rotated point at $\alpha = 0$. The maps in the bottom panels were calculated for two rotation points at $\alpha = 0$ and $\alpha = 1/12$ with the initial displacement field:
\begin{equation}
    \mathbf{u^i} = 
        \begin{cases}
            \alpha = 0,    & \text{for } 0\leq y\leq 7h/12;\\
            \mathbf{0},    & \text{for } 7h/12 < y\leq 2h/3;\\
            \alpha = 1/12, & \text{for } 3h/4 < y\leq h.
        \end{cases}
\end{equation} 
This initial displacement field realizes three areas upon reconstruction: two untwisted domains around points with $\alpha = 0$ and $\alpha = 1/12$ with zero initial displacement in between. The resulting patterns exhibit regions of both periodically reconstructed and rotated patterns. This procedure can be expanded to model realistic samples with complex reconstruction patterns as in Supplementary Fig.~\ref{SEM_H} and \ref{SEM_R} by taking into account multiple rotation points for a given sample geometry and strain distribution.

\clearpage

\begin{figure}[t!]
\includegraphics[scale=0.93]{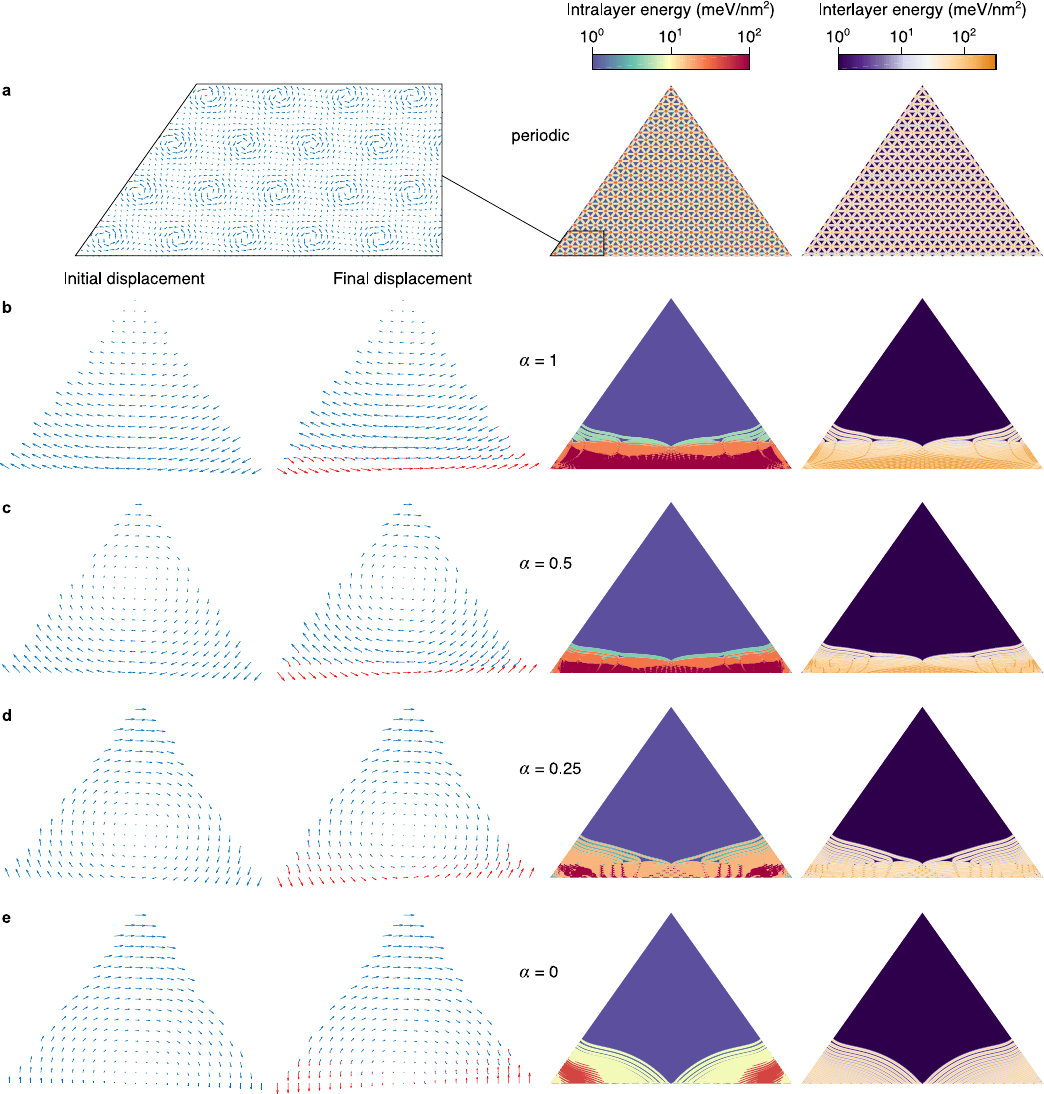}
\caption{\textbf{a -- e}, Initial and final displacement fields as well as intra- and interlayer energies of reconstructed patterns for R-type HBLs in Fig.~2a of the main text. For the periodic case, \textbf{a}, we show only the final displacement field. The panels \textbf{b -- e} relate to the initial displacement fields with rotation point at $\alpha = 1,0.5,0.25,0$.} 
\label{fig_R_vector}
\end{figure}

\clearpage

\begin{figure}[t!]
\includegraphics[scale=0.93]{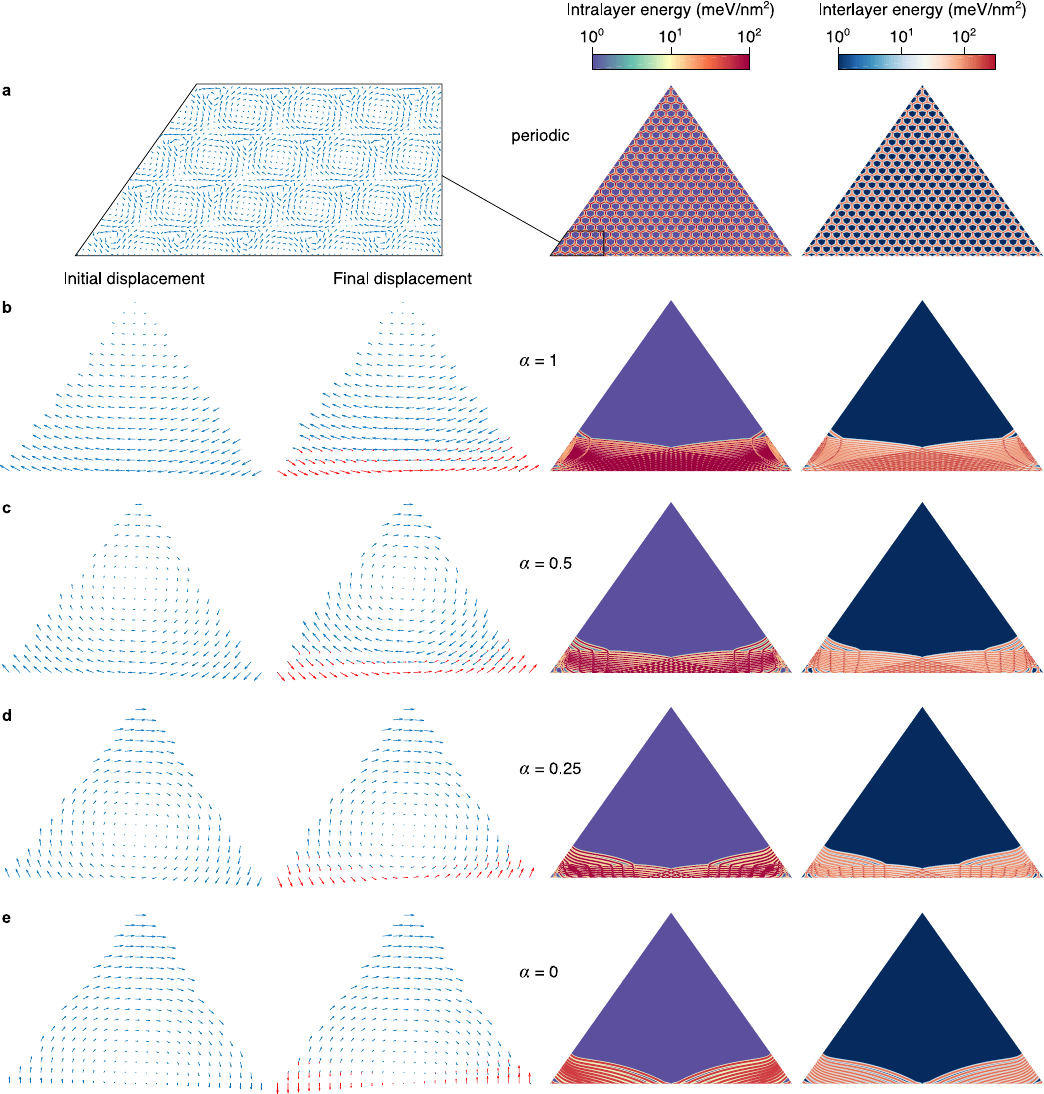}
\caption{\textbf{a -- e}, Initial and final displacement fields as well as intra- and interlayer energies of reconstructed patterns for H-type HBLs in Fig.~2d of the main text. For the periodic case, \textbf{a}, we show only the final displacement field. The panels \textbf{b -- e} relate to the initial displacement fields with rotation point at $\alpha = 1,0.5,0.25,0$.} 
\label{fig_H_vector}
\end{figure}

\clearpage

\begin{figure}[!t]
\includegraphics[scale=1]{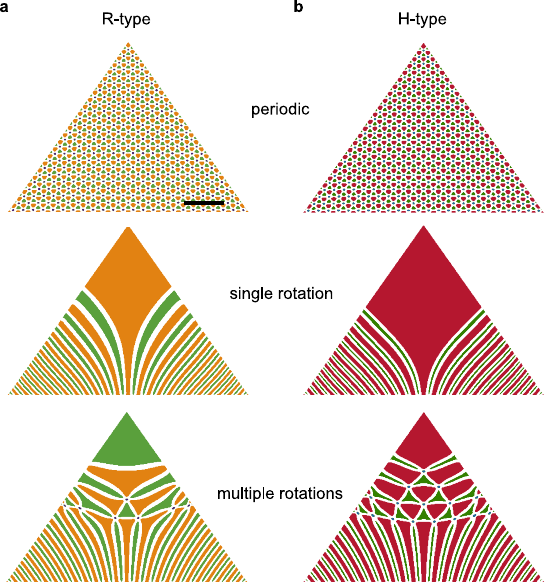}
\vspace{-6pt}
\caption{\textbf{a} and \textbf{b}, Reconstruction patterns in the tips of triangular R- and H-type HBLs with $\theta = 3^\circ$ twist angle for one rotation point at $\alpha = 0$ (central panel) and two rotation points at $\alpha = 0$ and $\alpha = 1/12$ (bottom panel). The top maps show periodic reconstruction without rotation for reference. The scale bar is $25$~nm.}
\label{tworotation}
\end{figure}

\clearpage

\noindent \textbf{Supplementary Note $\mathbf{3}$: Sample characteristics and spectral features}

Experimental studies were mainly performed on four samples with MoSe$_2$-WSe$_2$ HBLs encapsulated in hBN with characteristics shown in Supplementary Fig.~\ref{fig_PLmap}. Despite different geometries, all samples show two common features. First, aligned R-type HBLs are generally brighter than their H-type counterparts (as in sample 1 and 2 with nearly aligned R- and H-type HBLs), consistent with a factor of 10 difference in oscillator strengths of the lowest spin-singlet (R-type) and spin-triplet (H-type) interlayer exciton transitions. Second, all four samples exhibit substantial variations in PL intensity on micron scales across H- and R-type regions despite rather clean interfaces with only few interfacial bubbles. Moreover, spots of bright interlayer exciton PL are predominantly observed near HBL tips and edges, whereas sample cores are relatively dark. Based on observations in SEM and our reconstruction model, we interpret these spatial variations as arising from mesoscopic domain networks with extended 2D domains, 1D stripes and 0D domain arrays.

The spectral characteristics observed in cryogenic PL and differential reflectivity (DR) are shown in Supplementary Fig.~\ref{fig_PLspectra_1} - \ref{fig_PLspectra_4} for positions representative of 2D domains, 1D stripes and 0D domain arrays in each sample. Note that the signatures of 1D stripes with a high degree of linear polarization were observed only in R-type HBLs (sample 1 and 2 in Supplementary Fig.~\ref{fig_PLspectra_1} and Supplementary Fig.~\ref{fig_PLspectra_2}, respectively). Bright PL spots were consistently observed near line defects such as cracks and edges, whereas dark PL regions dominate the sample cores. In H-type samples, the PL characteristics of dark regions differ for nearly-aligned samples (samples 1, 2, and 4) and the sample with a twist angle of $\sim 3$°: Whereas the former exhibit quantum dot type features of localized excitons, the latter shows consistent exciton-polaron characteristics on different sample positions.           

\clearpage
\begin{figure}[!ht]
\includegraphics[scale=0.95]{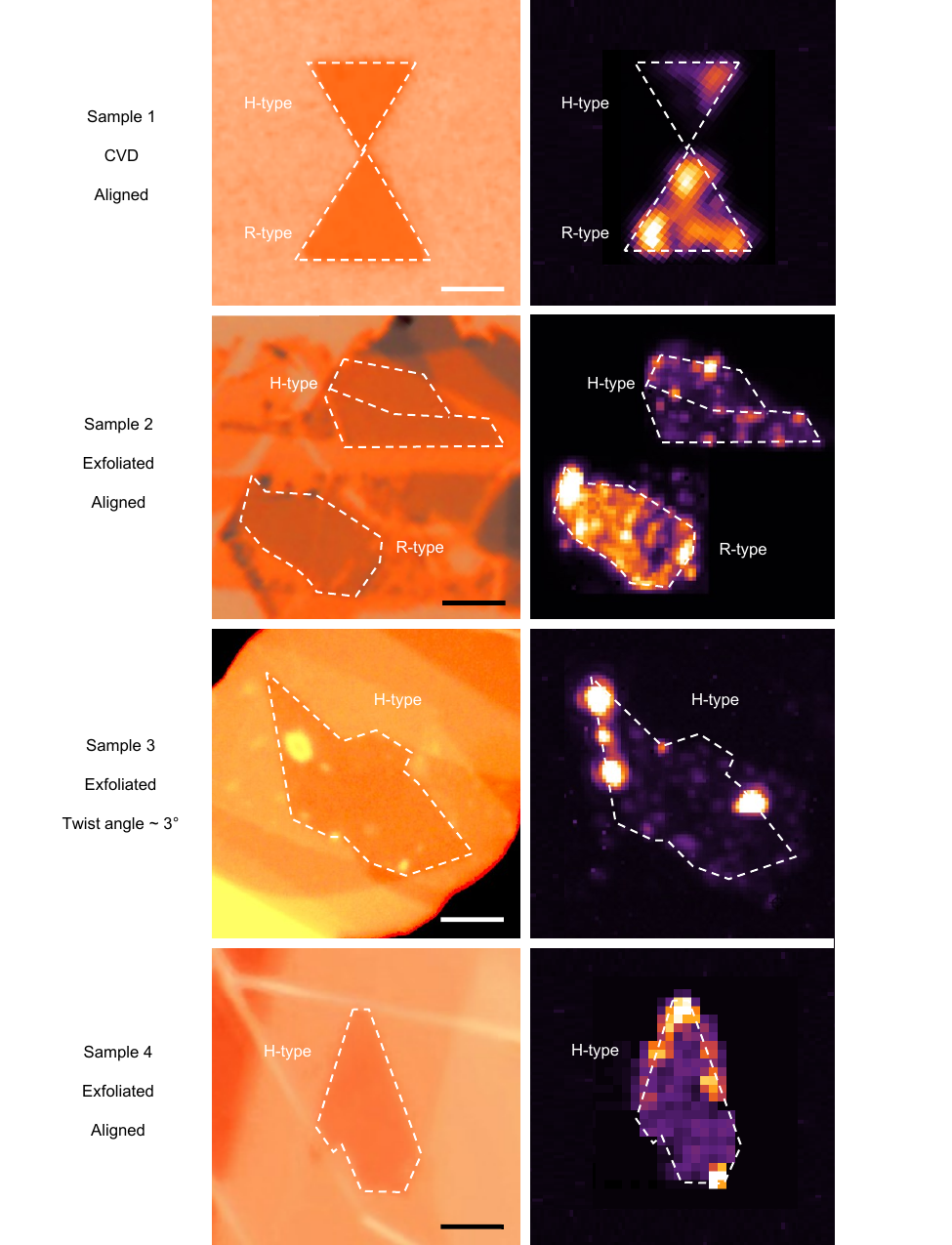}  
\vspace{-10pt}
\caption{Overview of hBN-encapsulated MoSe$_2$-WSe$_2$ HBL samples studied with optical spectroscopy (left panels: optical micrographs; right panels: raster-scan maps of PL intensity at $3.2$~K within the interlayer exciton band $1.2-1.5$~eV excited with a laser at $725$~nm and $2~\mu$W excitation power). The scale bars are $4$ and $2~\mu$m for samples $1 - 3$ and sample 4, respectively.}
\label{fig_PLmap}
\end{figure}

\clearpage

\begin{figure}[!ht]
\hspace*{-1cm}  
    \includegraphics[scale=1.1]{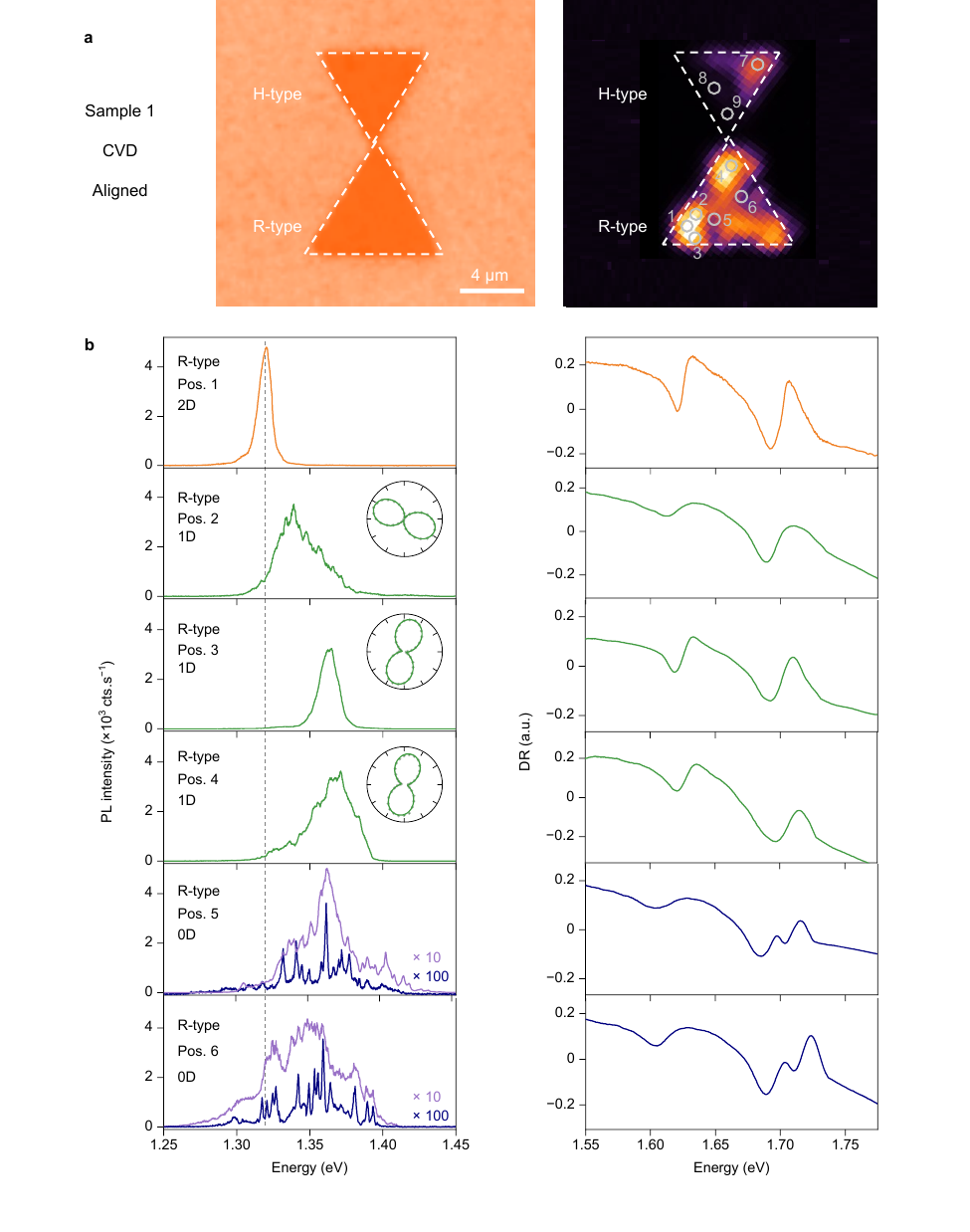}
\end{figure}
\begin{figure}[!ht]
\hspace*{-1cm}  
    \includegraphics[scale=1.1]{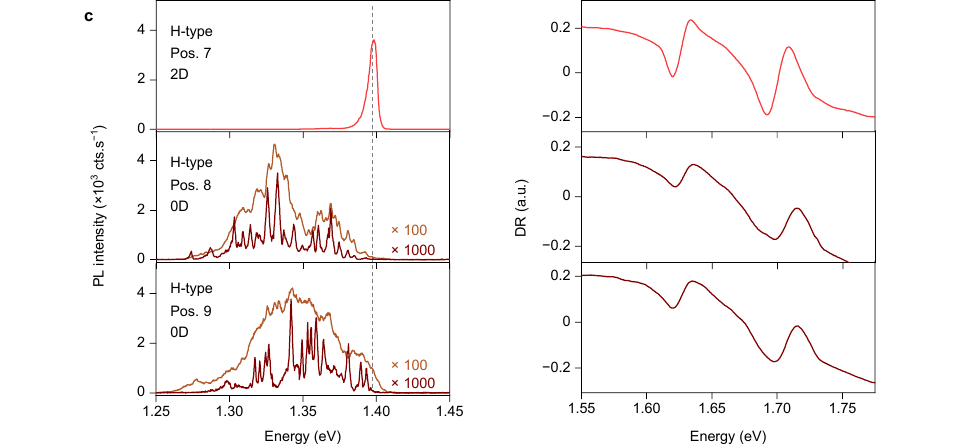}
    \caption{Characteristics of sample 1. \textbf{a}, Left panel: optical micrograph; right panel: map of interlayer exciton PL intensity with selected positions for PL and DR spectra in \textbf{b} and \textbf{c}. \textbf{b}, PL (left panel) and DR (right panel) characteristics of R-type stack assigned to 2D domains (orange), 1D stripes (green) and 0D domain arrays (purple). \textbf{c}, PL (left panel) and DR (right panel) characteristics of H-type stack assigned to 2D domains (red) and 0D domain arrays (brown). The PL peak positions of 2D domains are shown by dashed lines for reference. For regions assigned to 1D stripes, the orientation of linearly polarized PL is specified by the insets. The PL spectra were recorded with an excitation power of $2~\mu$W; for 0D arrays, additional spectra at $0.01~\mu$W excitation power are shown (light and dark purple for R-type, light and dark brown for H-type) and scaled for better visibility (factors of 10 and 100 for R-type, 100 and 1000 for H-type).} 
\label{fig_PLspectra_1}
\end{figure}

\clearpage

\begin{figure}[!ht]
\hspace*{-1cm}  
    \includegraphics[scale=1.1]{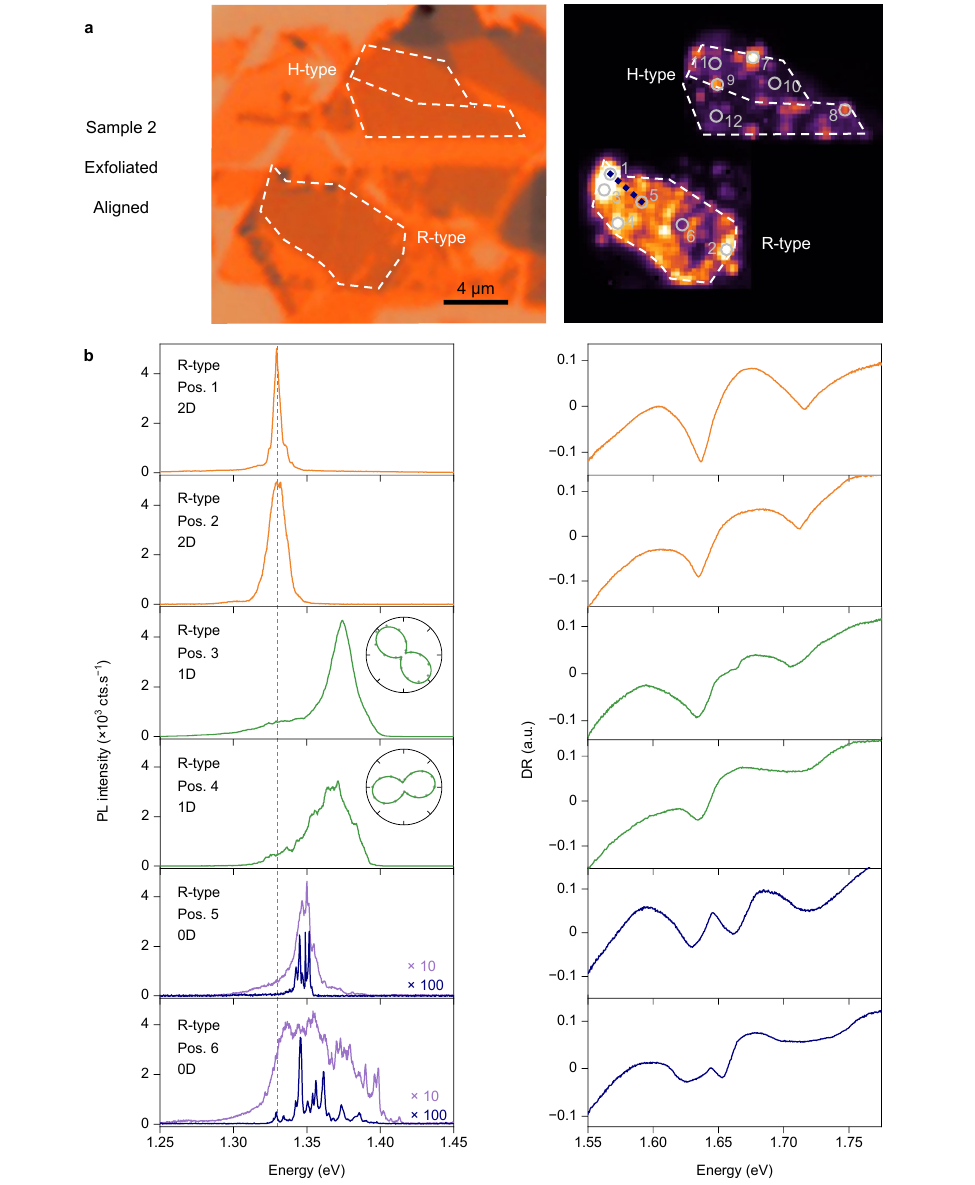}
\end{figure}
\begin{figure}[!ht]
\hspace*{-1cm}  
    \includegraphics[scale=1.1]{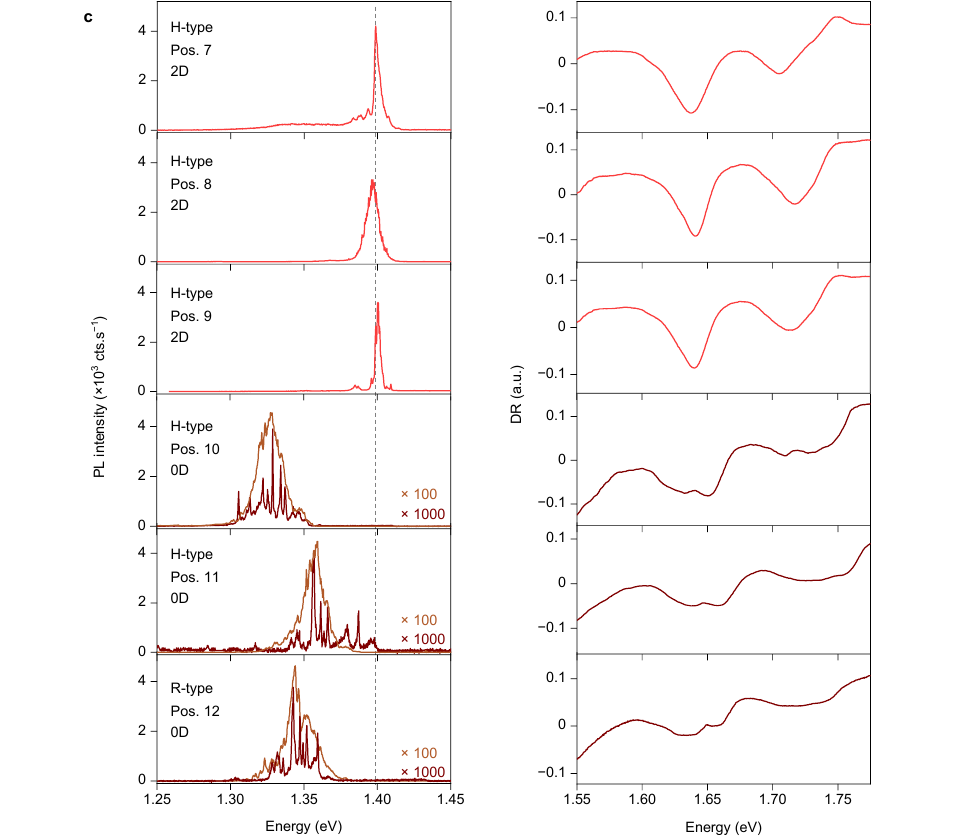}
    \caption{Same as Supplementary Fig.~\ref{fig_PLspectra_1} but for sample 2. The five black dots in the PL map in \textbf{a} indicate the corresponding positions of DR spectra in Fig.~3a of the main text.} 
    \label{fig_PLspectra_2}
\end{figure}

\clearpage

\begin{figure}[!ht]
    \vspace{-15pt}
    \hspace*{-1cm}  
    \includegraphics[scale=1.1]{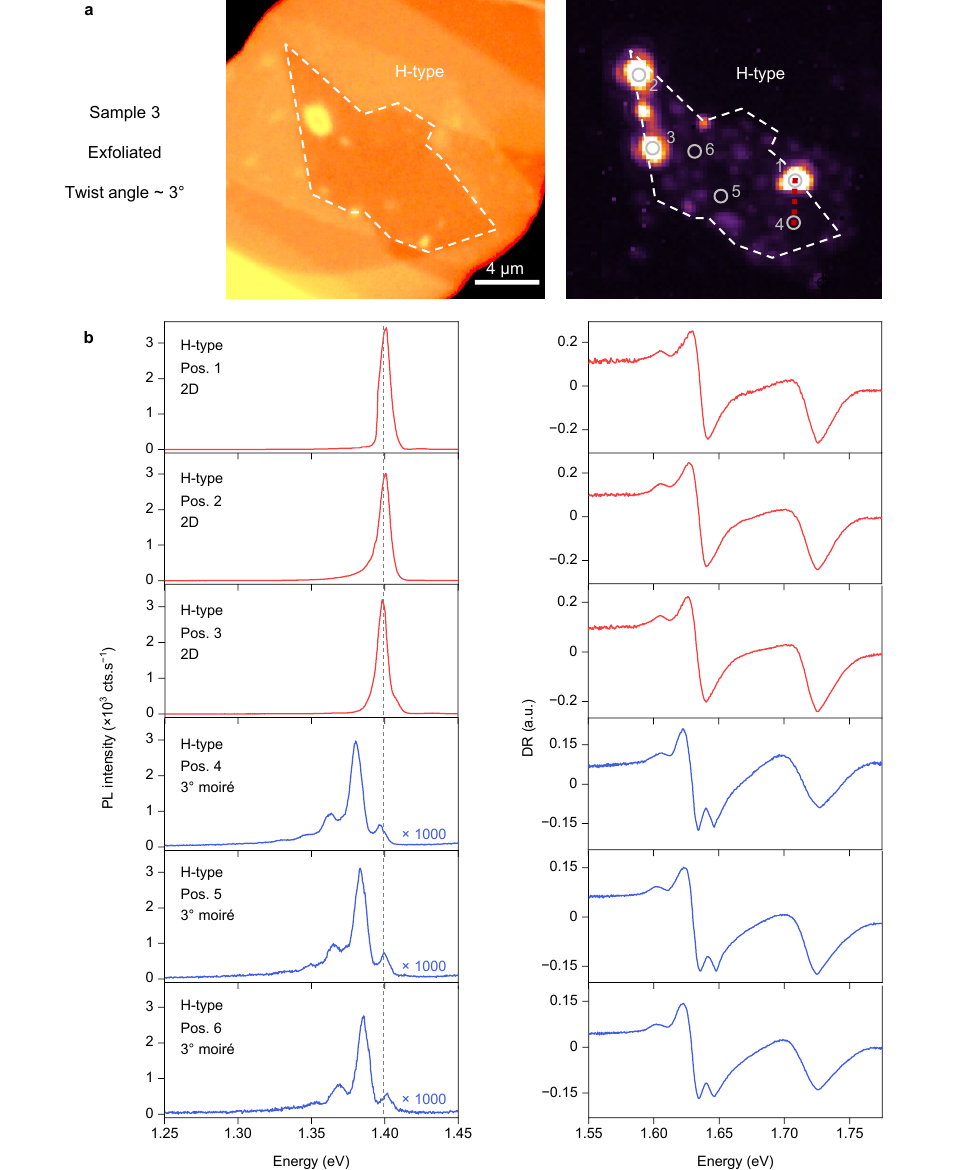}
    \vspace{-25pt}
    \caption{Same as Supplementary Fig.~\ref{fig_PLspectra_1} but for sample 3. Note the distinct PL features of exciton-polaron scaled by a factor of 1000. The five red dots in \textbf{a} indicate the positions of DR spectra in Fig.~4a of the main text.} 
    \label{fig_PLspectra_3}
\end{figure}

\clearpage

\begin{figure}[!ht]
\hspace*{-1cm}  
    \includegraphics[scale=1.1]{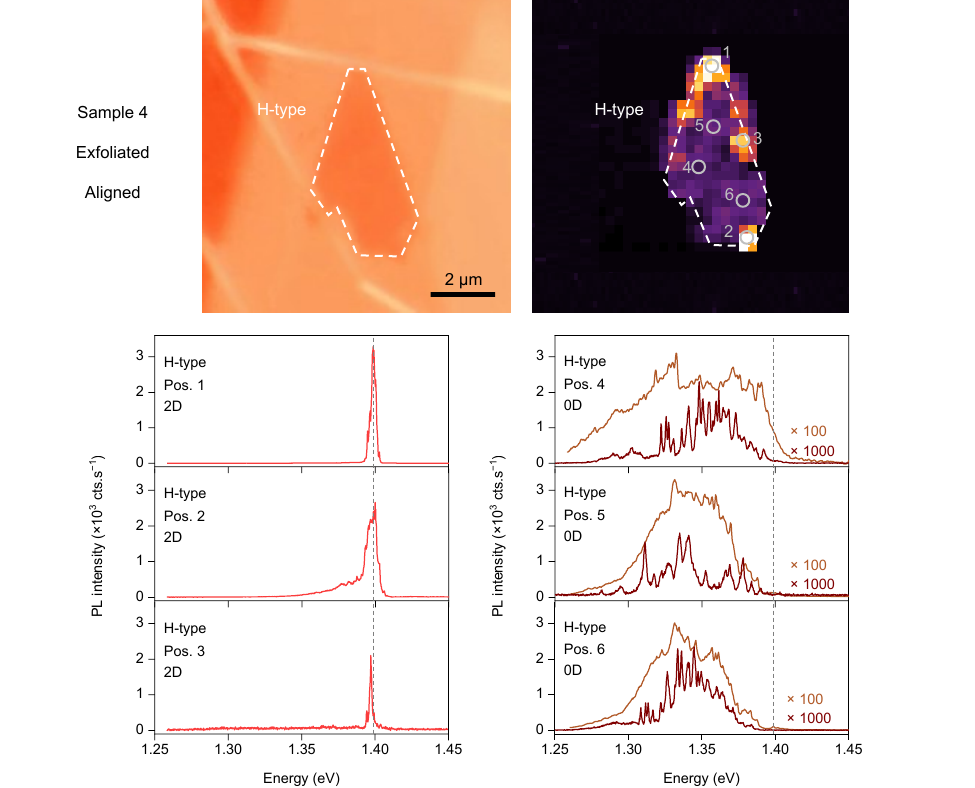}
    \caption{Characteristics of sample 4 without DR data.}
    \label{fig_PLspectra_4}
\end{figure}

\clearpage

\begin{figure}[!ht]
\includegraphics[scale=1.02]{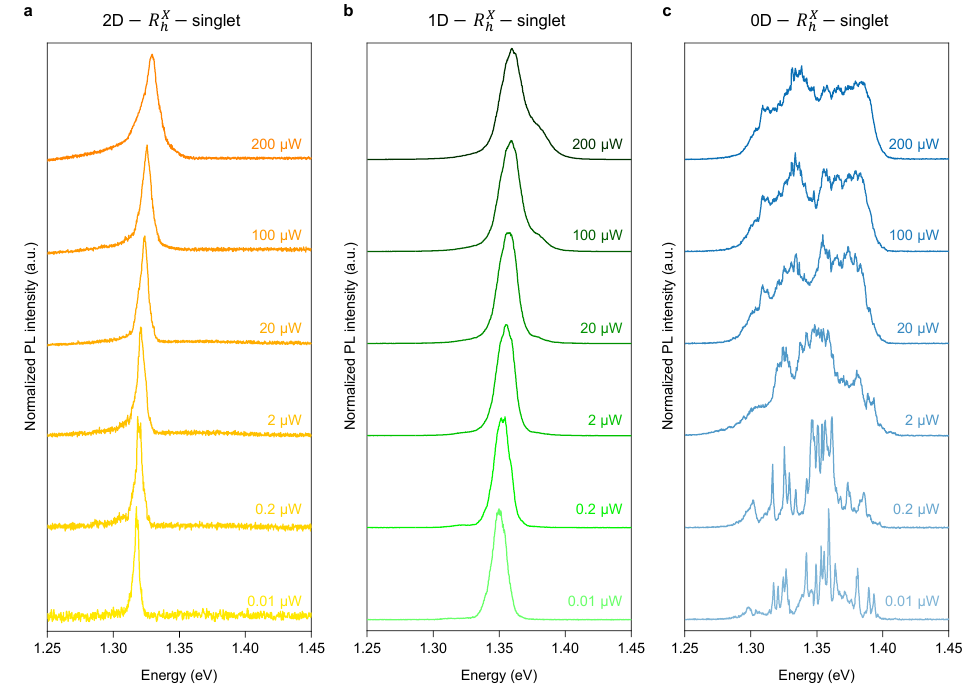}
\caption{Power-dependent interlayer PL of different domain types in nearly-aligned R-type HBL. \textbf{a -- c}, Normalized PL spectra at selected excitation powers for a 2D domain (\textbf{a}), a region with 1D stripes (\textbf{b}), and domains of 0D arrays (\textbf{c}). All data were acquired on sample 1 with excitation at $725$~nm.} 
\label{fig_PLpowerR-type}
\end{figure}

\clearpage

\begin{figure}[!ht]
\includegraphics[scale=1.02]{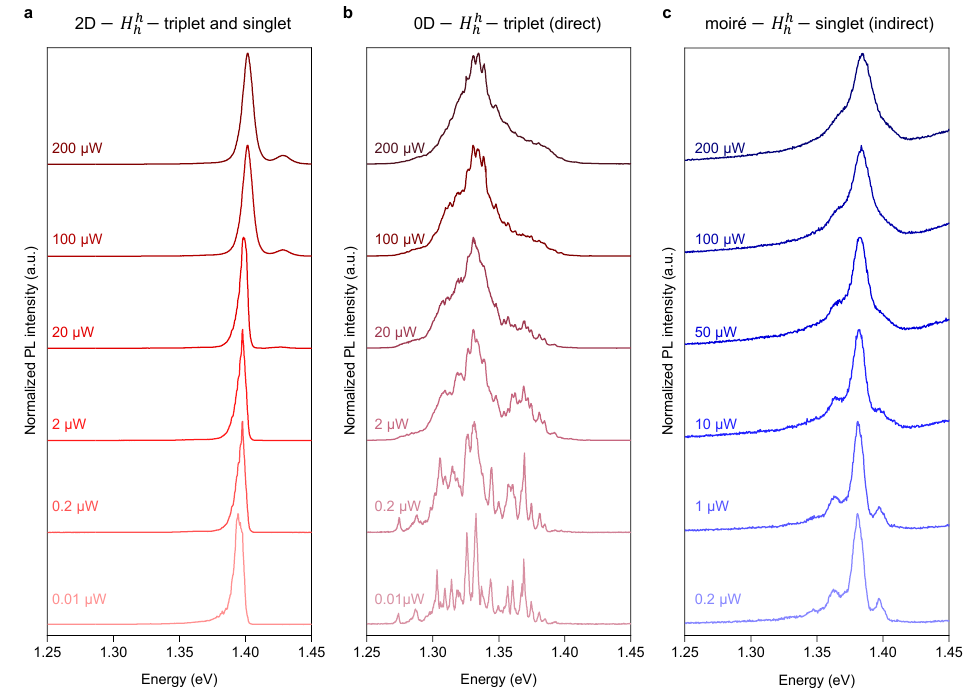}
\caption{Power-dependent interlayer PL of different domain types in H-type HBL. \textbf{a -- c}, Normalized PL spectra at selected excitation powers for a 2D domain (\textbf{a}), domains of 0D arrays (\textbf{b}), and a dark region in a HBL moir\'e superlattice with 3° twist angle (\textbf{c}). The data in \textbf{a} and \textbf{b} were acquired on sample 1, the data in \textbf{c} are from sample 3. For all data, the excitation wavelength was $725$~nm.} 
\label{fig_PLpowerH-type}
\end{figure}

\clearpage


\begin{figure}[!ht]
\vspace{-15pt}
\includegraphics[scale=0.985]{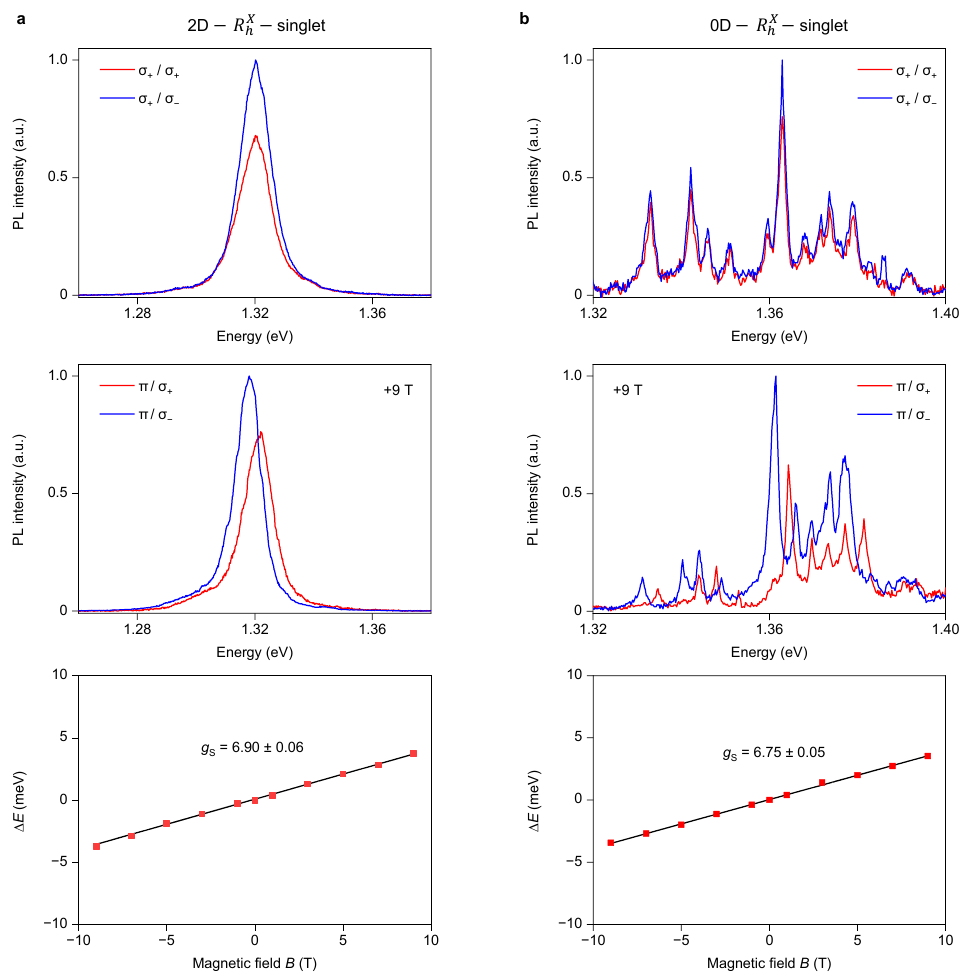}
\caption{Valley polarization and magneto-luminescence of interlayer excitons in nearly-aligned R-type HBL region of sample 1. 
with data in a 2D domain (\textbf{a}) and a 0D domain array (\textbf{b}). Top panels: Polarization-resolved PL (with $\sigma_+$ and $\sigma_-$ polarization shown in red and blue) with $\sigma_+$ polarized excitation at $0$~T. All peaks exhibit negative $P_\mathrm{c}$. Central panels: Valley Zeeman splitting between $\sigma_+$ and $\sigma_-$ polarized peaks in a magnetic field of $9$~T under linearly polarized excitation ($\pi$). Bottom panels: Valley Zeeman splitting $\Delta E$ as a function of magnetic field.
The solid lines are linear fits to the data with $g$-factors and error bars obtained from least-square best fits. The negative $P_\mathrm{c}$ and the positive $g$-factor assign the PL to singlet $R_h^X$ interlayer exciton. All data were recorded with excitation powers of $20$ and $0.01~\mu$W for 2D and 0D regions.} 
\label{fig_Pc+g-factor_R-type}
\end{figure}

\clearpage

\begin{figure}[!ht]
\includegraphics[scale=0.985]{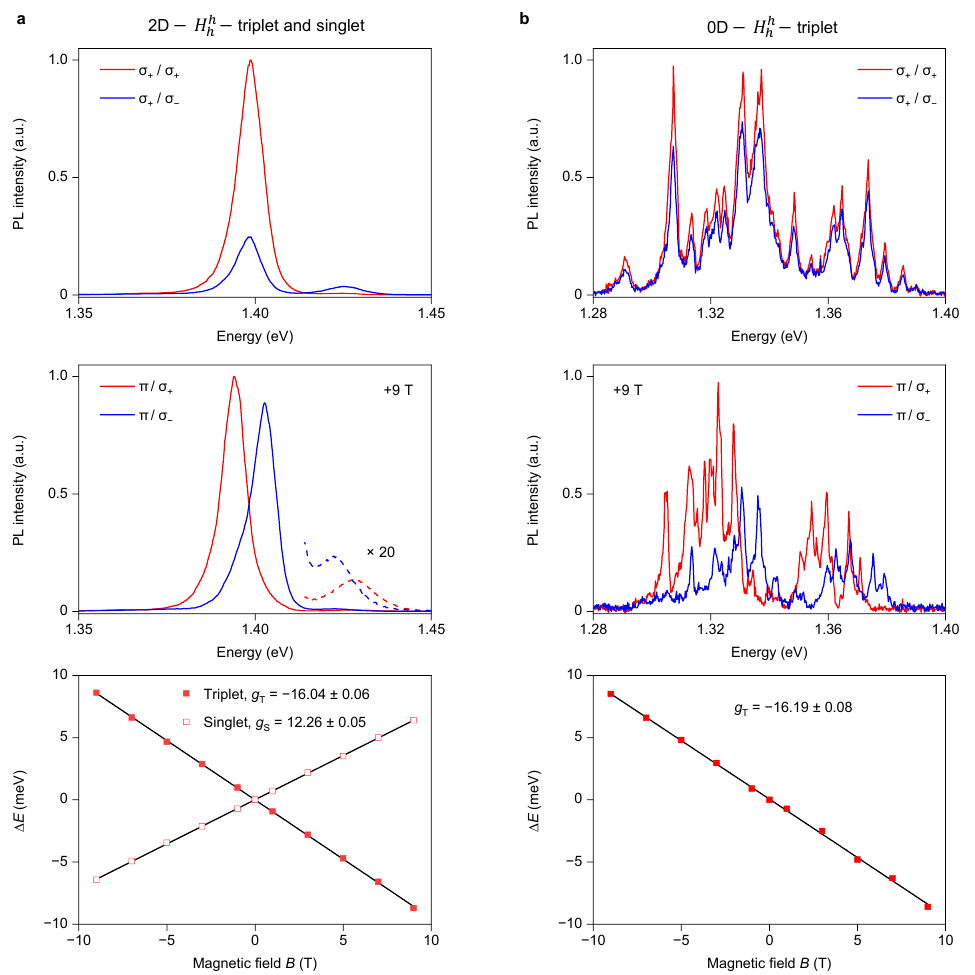}
\caption{Same as Supplementary Fig.~\ref{fig_Pc+g-factor_R-type} but for nearly-aligned H-type HBL in sample 1. All PL peaks exhibit positive $P_\mathrm{c}$ and $g$-factors of about $-16$, as obtained from linear fits to the data in the bottom panel with error bars from least-square best fits. Both features are characteristic of $H_h^h$ triplet interlayer excitons. In the regions of 0D arrays, $P_\mathrm{c}$ is smaller than in extended 2D regions due to reduced symmetry of the interlayer exciton wavefunctions in domains with distorted hexagonal shapes. The weak peak in the spectra of panel \textbf{a} with $25$~meV blue-shift from the main peak corresponds to $H_h^h$ singlet interlayer excitons with negative $P_\mathrm{c}$ and positive $g$-factor of $\sim 12$.} 
\label{fig_Pc+g-factor_H-type}
\end{figure}

\begin{figure}[!ht]
\includegraphics[scale=0.985]{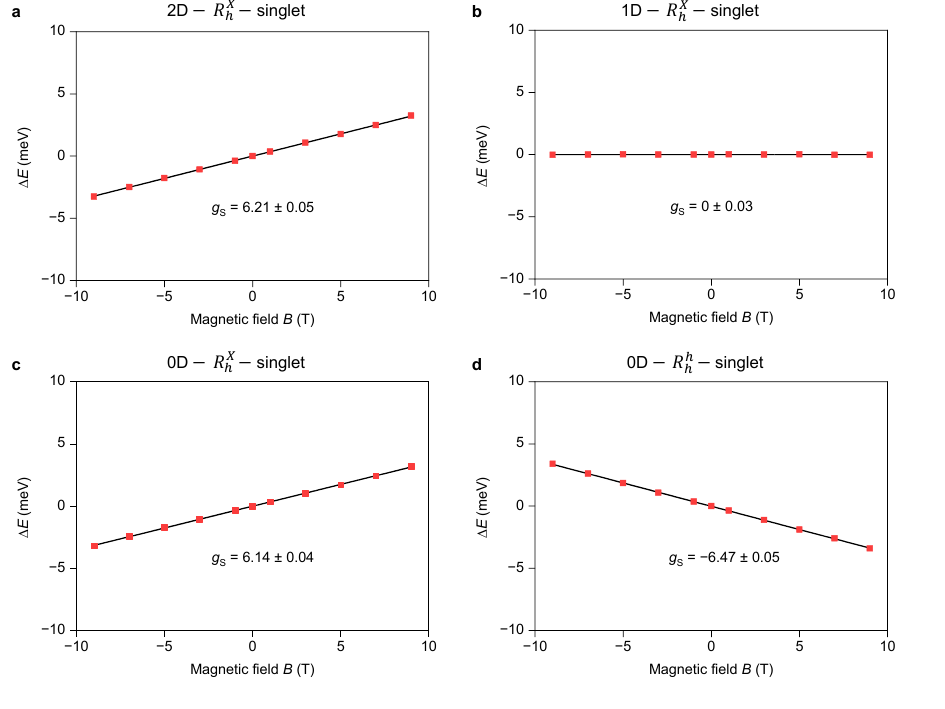}
\caption{Magneto-luminescence of interlayer excitons in nearly-aligned R-type HBL region of sample 2: valley Zeeman splitting $\Delta E$ as a function of magnetic field obtained as the energy difference between $\sigma_+$ and $\sigma_-$ polarized peaks of interlayer exciton PL under linearly polarized excitation. The solid lines are linear fits to the data with $g$-factors and error bars obtained from least-square best fits. The data in \textbf{a} and \textbf{b} correspond to the data in false-color representation in Fig.~3\textbf{d} and \textbf{g} of the main text; the data in \textbf{c} and \textbf{d} correspond to the data in Fig.~3\textbf{j} of the main text. All data were recorded on sample 2.} 
\label{fig_Pc+g-factor_main-R}
\end{figure}

\clearpage

\begin{figure}[!ht]
\includegraphics[scale=0.985]{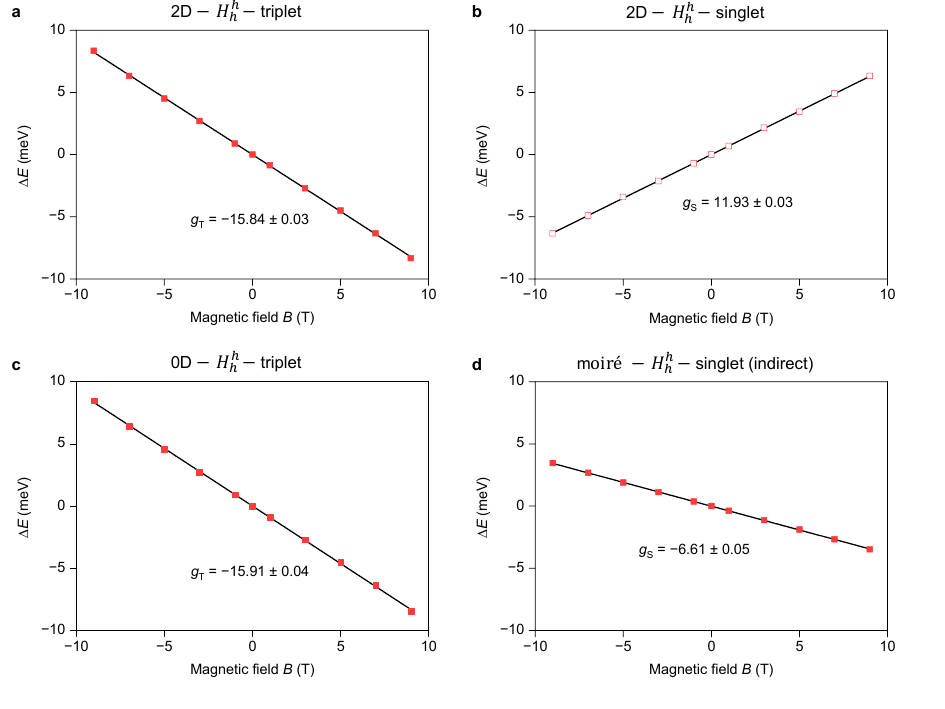}
\caption{Magneto-luminescence of interlayer excitons in H-type HBL of sample 2 and sample 3: valley Zeeman splitting $\Delta E$ as a function of magnetic field obtained as the energy difference between $\sigma_+$ and $\sigma_-$ polarized peaks of interlayer exciton PL under linearly polarized excitation. The solid lines are linear fits to the data with $g$-factors and error bars obtained from least-square best fits. The data in \textbf{a} and \textbf{b} correspond to the data in false-color representation in Fig.~4\textbf{d} of the main text; the data in \textbf{c} and \textbf{d} correspond to the data in Fig.~4\textbf{g} and \textbf{j} of the main text, respectively. Data in \textbf{a}, \textbf{b} and \textbf{c} were recorded on sample 2, and data in \textbf{d} are from sample 3.} 
\label{fig_Pc+g-factor_main-H}
\end{figure}

\clearpage

\begin{figure}[!ht]
\vspace{-10pt}
\includegraphics[scale=1.0]{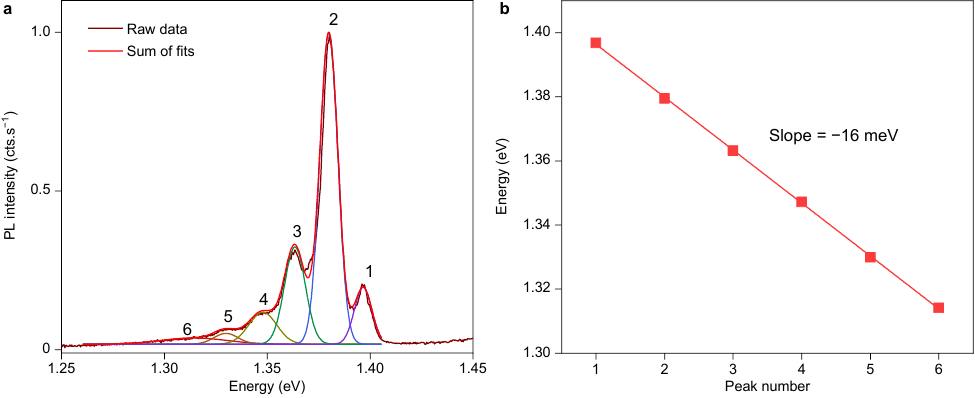}
\vspace{-20pt}
\caption{\textbf{a}, Representative PL spectrum from a dark region in H-type HBL moir\'e superlattice with 3° twist angle (sample 3). In contrast to 0D arrays of nearly-aligned samples, the spectrum is characterized by a series of six peaks (shown together with Gaussian fits). \textbf{b}, Energy of six Gaussian peaks in \textbf{a} plotted as a function of the peak number. The linear fit indicates equidistant peak spacing of $16$~meV.} 
\label{fig_PL-polaron}
\end{figure}


\noindent \textbf{Supplementary Note $\mathbf{4}$: Time-resolved photoluminescence of reconstructed domains}

The PL decay of interlayer excitons in different domain types was studied with time-resolved PL spectroscopy, with main results for sample 1 shown in Supplementary Fig.~\ref{fig_Lifetime} and 
Supplementary Table~\ref{tab_fit-lifetime} for an excitation power of $0.05~\mu$W at $725$~nm excitation wavelength. The set of data is complemented by power-dependent lifetime measurements of \RX singlet and \Hh triplet interlayer excitons in Supplementary Fig.~\ref{fig_Lifetime_power}. Focusing first on reconstructed 2D domains, where interlayer exciton reservoirs are limited to zero-momentum and momentum-dark exciton states in \Hh and \RX registries due to the absence of other stackings, we determine from best-fit analysis (taking into account the instrument response function) two decay channels for the \Hh singlet exciton and three decay channels for both the \Hh triplet and \RX singlet excitons. According to Supplementary Table~\ref{tab_fit-lifetime}, the \Hh singlet PL is characterized by two decay times - one below the resolution limit of $0.2$~ns and one with a decay time of $2.5$~ns. We ascribe the primary decay time to population loss of \Hh singlet excitons by rapid relaxation into the energetically lower reservoir of \Hh triplet excitons and other non-radiative decay processes including Auger-mediated population loss. The secondary decay channel with $2.5$~ns decay time reflects the characteristic lifetime of \Hh singlet interlayer exciton state which is proportional to the oscillator strength of the respective radiative transition. 

Using $2.5$~ns as the radiative lifetime of the \Hh singlet exciton reservoir, we estimate by scaling of oscillator strengths from Table~1 of the main text the radiative PL lifetimes of $65$~ns for the \Hh triplet interlayer exciton reservoir, as well as $5.3$ and $23$~ns for the \RX singlet and triplet exciton reservoirs, respectively. Experimentally, the secondary decay component of the \Hh triplet reservoir and the \RX singlet reservoir are determined to $48$~ns and $5$~ns, respectively, in good agreement with the scaling anticipated from theory, whereas the PL from the \RX triplet state is insufficient for time-resolved measurements. The contribution of Auger decay to the primary decay channel is evident for the \Hh triplet and \RX singlet PL decay from power-dependent data in Supplementary Fig.~\ref{fig_Lifetime_power}, where the amplitude of the primary component dominates at high excitation powers over the amplitudes of the two complementary channels with power-independent ratio.

With this assignment of primary and secondary PL decay components to non-radiative population loss processes (including Auger) and radiative decay, respectively, we attribute the third channel of \Hh triplet exciton PL decay on $440$~ns timescale and \RX singlet exciton state with $45$~ns decay time to population feeding from energetically higher-lying long-lived reservoirs, presumably constituted by momentum-dark configurations of valence band holes in WSe$_2$ and conduction band electrons in MoSe$_2$.  Note that the absence of such reservoirs could explain the lack of the third channel in the PL decay of the \Hh singlet state that is the top-most energy level according to our calculations summarized in Table~1 of the main text. 

\begin{figure}[t!]
\includegraphics[scale=1.0]{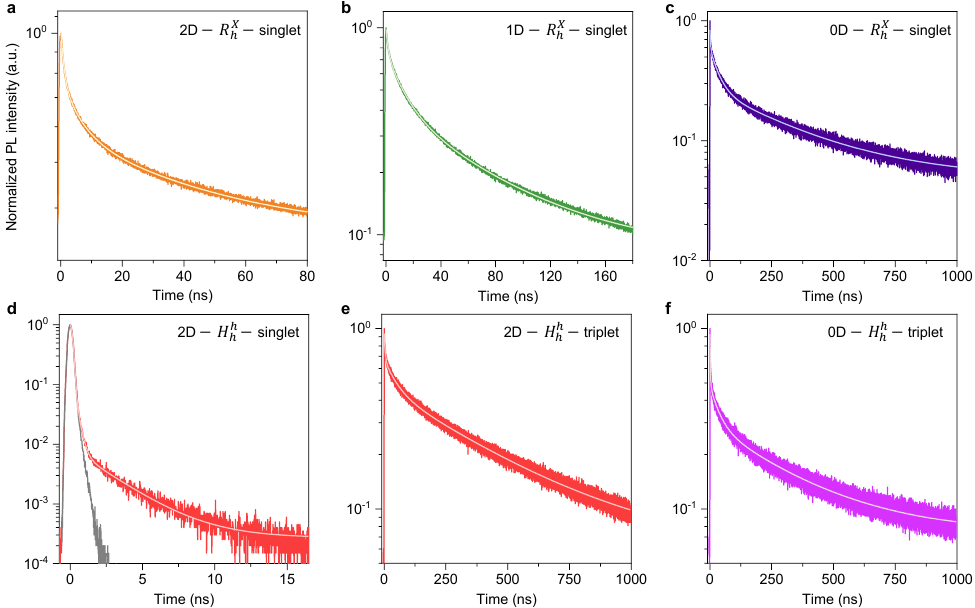}
\caption{Time-resolved PL of interlayer excitons in different types of domain. \textbf{a -- c}, Time-resolved PL decay of $R_h^X$ singlet exciton in a 2D domain (\textbf{a}) and regions of 1D stripes (\textbf{b}) and 0D arrays (\textbf{c}). \textbf{d -- f}, Time-resolved PL decay of $H_h^h$ singlet exciton (\textbf{d}) and triplet exciton (\textbf{e}) in a 2D domain region, as well as $H_h^h$ triplet exciton in a 0D domain region (\textbf{f}). The instrument response function (IRF) is shown in \textbf{d} in gray. The solid lines show in each panel best fits to the data obtained by a convolution of the IRF and a tri-exponential decay. The best-fit parameters are listed in Supplementary Table~\ref{tab_fit-lifetime}. All data were recorded on sample 1 with excitation at $725$~nm and $0.05~\mu$W power.} 
\label{fig_Lifetime}
\end{figure}

For domains with reduced dimensionality, we observe similar behavior in time-resolved PL. With increasing quantum confinement effects due to reduced system dimensionality, \RX states show the trend of prolonged lifetimes for 1D stripe and 0D array regions. The overall ratio of the three components shows no strong dependence on dimensionality. The trend is opposite in 0D arrays of \Hh triplet excitons, where all decay components decrease as compared to the 2D domain limit, consistent with the absence of quantum confinement effects in reconstructed domains of \Hh stacking and potentially related to population loss into surrounding \HM and \HX domain walls with lower-energy interlayer exciton states.


\begin{table*}[t!]
\begin{ruledtabular}
\begin{tabular}{lllRRRRRR}
Domain & Stacking & Spin configuration & \tau_{1}  & A_{1} & \tau_{2} & A_{2} & \tau_{3} & A_{3} \\ 
\hline
2D &\RX & singlet                 & 1.1~\text{ns} &  34\% & 5~\text{ns} & 30\% & 45~\text{ns}& 36\%     \\
1D &\RX &  singlet                & 1.5~\text{ns}&  25\% & 13~\text{ns}& 37\% &  70~\text{ns}& 38\%     \\
0D&\RX &   singlet               & 2.2~\text{ns}&   32\% & 34~\text{ns}& 40\% & 335~\text{ns}& 28\%     \\
\hline
2D &\Hh& singlet         & <0.2~\text{ns}& 98\% & 2.5~\text{ns}& 2\% &  &               \\
2D &\Hh& triplet         & 2.4~\text{ns}&  29\% & 48~\text{ns}& 24\% & 440~\text{ns}& 47\%     \\ 
0D &\Hh& triplet    & 1.7~\text{ns}& 52\% & 29~\text{ns}& 22\% & 305~\text{ns}& 26\%\ \end{tabular}
\caption{Best-fit parameters obtained for the data in Supplementary Fig.~\ref{fig_Lifetime} using tri-exponential decay with characteristic lifetime ($\tau_{i}$) and amplitude ($A_{i}$) of each decay channel.}
\label{tab_fit-lifetime}
\end{ruledtabular}
\end{table*}


\begin{figure}[!ht]
\includegraphics[scale=1.03]{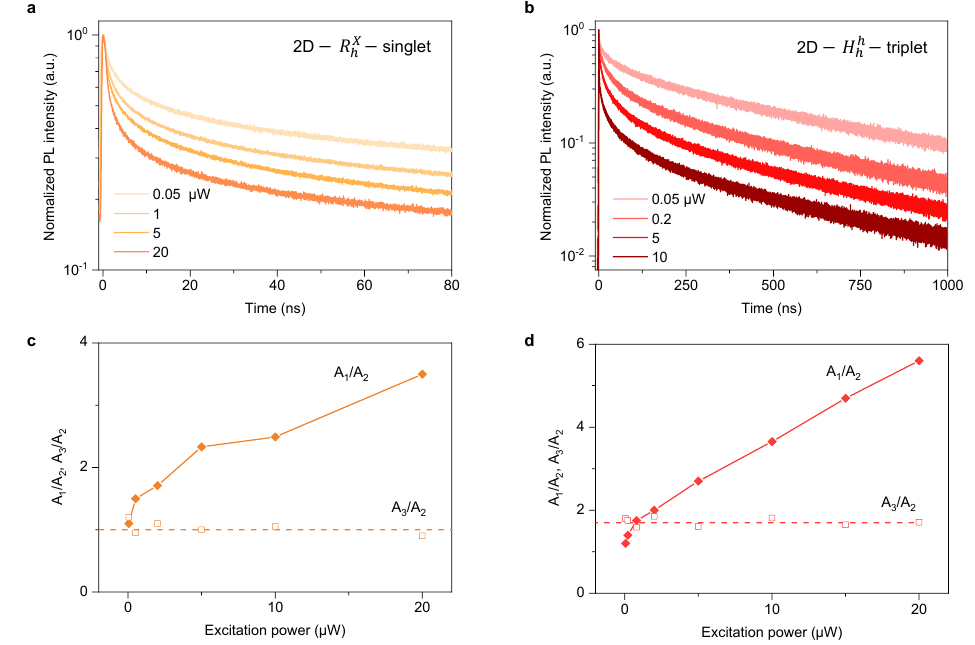}
\caption{Power dependent PL decay. \textbf{a} and \textbf{b}, Time-resolved PL decay of $R_h^X$ singlet (\textbf{a}) and $H_h^h$ triplet (\textbf{b}) interlayer exciton in 2D domains for different excitation powers. \textbf{c} and \textbf{d}, The amplitude ratios between the fast and intermediate decay components ($A_{1}/A_{2}$, shown by filled diamonds) and between the intermediate and long decay components ($A_{2}/A_{3}$, shown by empty squares) as a function of excitation power. According to best-fits, the three characteristic decay times are independent of the excitation power. However, the fast decay component becomes dominant at high excitation powers as $A_{1}/A_{2}$ increases but $A_{2}/A_{3}$ remains constant, ascribing the fast decay channel to Auger-mediated processes.}
\label{fig_Lifetime_power}
\end{figure}

\clearpage

\noindent \textbf{Supplementary Note $\mathbf{5}$: Direct correlation between domain patterns and optical properties}

To provide a direct evidence for correlations between mesoscopically reconstructed domain dimensionality and respective optical features, we performed SEM imaging and cryogenic optical spectroscopy on the same MoSe$_2$-WSe$_2$ HBL. To this end, we fabricated one H-type and one R-type fully hBN encapsulated samples with thin hBN top layers below 5~nm to ensure sufficient transparency for secondary electrons. Since the thin hBN layer is too fragile to pick up additional TMD and hBN layers of the heterostructure, the sample fabrication process was modified. We first picked up a thick hBN flake with the PC/PDMS stamp. This hBN layer was then used to pick up successively MoSe$_2$ and WSe$_2$ monolayers and a thin hBN flake. The PC film together with the stacked sample was subsequently cut off from the PDMS stamp, flipped over and transferred onto a $300$~nm SiO$_2$/Si chip. After high vacuum annealing at $450\degree$ for $2$~h, the wafer was carefully immersed in chloroform for $5$~min to remove the PC film. With this procedure, we obtained a clean heterostack with the thin hBN layer on top.

\begin{figure}[!t]
\includegraphics[scale=0.295]{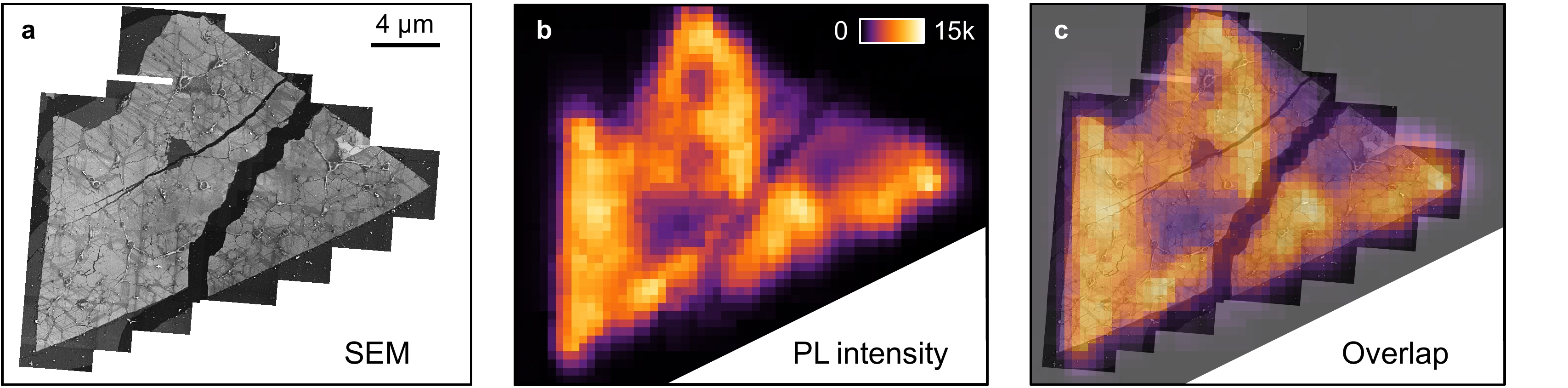}
\caption{\textbf{a}, Scanning electron micrograph of the H-type sample, composed of individual small-area images, with characteristic features of sample imperfections and mesoscopic reconstruction. \textbf{b}, PL intensity map in the spectral band of interlayer excitons at $3.2$~K. \textbf{c}, Overlay of the SEM image in \textbf{a} and the PL map in \textbf{b}.}
\label{fig_SEM_correlation}
\end{figure}

Supplementary Fig.~\ref{fig_SEM_correlation} \textbf{a} and \textbf{b} show the secondary electron modulated SEM image and interlayer exciton PL intensity map of the H-type heterostack. The SEM image shows coexisting 2D, 1D and 0D domains. Consistently, the PL map exhibits sizable intensity variations across the sample. By co-aligning the PL map and the SEM image (Supplementary Fig.~\ref{fig_SEM_correlation} \textbf{c}), we correlate the spectral features with the local characteristics of domain geometry. All observations and conclusions from direct correlation between reconstructed domain structures and optical properties are consistent with the analysis provided in the main text. 

For large 2D domains, exemplified in Supplementary Fig.~\ref{PL_SEM_PL_DR} \textbf{a} and \textbf{b}, the interlayer PL is bright and characterized by a single peak at $\sim$1.40~eV of the triplet exciton state in \Hh stacking. Consistently, the corresponding DR spectra feature single-peak resonances of intralayer excitons in MoSe$_2$ and WSe$_2$. In these domains, the degree of circular polarization is high and the degree of linear polarization is zero, as indicated by numbers in Supplementary Fig.~\ref{PL_SEM_PL_Pc} \textbf{a} and \textbf{b}.   

\begin{figure}[!ht]
\vspace{-25pt}
\includegraphics[scale=0.294]{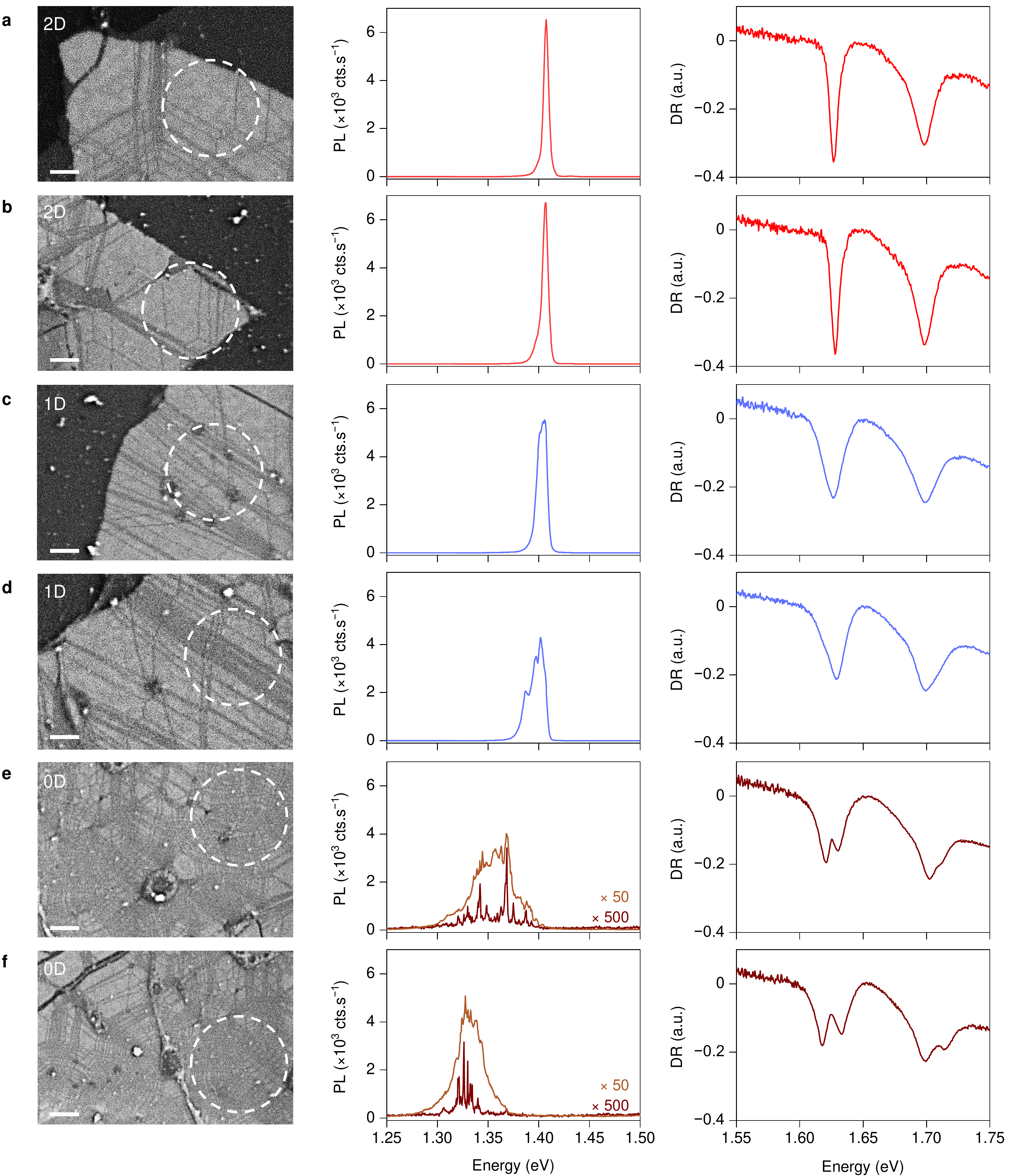}
\vspace{-30pt}
\caption{Examples of reconstruction patterns and the corresponding optical spectra in the H-type sample. \textbf{a} -- \textbf{f}, Left panels show representative areas of 2D (\textbf{a} and \textbf{b}), 1D (\textbf{c} and \textbf{d}) and 0D (\textbf{e} and \textbf{f}) domains identified in SEM imaging (the scale bars are $200$~nm). Central and right panels show the corresponding PL and DR spectra within the optical spot delimited by dashed circles. The PL spectra were excited with $2~\mu$W laser power. For 0D domains, the PL spectra at $0.01~\mu$W are shown (dark brown) with respective scaling factors in addition to the spectra recorded at $2~\mu$W (light brown).}
\label{PL_SEM_PL_DR}
\end{figure}

\begin{figure}[!ht]
\vspace{-25pt}
\includegraphics[scale=0.294]{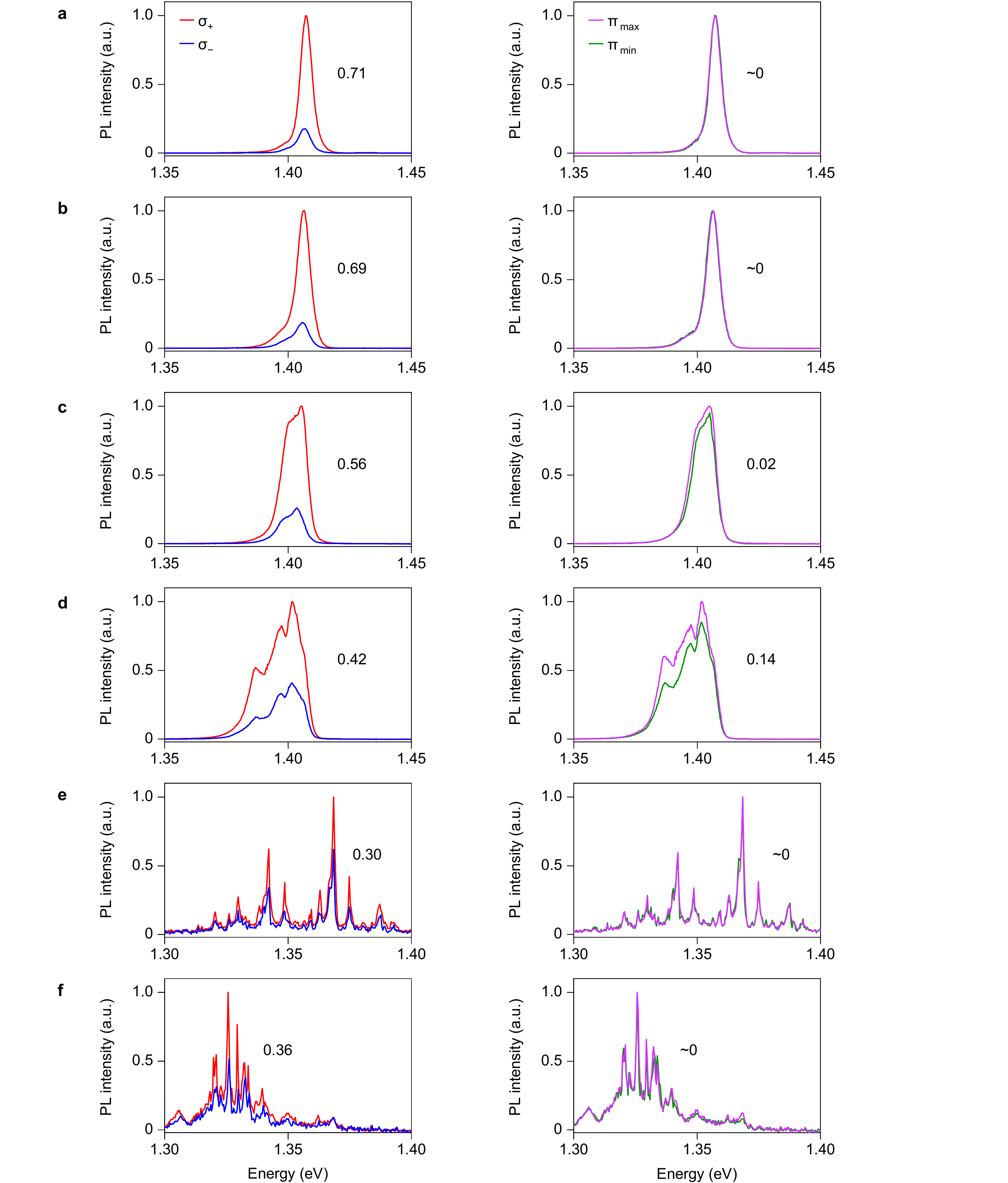}
\vspace{-30pt}
\caption{\textbf{a} -- \textbf{f}, PL spectra recorded in circular (left panels, with $\sigma_+$ and $\sigma_-$ polarized detection shown in red and blue, respectively) and linear (right panels, with two orthogonal orientations of linear polarization for maximum and minimum PL intensity shown in purple and green, respectively) basis from the same regions as in Supplementary Fig.~\ref{PL_SEM_PL_DR} \textbf{a} -- \textbf{f}. The numbers in each panel denote the respective degrees of circular and linear polarization, $P_\mathrm{c}$ and $P_\mathrm{\,l}$.} 
\label{PL_SEM_PL_Pc}
\end{figure}

For regions with elongated 1D domains, exemplified in Supplementary Fig.~\ref{PL_SEM_PL_DR} \textbf{c} and \textbf{d}, the interlayer exciton PL is detected around 1.40~eV or with slightly red-shifted additional peaks. In contrast to 1D stripes in R-type HBLs, the PL preserves a sizable degree of circular polarization, whereas the degree of linear polarization is either negligible or very small where the density of 1D stripes is high (Supplementary Fig.~\ref{PL_SEM_PL_Pc} \textbf{c} and \textbf{d}). As argued in the main text, pronounced effects of 1D quantum confinement are expected in R-type stacks where \RM domains represent potential wells for \RX exciton states. In H-type stacks, on the contrary, \Hh excitons are energetically highest and thus deprived of efficient quantum confinement by surrounding domains of alternative registries. In these regions, structured PL and broadened DR spectra reflect variations in the size of 1D domains.  

For 0D array regions, finally, exemplified in Supplementary Fig.~\ref{PL_SEM_PL_DR} \textbf{e} and \textbf{f}, the interlayer exciton PL is very low and red-shifted by several tens of meV compared to 2D domains. This red-shift is size-dependent, with smaller domains exhibiting larger red-shifts. Moreover, for excitation powers below $100$~nW, the PL spectra develop into quantum dot like spectrally narrow peaks. As shown in Supplementary Fig.~\ref{PL_SEM_PL_Pc} \textbf{e} and \textbf{f}, the degree of linear polarization is zero in the regions of 0D domains, and the degree of circular polarization is reduced as compared to extended 2D domain regions because of lower symmetry of the interlayer exciton wave functions in the presence of distorted hexagonal lattice of 0D arrays. The corresponding DR spectra exhibit for both MoSe$_2$ and WSe$_2$ intralayer exciton resonances characteristic splittings and broadenings which depend on the local homogeneity of domain sizes and shapes. Smaller domains feature larger energy splittings in the DR spectra, consistent with the analysis in the subsequent Supplementary Note 6.

\clearpage

\begin{figure}[!t]
\includegraphics[scale=0.118]{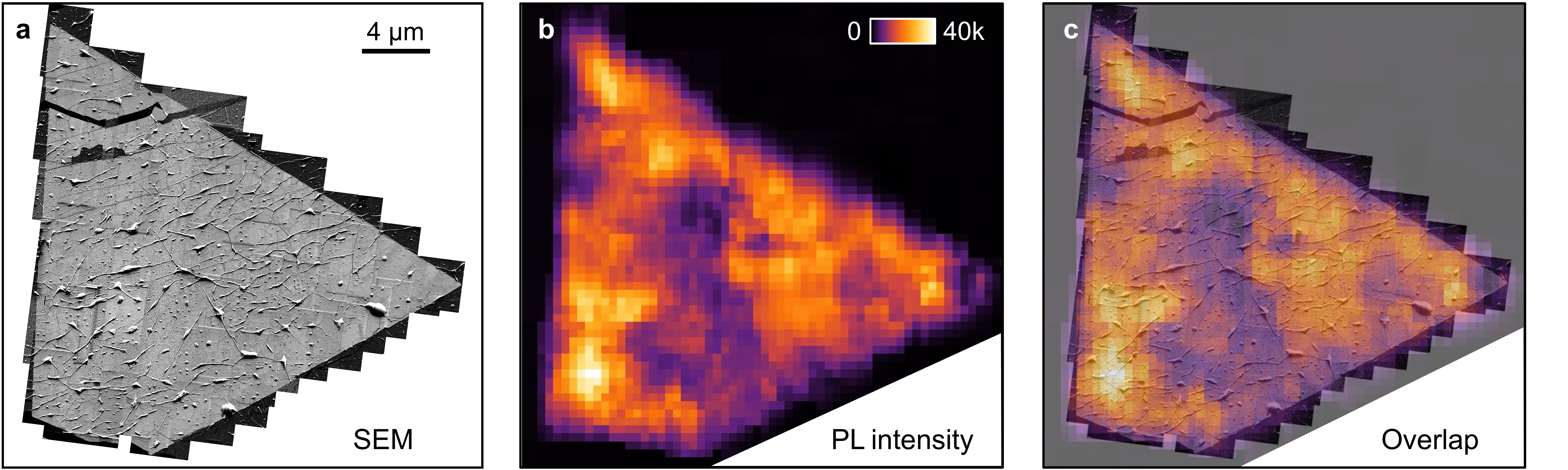}
\vspace{-20pt}
\caption{Same as Supplementary Fig.~\ref{fig_SEM_correlation} but for the R-type sample.}
\label{fig_SEM_correlation_R}
\end{figure}

In the same manner, we performed secondary electron modulated SEM and PL hyperspectral imaging on the R-type sample and co-aligned the two images (Supplementary Fig.~\ref{fig_SEM_correlation_R}). From the direct correlation between SEM and PL images, we again obtained the results consistent with the analysis in the main text on reconstruction patterns and their optical characteristics of R-type samples. For 2D domains of \RX stacking, exemplified in Supplementary Fig.~\ref{PL_SEM_PL_DR_R} \textbf{a} and \textbf{b}, their interlayer exciton PL is bright and the spectra feature a single peak at $\sim$1.33~eV from the singlet exciton state. Consistently, their DR spectra show single-peak resonances of intralayer excitons in MoSe$_2$ and WSe$_2$. The interlayer exciton PL from these 2D domains has negative degree of circular polarization and zero degree of linear polarization (Supplementary Fig.~\ref{PL_SEM_PL_Pc_R} \textbf{a} and \textbf{b}). For 1D stripe regions, exemplified in Supplementary Fig.~\ref{PL_SEM_PL_DR_R} \textbf{c} and \textbf{d}, the interlayer exciton PL is blue-shifted and broadened compared to that from 2D domains of \RX stacking. This PL features almost zero degree of circular polarization but high degree of linear polarization along the elongated direction of the stripes. Finally, for 0D array regions, exemplified in Supplementary Fig.~\ref{PL_SEM_PL_DR_R} \textbf{e} and \textbf{f}, the interlayer exciton PL is reduced and blue-shifted by several tens of meV compared to 2D domains. In contrast to H-type, the smaller 0D domains show larger blue-shifts because of the quantum confinement. When excitation powers are below $100$~nW, the PL spectra of 0D domains have the characteristic of quantum dots, showing narrow peaks with negative degree of circular polarization and zero degree of linear polarization (Supplementary Fig.~\ref{PL_SEM_PL_Pc_R} \textbf{e} and \textbf{f}). The DR spectra of 0D regions in R-type exhibit for both MoSe$_2$ and WSe$_2$ intralayer exciton resonances characteristic splittings and broadenings with smaller domains featuring larger energy splittings, which is similar to H-type.
\clearpage

\begin{figure}[!ht]
\vspace{-25pt}
\includegraphics[scale=0.294]{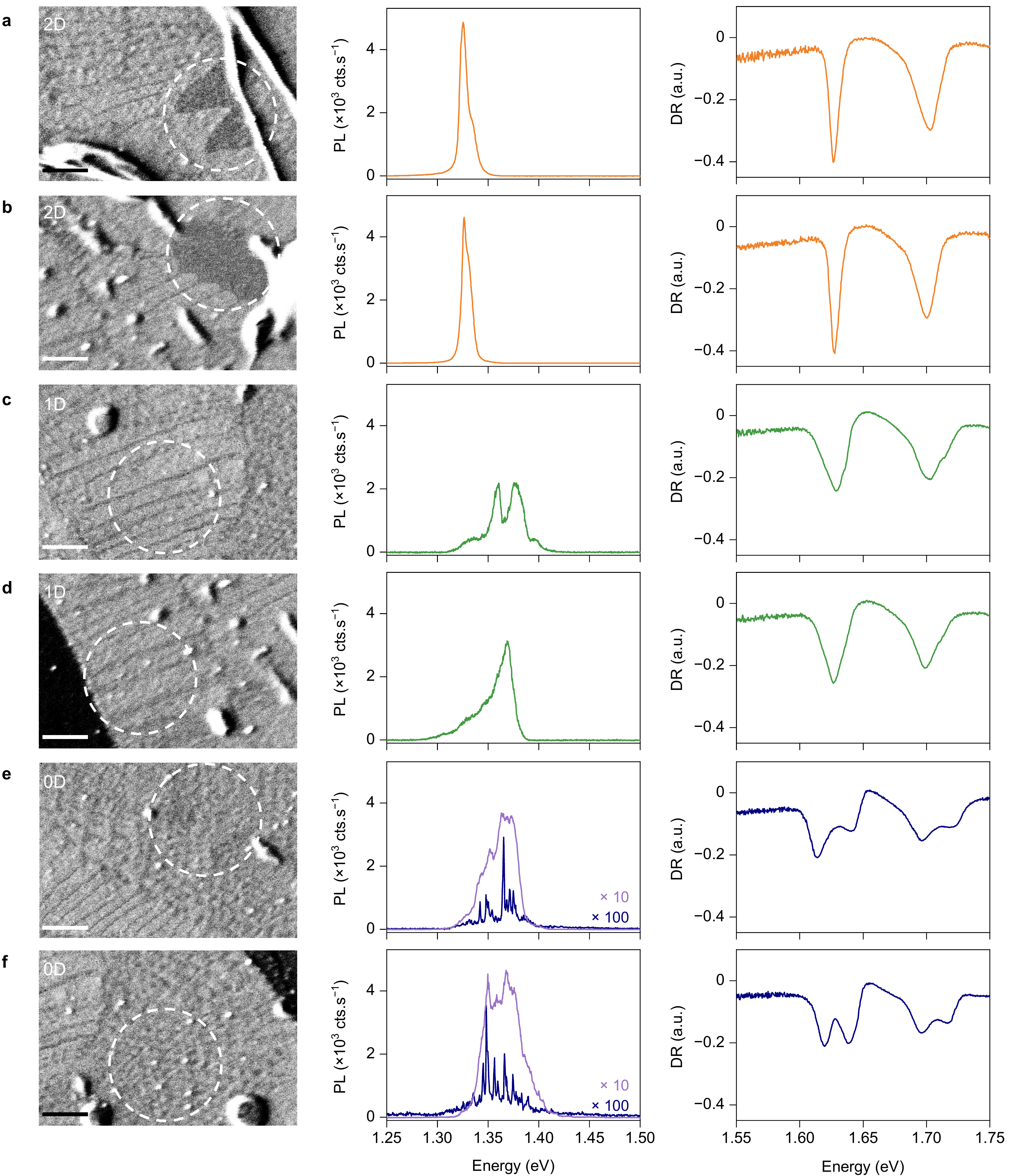}
\vspace{-30pt}
\caption{Examples of reconstruction patterns and the corresponding optical spectra in the R-type sample. \textbf{a} -- \textbf{f}, Left panels show representative areas of 2D (\textbf{a} and \textbf{b}), 1D (\textbf{c} and \textbf{d}) and 0D (\textbf{e} and \textbf{f}) domains identified in SEM imaging (the scale bars are $250$~nm). Central and right panels show the corresponding PL and DR spectra within the optical spot delimited by dashed circles. The PL spectra were excited with $2~\mu$W laser power. For 0D domains, the PL spectra at $0.01~\mu$W are shown (dark purple) with respective scaling factors in addition to the spectra recorded at $2~\mu$W (light purple).}
\label{PL_SEM_PL_DR_R}
\end{figure}

\begin{figure}[!ht]
\vspace{-25pt}
\includegraphics[scale=0.294]{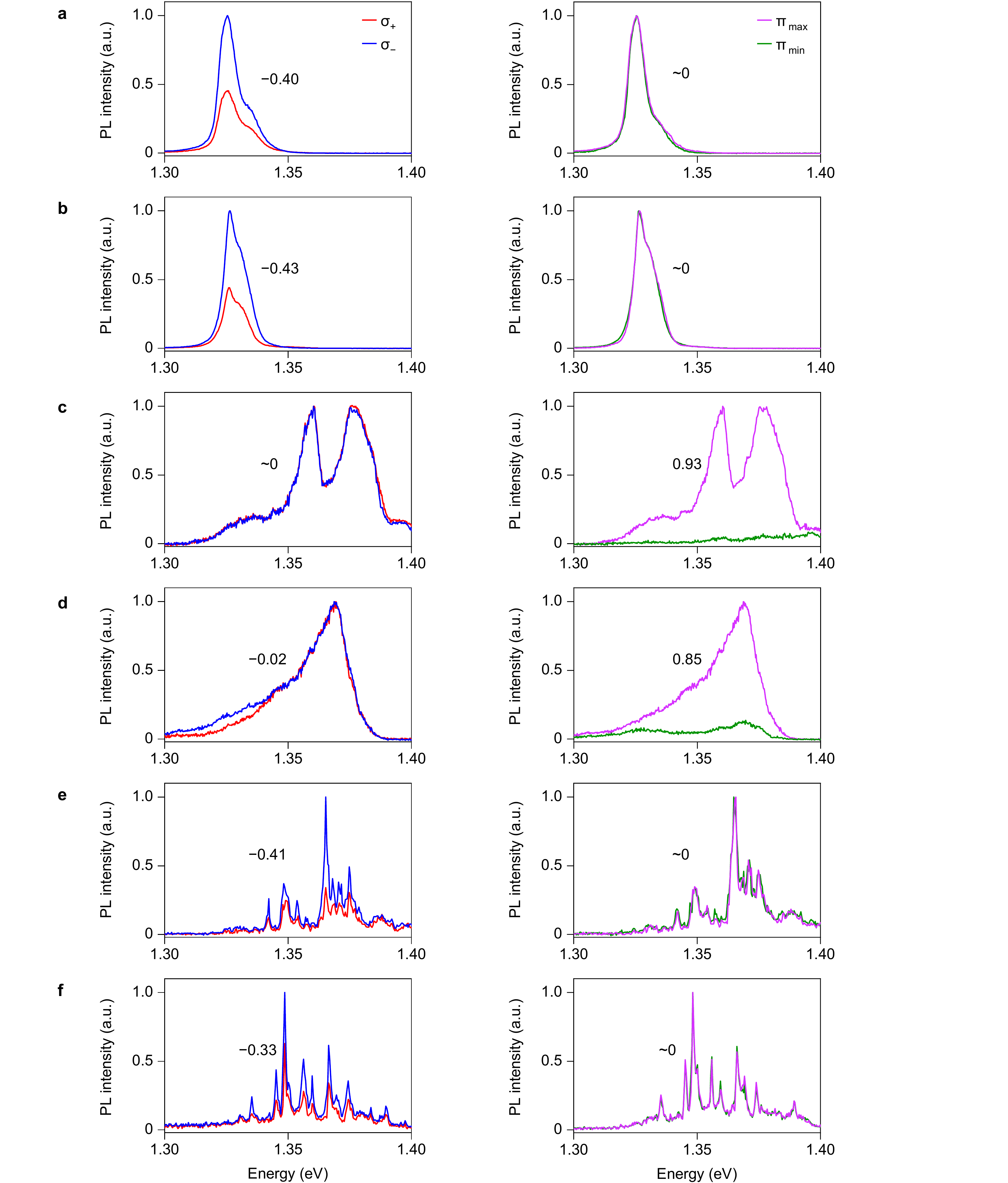}
\vspace{-30pt}
\caption{\textbf{a} -- \textbf{f}, PL spectra recorded in circular (left panels, with $\sigma_+$ and $\sigma_-$ polarized detection shown in red and blue, respectively) and linear (right panels, with two orthogonal orientations of linear polarization for maximum and minimum PL intensity shown in purple and green, respectively) basis from the same regions as in Supplementary Fig.~\ref{PL_SEM_PL_DR_R} \textbf{a} -- \textbf{f}. The numbers in each panel denote the respective degrees of circular and linear polarization, $P_\mathrm{c}$ and $P_\mathrm{\,l}$.} 
\label{PL_SEM_PL_Pc_R}
\end{figure}

\clearpage
\noindent \textbf{Supplementary Note $\mathbf{6}$: Analysis of intralayer exciton absorption spectra in reconstructed domains}

In the following, we analyze the DR spectra of intralayer excitons in Fig.~3a and 4a (main text) recorded on different positions of reconstructed R- and H-stacks. First, we note that the DR spectra exhibit asymmetric lineshapes due to multiple interferences at the interfaces of samples containing stacks of hBN, TMD, SiO$_2$ and Si. To simplify the lineshape analysis, we transform the DR spectra into absorption spectra by adapting the numerical method developed in Ref.~\cite{Back2017}.

We begin with a simple case of two interfaces provided by a thin TMD layer on top of SiO$_2$ in vacuum. Since the thickness of TMD layer $d$ is much smaller than the wavelength of light $\lambda$, DR (defined as $\Delta \mathrm{R}/\mathrm{R}$) can be written as~\cite{McIntyre1971}:
\begin{equation}
\mathrm{DR}=-\frac{8 \pi d }{\lambda} \operatorname{Im}\left(\frac{\epsilon_{1}-\epsilon_{2}}{\epsilon_{1}-\epsilon_{3}}\right),
\label{eq-DR}
\end{equation} 
where $\epsilon_{1,2,3}$ is the dielectric function of vacuum, TMD layer and SiO$_2$. In linear response approximation, $\epsilon_{1,2,3} = 1 + \chi_{1,2,3}$ with dielectric susceptibility of each layer $\chi_{1,2,3}$. With vanishing absorption in vacuum and SiO$_2$ at the wavelengths of relevance in the spectral window of intralayer exciton transitions in MoSe$_2$ and WSe$_2$, $\operatorname{Im}(\epsilon_{1}) =\operatorname{Im}(\epsilon_{3})=0$, and thus DR in Eq.~\ref{eq-DR} is simply proportional to the imaginary part of the optical susceptibility $\chi_2''$ of the thin TMD layer, which in turn corresponds to its absorption. 

\begin{figure}[!ht]
\includegraphics[scale=0.6]{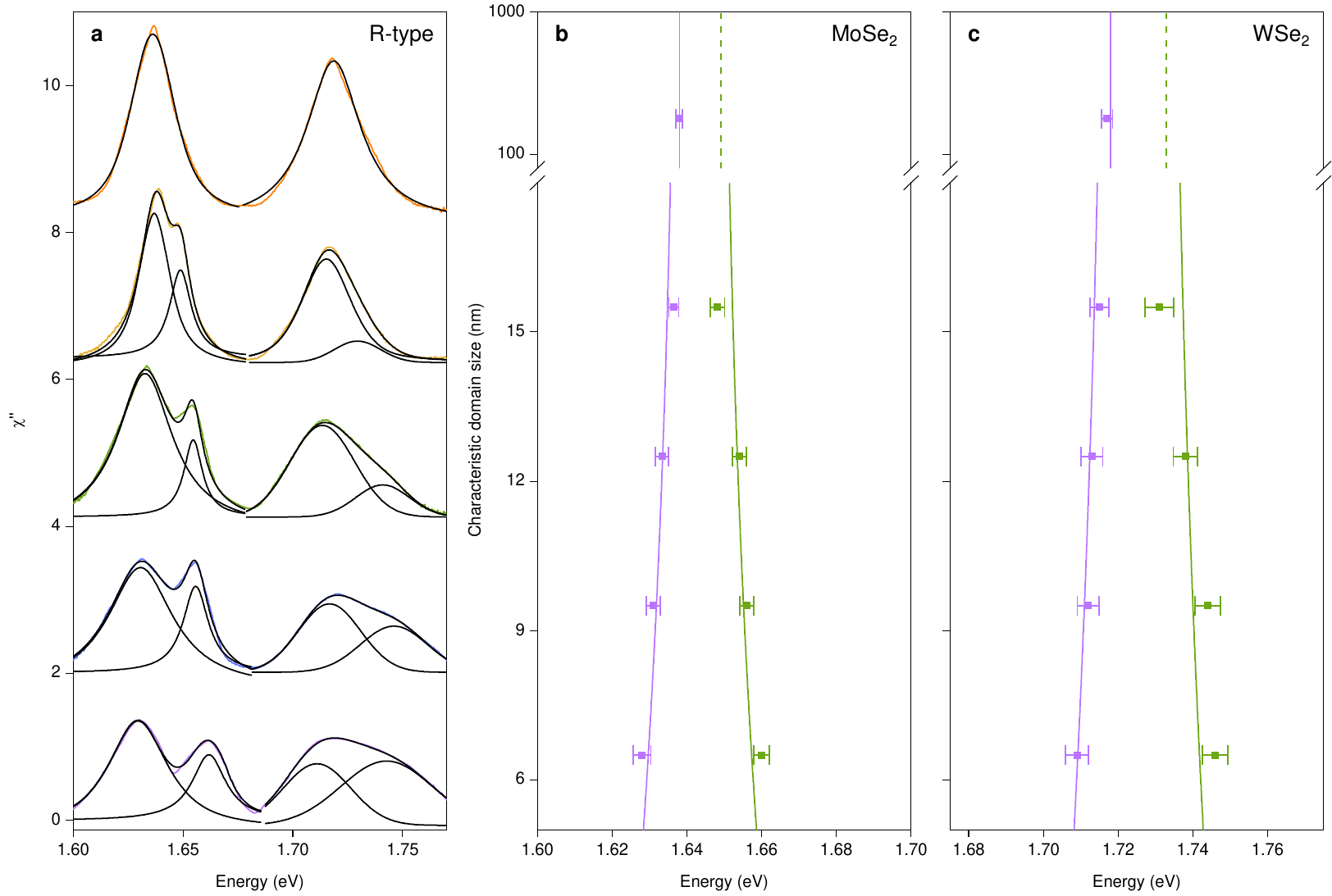}
\caption{\textbf{a}, Absorption $\chi''$ spectra (colored lines) of MoSe$_2$ and WSe$_2$ intralayer excitons upon gradual displacement from a bright to a dark region in R-type HBL (shown from top to bottom), obtained from the DR spectra of Fig.~3a of the main text and fitted to multiple peaks with a Voigt profile (black lines). The top-most spectrum exhibits single-peak resonances in large 2D domains; the spectra below feature two-peak resonances with a splitting sensitive to the size of reconstructed domains. \textbf{b} and \textbf{c}, Model analysis (solid lines) of the peak splitting in the intralayer exciton transition doublets of MoSe$_2$ and WSe$_2$, respectively, as a function of the characteristic size of reconstructed domains. The data points (dots) and their error bars are extracted from the multiple-peak fit of the corresponding spectrum in \textbf{a}.}
\label{fig-DR-R}
\end{figure}

\begin{figure}[!ht]
\includegraphics[scale=0.6]{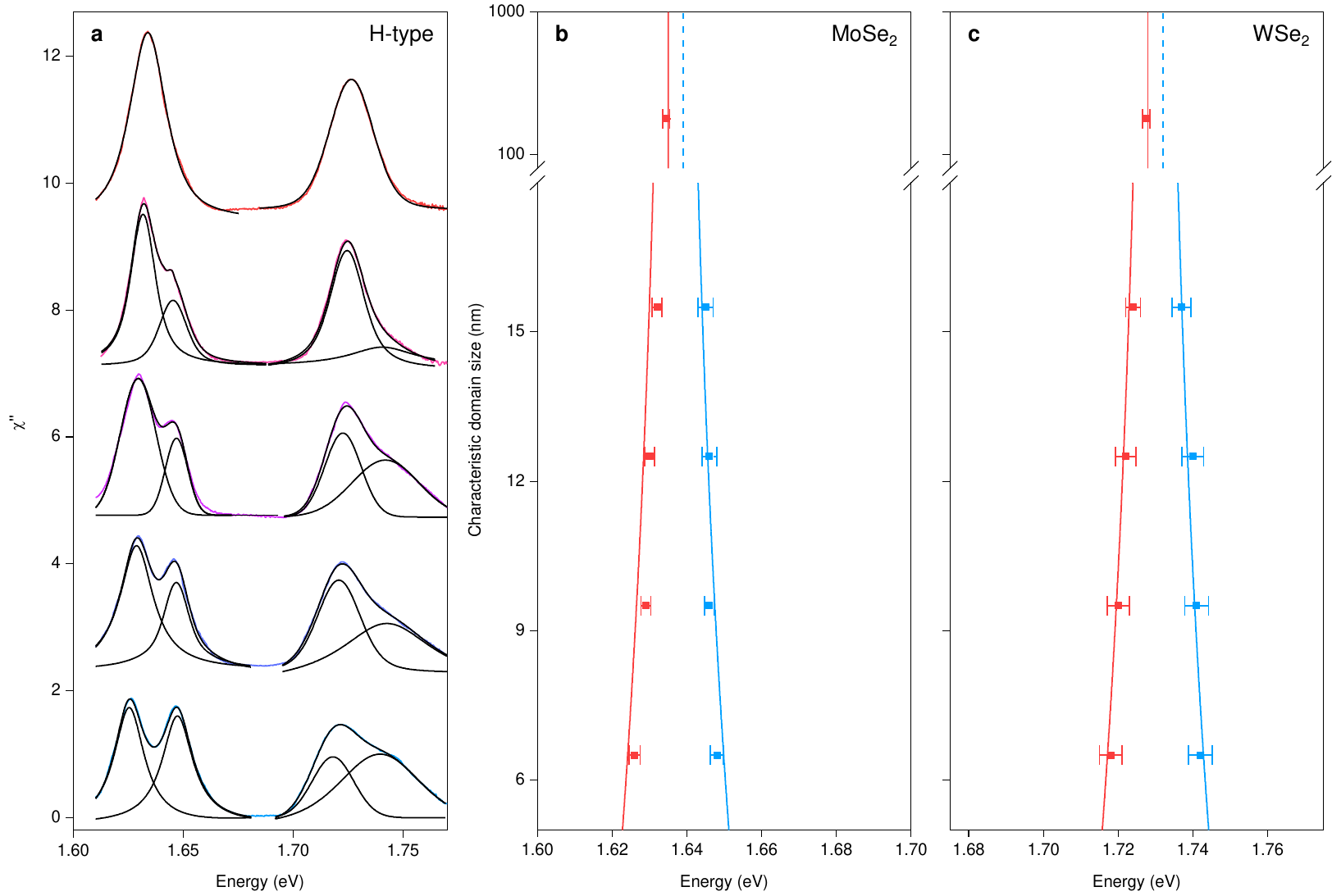}
\caption{\textbf{a}, Absorption $\chi''$ spectra (colored lines) of MoSe$_2$ and WSe$_2$ intralayer excitons upon gradual displacement from a bright to a dark region in H-type HBL (shown from top to bottom), obtained from the DR spectra of Fig.~4a of the main text and fitted to multiple peaks with a Voigt profile (black lines). The top-most spectrum exhibits single-peak resonances in large 2D domains; the spectra below feature two-peak resonances with a splitting sensitive to the size of reconstructed domains. \textbf{b} and \textbf{c}, Model analysis (solid lines) of the peak splitting in the intralayer exciton transition doublets of MoSe$_2$ and WSe$_2$, respectively, as a function of the characteristic size of reconstructed domains. The data points (dots) and their error bars are extracted from the multiple-peak fit of the corresponding spectrum in \textbf{a}.}
\label{fig-DR-H}
\end{figure}

To model the DR response of our samples with two hBN layers that embed the TMD layer on the top and bottom sides and $\sim 500~\mu$m Si substrate underneath the SiO$_2$ layer, we account for additional interfaces between the top hBN layer and the TMD layer, the TMD layer and the bottom hBN layer, its interface to SiO$_2$, and the SiO$_2$-Si interface by a phase factor $e^{i\alpha}$~\cite{Arora2013} in the effective susceptibility $\tilde{\chi}_{2} = e^{-i\alpha_0}\chi_2$, where we approximate $\alpha$ by a constant $\alpha_0$ in the relevant wavelength range. 
By decomposing $\tilde{\chi}_{2}$ into real and imaginary parts as $\tilde{\chi}_{2}' + i\tilde{\chi}_{2}''$, the imaginary part of the effective susceptibility $\tilde{\chi}_{2}''$ can thus be expressed as $\chi_{2}'' =\cos(\alpha_0)\tilde{\chi}_{2}'' + \sin(\alpha_0)\tilde{\chi}_{2}'$. Finally, as DR is proportional to $\tilde{\chi}_{2}''$, using Kramers-Kronig relations we obtain:
\begin{eqnarray}
\nonumber
\chi_{2}''(\omega) = \cos \left(\alpha_{0}\right) & \tilde{\chi}_{2}^{\prime \prime}(\omega)+\sin \left(\alpha_{0}\right) \frac{2}{\pi} \mathcal{P} \int_{0}^{\infty} \frac{\omega^{\prime} \tilde{\chi}_{2}^{\prime\prime}\left(\omega^{\prime}\right)}{\omega^{\prime 2}-\omega^{2}} d \omega^{\prime} \\ 
\propto \cos \left(\alpha_{0}\right) & \mathrm{DR} (\omega)+\sin \left(\alpha_{0}\right) \frac{2}{\pi} \mathcal{P} \int_{0}^{\infty} \frac{\omega^{\prime} \mathrm{DR}(\omega^{\prime})}{\omega^{\prime 2}-\omega^{2}} d \omega^{\prime},
\label{eq-KK}
\end{eqnarray}
where $\omega$ denotes the angular frequency. Using Eq.~\ref{eq-KK}, we compute $\chi_{2}''(\omega)$ of the exciton transitions in the regions of MoSe$_2$ and WSe$_2$ monolayers from the respective DR spectra in the charge-neutral regime for different values of the phase shift $\alpha_0$ from $-\pi$ to $\pi$. Since the exciton absorption of monolayer TMD is expected to have a Lorentzian lineshape in the absence of free charge carriers~\cite{Scuri2018}, we choose the value of $\alpha_0$ such that the obtained absorption spectra $\chi_{2}''(\omega)$ can be accurately fitted by a Lorentzian. Successively, we use this value of $\alpha_0$ to compute the absorption spectra of intralayer excitons in HBL regions as the transition energies are similar to monolayer exciton transitions. The respectively obtained absorption spectra $\chi_2''(\omega)$ in Supplementary Fig.~\ref{fig-DR-R} and \ref{fig-DR-H} are predominantly positive, providing confidence in our approach. 

In the following analysis of absorption spectra, we assume intralayer exciton transition energies are sensitive to the local stacking and domain size. In regions of two proximal stackings A and B ($R_h^X$ and $R_h^M$, and $H_h^h$ and $H_h^X$ in the case of 0D reconstructed domains of R- and H-stacks, respectively) with the corresponding energy minima, coupling of the intralayer exciton states via tunneling gives rise to a doublet peak structure. By introducing tunnel-hopping between the domains of different registries, we model the eigenenergies of the coupled two-level system with the following Hamiltonian:
\begin{equation}
H=  \begin{pmatrix}
    E_A & t e^{-\beta a_m} \\
    t e^{-\beta a_m} & E_B
  \end{pmatrix},
\end{equation}
where $E_A$ and $E_B$ are the intralayer exciton energies for the respective commensurate stackings, $t$ and $\beta$ are the hopping parameters, and $a_m$ is the domain size (or moir\'e period). Using $t=20$~meV and $\beta=0.07$~nm$^{-1}$ estimated from the calculations of Yu et al. \cite{Yu2017}, we compute the energies (in meV) in Supplementary Tab.~\ref{fit-abs} by fitting the peak evolution in Supplementary Fig.~\ref{fig-DR-R} and \ref{fig-DR-H}.

\vspace{10pt}
\begin{table}[!h]
   \begin{ruledtabular}
        \begin{tabular}{ccccc}
            & R, MoSe$_2$  & R, WSe$_2$  & H, MoSe$_2$  & H, WSe$_2$ \\
            \hline
            $E_A$ & 1638 & 1718 & 1635 & 1728 \\
            $E_B$ & 1649 & 1733 & 1639 & 1732 \\
        \end{tabular}
        \end{ruledtabular}
\caption{Energies (in meV) of intralayer excitons derived from absorption spectra analysis in the framework of a tunnel-coupled two-level system. $E_A$ and $E_B$ refer to the interlayer excitons energies in $R_h^X$ and $R_h^M$ as well as in $H_h^h$ and $H_h^X$ registries of R- and H-stacks, respectively.}
    \label{fit-abs}
\end{table}


\noindent \textbf{Supplementary Note $\mathbf{7}$: Density functional theory calculations}

We used density functional theory (DFT) to calculate exciton $g$-factors and oscillator strengths of interlayer excitons in R- and H-type MoSe$_2$-WSe$_2$ HBLs as described in detail in Refs.~\cite{Foerste2020,Foerg2021}. DFT calculations were performed with the PBEsol exchange-correlation functional \cite{Csonka_2009_PRB_79_155107} as implemented in the Vienna ab- initio simulation package (VASP) \cite{Kresse1996}. Van der Waals interactions were included with the DFT-D3 method \cite{Grimme_2010_JCP_132_154104} and Becke-Johnson damping \cite{Grimme_2011_JCC_32_1456}. Moreover, spin-orbit interactions were included at all stages. Elementary cells with thickness of 35~\AA{} in the $z$-direction were used to minimize the interactions between periodic images. The atomic positions were relaxed with a cutoff energy of $400$~eV until the total energy change was less than $10^{-6}$~eV. Calculations were performed for high-symmetry points of HBL moir\'e patterns in R- and H-type stackings on the $\Gamma$-centered $\mathbf k$ grid of $6 \times 6$ divisions with 600 bands and the cutoff energy of $300$~eV.

\begin{figure}[t]
\includegraphics[scale=1.27]{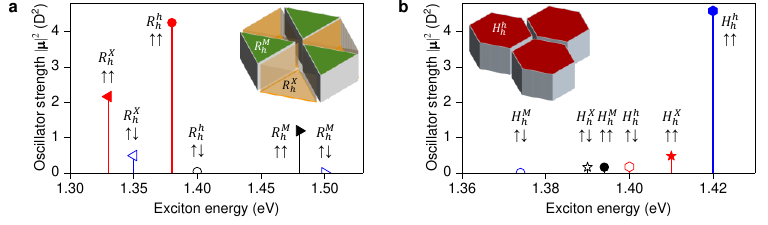}
\caption{\textbf{a} and \textbf{b}, Oscillator strengths of interlayer exctons in R- and H-type MoSe$_2$-WSe$_2$ HBLs. The dipolar selection rules are represented by colors (red: $\sigma_+$; blue: $\sigma_-$; black: $z$-polarized in-plane emission), filled and empty symbols indicate the corresponding singlet and triplet exciton states. The inset schematics show the energy landscape in the presence of periodic reconstruction, with excitons in triangular $R_h^X$ domains confined by the potential walls of $R_h^M$ domains, and hexagonal $H_h^h$ domains with highest-energy excitons.}
\label{IX_energy}
\end{figure}

The theoretical $g$-factors and oscillator strengths of interlayer excitons obtained from our DFT calculations for R- and H-type MoSe$_2$-WSe$_2$ HBLs are listed in Table.~1 of the main text. For completeness, we list in Supplementary Table~\ref{gfactor2} interlayer exciton $g$-factors in different stackings and momentum-dark spin-valley configurations. Supplementary Fig.~\ref{IX_energy} gives a graphical representation of the energetic ordering of singlet and triplet interlayer exciton states with respective dipolar selection rules and oscillator strengths, labelled by the according spin configuration and atomic registry of high symmetry stackings in R- and H-type HBLs. The insets of Supplementary Fig.~\ref{IX_energy} a and b show the exciton energy landscape in R- and H-type stacking in the presence of periodic reconstruction, with excitons in triangular $R_h^X$ domains confined by the surrounding potential walls of $R_h^M$ domains, and domains of highest-energy $H_h^h$ excitons in hexagonal tiling. Note that in such potential landscapes, the energy of $R_h^X$ excitons will increase in nanosized domains due to quantum confinement, whereas a decrease is expected for $H_h^h$ excitons due to decreasing potential energy in the surrounding domains. This distinction explains both the blue-shift of quantum dot type peaks for 0D arrays with respect to the $R_h^X$ singlet emission in R-type 2D domains and the red-shift of $H_h^h$ quantum dot features below the emission peak of H-type 2D domains.

\begin{table*}[!hb]
\begin{ruledtabular}
\begin{tabular}{llRRRRRR}
& \multicolumn{3}{r}{R-type} & \multicolumn{3}{r}{H-type}\\
            \cline{3-5} \cline{6-8}
Exciton &  Spin configuration & \text{$R_h^X$} & \text{$R_h^h$} & \text{$R_h^M$} & \text{$H_h^M$} & \text{$H_h^X$} & \text{$H_h^h$} \\ 
            \hline 
            $K'K$ ($KK$) &  singlet&    13.1    &13.0   &13.0   &6.5    &6.4    &5.7\\
            $K'K$ ($KK$) &  triplet&    17.8    &17.6   &17.6   &11.1   &11.0   &10.4\\
            $QK$  & singlet&    8.6     &9.0    &8.7    &10.4   &10.0   &9.6\\
            $QK$  & triplet&    13.0    &13.3   &12.9   &13.9   &13.6   &13.2\\
            $Q'K$ & singlet&    10.6    &10.7   &11.0   &10.3   &10.5   &10.0\\
            $Q'K$ & triplet&    14.9    &15.0   &15.3   &13.8   &14.2   &13.7\\
            $K\Gamma$  &    singlet&    4.0     &3.6    &3.7    &3.0    &3.1    &3.3\\
            $K\Gamma$  &    triplet&    0.7     &1.0    &1.0    &7.6    &7.6    &7.9\\
            $K'\Gamma$ &    singlet&    3.3     &3.0    &3.0    &3.6    &3.6    &3.9\\
            $K'\Gamma$ &    triplet&    8.0     &7.6    &7.7    &1.0    &0.9    &0.7\\
            $Q\Gamma$  &    singlet&    1.1     &1.0    &1.3    &0.3    &0.1    &0.0\\
            $Q\Gamma$   &   triplet&    3.2     &3.3    &2.9    &3.8    &3.5    &3.6\\
            $Q'\Gamma$  &   singlet&    0.8     &0.7    &1.1    &0.2    &0.5    &0.4\\
            $Q'\Gamma$  &   triplet&    5.1     &5.0    &5.3    &3.7    &4.1    &4.0  
\end{tabular}
\caption{Land\'e $g$-factor values of momentum-indirect interlayer excitons in R- and H-type MoSe$_2$-WSe$_2$ HBL calculated from DFT. We restrict the table to lowest-energy interlayer exciton states formed between conduction band electrons in $K$, $K'$, $Q$ or $Q'$ valleys of MoSe$_2$ and valence band holes at $K$ in WSe$_2$ or at $\Gamma$ in the hybrid band of MoSe$_2$-WSe$_2$ according to the electronic band structure~\cite{Gillen2018}. For each state, the absolute $g$-factor values of both singlet and triplet spin configurations are listed for distinct atomic registries. Note that due to different symmetry in R- and H-type MoSe$_2$-WSe$_2$ HBLs, the spin-valley configurations are distinct~\cite{Forg2019}: In R-type (H-type), $KK$ ($K'K$) interlayer exciton states are momentum-direct with lowest energy states in singlet (triplet) configuration, whereas $K'K$ ($KK$) states are momentum-indirect with triplet (singlet) lowest-energy states.}
\label{gfactor2}
\end{ruledtabular}
\end{table*}

%